\documentclass[a4paper,fleqn,usenatbib]{mnras}

\usepackage[T1]{fontenc}
\usepackage{ae,aecompl}

\usepackage{graphicx}	
\usepackage{amsmath}	
\usepackage{amssymb}	
\usepackage{relsize}

\newcommand{\msun}{\hbox{M$_{\odot}$}}
\newcommand{\kms}{\hbox{km s$^{-1}$}}
\newcommand{\re}{\hbox{${\rm R}_{\rm e}$}}

\title[Timing the formation of ETGs]{Timing the formation and assembly of early-type galaxies via spatially resolved stellar populations analysis}

\author[I. Mart\'in-Navarro]{Ignacio Mart\'in-Navarro$^{1,2}$\thanks{E-mail: imartinn@ucsc.edu}, Alexandre Vazdekis$^{3,4}$,  Jes\'us Falc\'on-Barroso$^{3,4}$, \newauthor Francesco La Barbera$^{5}$, Ak{\i}n Y{\i}ld{\i}r{\i}m$^{6}$, and Glenn van de Ven$^{2,7}$
\\
$^{1}$University of California Observatories, 1156 High Street, Santa Cruz, CA 95064, USA\\
$^{2}$Max-Planck Institut f\"{u}r Astronomie, Konigstuhl 17, D-69117 Heidelberg, Germany\\
$^{3}$Instituto de Astrof\'isica de Canarias, E-38200 La Laguna, Tenerife, Spain\\
$^{4}$Departamento de Astrof\'isica, Universidad de La Laguna, E-38205 La Laguna, Tenerife, Spain\\
$^{5}$INAF - Osservatorio Astronomico di Capodimonte, Napoli, Italy\\
$^{6}$Max-Planck Institut f\"{u}r Astrophysics, Karl-Schwarzschild-Str. 1, 85741 Garching, Germany\\
$^{7}$European Southern Observatory, Karl-Schwarzschild-Str. 2, 85748 Garching b. M\"unchen, Germany
}

\date{Accepted XXX. Received YYY; in original form ZZZ}

\pubyear{2017}

\begin{document}
\label{firstpage}
\pagerange{\pageref{firstpage}--\pageref{lastpage}}
\maketitle

\begin{abstract}
To investigate star formation and assembly processes of massive galaxies, we present here a spatially-resolved stellar populations analysis of a sample of 45 elliptical  galaxies (Es) selected from the CALIFA survey. We find rather flat age and [Mg/Fe] radial gradients, weakly dependent on the effective velocity dispersion of the galaxy within half-light radius. However, our analysis shows that metallicity gradients become steeper with increasing galaxy velocity dispersion. In addition, we have homogeneously compared the stellar populations gradients of our sample of Es to a sample of nearby relic galaxies, i.e., local remnants of the high-$z$ population of {\it red nuggets}. This comparison indicates that, first, the cores of present-day massive galaxies were likely formed in gas-rich, rapid star formation events at high redshift ($z\gtrsim2$). This led to radial metallicity variations steeper than observed in the local Universe, and positive [Mg/Fe] gradients. Second, our analysis also suggests that a later sequence of minor dry mergers, populating the outskirts of early-type galaxies (ETGs), flattened the pristine [Mg/Fe] and metallicity gradients. Finally, we find a tight age-[Mg/Fe] relation, supporting that the duration of the star formation is the main driver of the [Mg/Fe] enhancement in massive ETGs. However, the  star formation time-scale alone is not able to fully explain our [Mg/Fe] measurements. Interestingly, our results match the expected effect that a variable stellar initial mass function would have on the [Mg/Fe] ratio.
\end{abstract}



\section{Introduction} \label{sec:intro}

The hierarchical assembly of galaxies is a direct consequence of the favored $\Lambda$-CDM interpretation of the Universe. Structures within this paradigm grow via gravitational instabilities, in remarkable agreement with the observed large scale structure of the Universe \citep[e.g.][]{Geller89,Hernquist96,Dave99,Colless01,Cole05,Springel05,Springel06}. Given the initial conditions provided by the cosmic microwave background \citep{Bernardis00,wmap13,Planck14}, the later evolution of structures is mostly driven by the dark matter gravitational field \citep{White76,Frenk88,Bond96}, whose dynamics is well described by Newtonian gravity \citep{Adamek13}.

Understanding the formation and evolution of galaxies, however, is a more complex task, as cosmological effects are generally overshadowed by baryonic processes. The evolution of the color magnitude relation \citep{Bower92,Ellis97,vdk00,vdk01,Blakeslee03,Bernardi05} or the tight scaling relations between stellar populations properties and galaxy mass \citep{Faber73,Peletier,Worthey92,Bender,Trager00,Thomas05} seem to question the hierarchical nature of the  $\Lambda$-CDM Universe. Ultimately, these difficulties to reconcile simulations and observations of galaxies arise from numerically unresolvable physical scales. Cosmological simulations including coupling between electromagnetic fields and baryonic matter, stellar physics and even parsec-scale star formation are not yet feasible \citep[see e.g.][]{Baugh06}. Thus, all these processes are usually simplified, using analytic recipes tuned up to match observations \citep[e.g.][]{Rees77,White91,Kauffmann93,Somerville99,dL07,dL12,Guo16}.

Arguably, the most paradigmatic example of the tension between cosmological simulations and actual observations is the apparent downsizing of nearby galaxies, which become older, more metal rich and formed more rapidly with increasing galaxy mass \citep{Trager00,Thomas05,Gallazzi05,Gallazzi06,Yamada06,Kuntschner10,McDermid15}. Naively, and supported by the first cosmologically-motivated numerical simulations, one would expect that massive galaxies in a $\Lambda$-CDM Universe form through successive mergers of smaller galaxies. This would lead to young massive galaxies at $z\sim0$, in sharp contrast with observations. To overcome this so called over-cooling problem, feedback from the central super-massive black hole is usually invoked \citep[e.g.][]{dM05,dL06,Vogelsberger14,Schaye15}. 

Among all the observables that hold information about numerically unresolved scales, the overabundance of $\alpha$-elements compared to Iron-peak elements, i.e., the [$\alpha$/Fe] ratio, stands as one of the most relevant. Under the standard interpretation, the [$\alpha$/Fe] ratio is set by the time delay between core-collapse and Type Ia supernovae (SNe) \citep{Tinsley79,Vazdekis96,Thomas99}. The bulk of $\alpha$-elements is produced in short-lived core-collapse SN, whereas the amount of Iron-peak elements accumulates as Type Ia SNe keep polluting the interstellar medium in scales of a few Gyr. Therefore, the longer star formation lasts in a galaxy, the more Iron is produced, and the lower is the observed [$\alpha$/Fe] ratio. The fact that massive nearby galaxies exhibit enhanced [$\alpha$/Fe] has been then interpreted as a consequence of a rapid formation process \citep{Peletier,Worthey92,Thomas05,dlr,cgvd,McDermid15}.

However, whether this standard picture is enough to explain the observed properties of galaxies remains unclear. \citet{Segers16} have recently claimed that feedback from active galactic nuclei solving the over-cooling problem, explains also their enhanced [$\alpha$/Fe] ratios, as star formation is rapidly quenched in massive galaxies. Such scenario seems to be supported by the relation between black holes and star formation \citep{MN16,Terrazas16}, and by the star formation histories of nearby galaxies \citep{dlr,McDermid15}. In contrast, it has also been suggested that a variable stellar initial mass function (IMF) might be necessary to explain the observed [$\alpha$/Fe] dependence on galaxy mass \citep{Arrigoni,Gargiulo15,dL17,Fontanot17}. The impact of a variable IMF in the interpretation of the [$\alpha$/Fe] ratio \citep{MN16b} has to be carefully considered given the empirical relation between IMF slope and galaxy mass \citep{Cenarro03,Treu,Thomas11,Cappellari,cgvd,ferreras,labarbera,Spiniello2013}. Moreover, IMF may also vary in time, as expected from observations, both locally and at high-$z$ \citep{Vazdekis96,Vazdekis97,vdk08,Wang11,weidner:13,Ferreras15,MN15}.

Understanding whether the abundance pattern probes the formation time scale of a galaxy, or whether it is set by the (time-varying) high-mass end of the IMF, needs for detailed stellar populations analysis. The relatively simple star formation histories of early-type galaxies (ETGs) make them ideal benchmark for stellar populations studies. Dominating the high-mass end of the galaxy mass function \citep{Bell03}, ETGs are expected to be among the most strongly AGN-quenched objects \citep{Schawinski07}, exhibiting also the strongest IMF variations \citep{vdk10} and highest [$\alpha$/Fe] ratios \citep{Thomas05}.  

\subsection{Stellar populations gradients}
The aforementioned relation between galaxy mass (or stellar velocity dispersion) and [$\alpha$/Fe] is observationally well established for ETGs \citep{Faber73,Peletier,Worthey92,Trager00,Thomas05,Graves09,Graves09b,cgvd,Walcher15}. However,these studies were focused on the global properties of ETGs, i.e., comparing quantities integrated over a fixed radial aperture. A few studies have taken a step forward analyzing the radial [$\alpha$/Fe] gradients in different ETG samples. \citet{Mehlert03} analyzed a sample of 35 ETGs in the Coma cluster and found that, on average, [$\alpha$/Fe] gradients were flat. This was latter confirmed by \citet{SB07}, who studied the radial stellar populations gradients of 11 ETGs covering a wide range of masses, although individual galaxies could show slightly positive or negative gradients. Because of the absence of strong positive [$\alpha$/Fe] gradients, \citet{SB07} claimed that the properties of stellar populations in ETGs are not compatible a purely outside-in collapse \citep{Pipino04,Pipino06}. Conversely, \citet{SB07} suggested that the correlation between structural parameters (in particular a$_4$, which quantifies isophotal deviations from a perfect ellipse) and stellar populations gradients was a signature of merger driven formation scenario for ETGs. Spatially unresolved studies also show clear relations between galaxy structure and stellar populations properties \citep{Vazdekis04}. Similar trends as those described by \citet{SB07} were also found by \citet{Kuntschner10} studying 48 SAURON ETGs, and latter revisited by \citet{McDermid15} with the larger ATLAS$^\mathrm{3D}$ sample. \citet{Kuntschner10} proposed that the observed relation between galaxy mass and [$\alpha$/Fe] is due to metal-rich, young and mildly [$\alpha$/Fe] enhanced stellar populations, which become more prominent in low-mass ETGs. In agreement with a time-scale dependent origin for the abundance pattern, these young and no [$\alpha$/Fe] enhanced stellar populations can also be seen in extreme kinematically decoupled components \citep{Kleineberg11}, although it is not a common feature \citep{Kuntschner10}.

In this paper we observationally address the origin of the stellar populations gradients within ETGs by studying the two-dimensional age, metallicity and [Mg/Fe] maps of 45 ETGs drawn from the CALIFA sample \citep{califa,mother}. In addition, we trace back the evolution of the stellar population gradients by comparing our CALIFA galaxies to a sample of massive relic galaxies \citep{akin17}, discussing the implications for the evolutionary paths of ETGs. The outline of this paper is as follows. Data and sample are presented in \S~\ref{sec:data}. The ingredients of the stellar population modeling are presented in \S~\ref{sec:miles}, and the analysis is detailed in \S~\ref{sec:ana}. The results presented in \S~\ref{sec:resu} are discussed in \S~\ref{sec:blabla}. Finally, the conclusions of the paper are summarized in \S~\ref{sec:fin}.

\section{Data and sample selection} \label{sec:data}

Our analysis is based on the Calar Alto Legacy Integral Field Area (CALIFA) survey \citep{califa}, that was intended to observe a diameter-selected sample of nearby galaxies with the PMAS/PPak integral-field spectrograph \citep{Verheijen04,Roth05,Kelz06}. Due to its wider wavelength coverage, from 3700 to 7000 \AA \ (rest-frame), our analysis uses the V500 CALIFA setup, which provides an intermediate spectral resolution of 6 \AA. The hexagonal CALIFA field of view covers $\sim 1.3$ arcmin$^2$, with 1 arcsec$^2$ spaxel size. The exposure time was fixed to 900 sec for the V500 setup, and allows us to typically reach up to 1 \re \ (see \S~\ref{sec:ana}).

Our preliminary sample encompassed every ETG both from the CALIFA mother sample \citep{mother}, and from the extended CALIFA sample presented in \citet{sanchez16}. S0 were excluded from the analysis since they may represent an intermediate galaxy type, whose gaseous and stellar population properties are distinct from elliptical galaxies (Es) \citep[e.g.][]{Temi09,sarzi10,Sarzi13,Kuntschner10,Amblard14}. However, some confusion between Es and S0s is expected for the latests morphological types in our sample. In \S~\ref{sec:mgrad} we discuss the impact of morphological subtypes on our results.

We found a strong systematic artifact at a rest-frame wavelength of $\lambda \sim 5460$\AA. Since we were interested on the radial variations of the stellar population properties, we had to apply a redshift cut to our sample of elliptical galaxies to ensure that our analysis was not affected by this systematic. Therefore, our final sample consisted only of those Es within the CALIFA survey with redshift $z<0.026$, 45 in total. The criterion for this redshift limit, based on the details of our stellar populations analysis, is further explained in \S~\ref{sec:pop}

\begin{figure}
\begin{center}
\includegraphics[width=7.5cm]{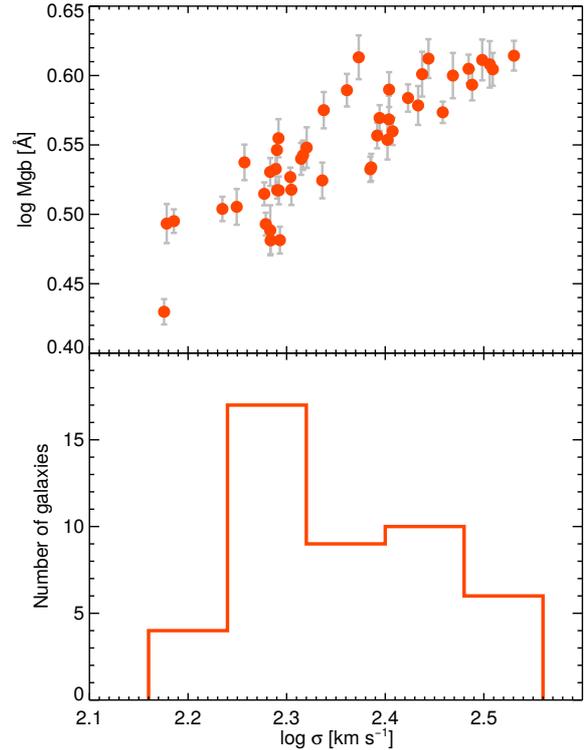}
\end{center}
\caption{Mgb -- $\sigma$ relation. The top panel shows the measured Mgb line-strength (convolved to a common $\sigma=350$) as a function of the effective velocity dispersion at 0.5 \re \ for our sample of galaxies. The bottom panel shows the $\sigma$ distribution for the 45 objects in the sample. The median value is $\sigma = 217$ \kms, or conversely, $9.8\times10^{10}$ \msun    \ \citep{mother}.}
\label{fig:sample}
\end{figure}

Effective velocity dispersions in our sample, measured within half \re, range from $\sigma \sim 150$ to $\sigma \sim 350$ \kms, with a median value of $\sigma = 217$ \kms. According to \citet{mother}, this translates to a median stellar mass of $\sim10^{11}$ \msun, mass range where the CALIFA sample is representative of the local population of galaxies. In Fig.~\ref{fig:sample} we show the classical relation between Mgb line-strength and stellar velocity dispersion for our 45 Es. The basic properties of the sample are listed in Table~\ref{tab:sample}.

For comparison, we also re-analyze the sample of massive compact relic galaxies presented in \citet{akin17}. These objects were also observed using the PMAS/PPak integral-field spectrograph, with the same instrumental setup as the V500 CALIFA data. This sample of nearby {\it red nuggets} (see \S~\ref{sec:relics}), consists of 14 galaxies with a typical velocity dispersion of $\sigma\sim275$\kms. Their individual kinematical and stellar populations properties are fully described in \citet{akin17}.

\begin{table*}
\centering
\begin{center}\begin{tabular}{c c c c c c}

\hline
Galaxy & Morphology & M$_\star$      & $\sigma_{0.5\mathrm{R_e}}$ & \re       & $\epsilon$ \\
       &            & [$\log$ \msun] & [km s$^{-1}$]              & [arcsec]  &            \\
\hline 
NGC\,0155    &  E1     &  10.8   &    177.5     &       15.8            &        0.14         \\
NGC\,0364    &  E7     &  10.8   &    229.7     &       15.8            &        0.27         \\
NGC\,0499    &  E5     &  11.1   &    277.9     &       21.3            &        0.32         \\
NGC\,0529    &  E4     &  11.0   &    247.9     &       12.6            &        0.09         \\
NGC\,0731    &  E1     &  10.7   &    171.7     &       11.4            &        0.07         \\
NGC\,0938    &  E3     &  10.7   &    192.2     &       13.8            &        0.25         \\
NGC\,0962    &  E3     &  10.9   &    180.8     &       13.8            &        0.22         \\
NGC\,1026    &  E2     &  10.7   &    206.4     &       17.4            &        0.24         \\
NGC\,1060    &  E3     &  11.4   &    322.9     &       27.3            &        0.18         \\
NGC\,1132    &  E6     &  11.1   &    252.5     &       32.8            &        0.40         \\
NGC\,1349    &  E6     &  11.0   &    209.1     &       17.0            &        0.11         \\
NGC\,1361    &  E5     &  10.6   &    194.5     &       16.2            &        0.32         \\
NGC\,1656    &  E6     &  10.5   &    149.8     &       17.4            &        0.44         \\
NGC\,2513    &  E2     &  11.2   &    307.7     &       26.5            &        0.26         \\
NGC\,2592    &  E4     &  10.5   &    236.1     &       9.9             &        0.21         \\
NGC\,2880    &  E7     &  10.5   &    153.3     &       18.2            &        0.35         \\
NGC\,2918    &  E6     &  11.2   &    243.1     &       12.2            &        0.30         \\
NGC\,3158    &  E3     &  11.5   &    339.3     &       32.4            &        0.18         \\
NGC\,3615    &  E5     &  11.3   &    287.3     &       15.4            &        0.41         \\
NGC\,4874    &  E0     &  11.3   &    273.8     &       55.0            &        0.23         \\
NGC\,5029    &  E6     &  11.3   &    271.2     &       25.3            &        0.39         \\
NGC\,5198    &  E1     &  10.8   &    217.6     &       16.2            &        0.12         \\
NGC\,5216    &  E0     &  10.4   &    150.8     &       20.1            &        0.32         \\
NGC\,5485    &  E5     &  10.7   &    195.0     &       31.6            &        0.32         \\
NGC\,5513    &  E6     &  11.0   &    189.3     &       15.8            &        0.44         \\
NGC\,5546    &  E3     &  11.3   &    320.6     &       17.4            &        0.12         \\
NGC\,5557    &  E4     &  11.3   &    255.5     &       24.5            &        0.20         \\
NGC\,5580    &  E2     &  10.6   &    195.7     &       15.0            &        0.13         \\
NGC\,5598    &  E6     &  10.9   &    192.0     &       9.10            &        0.32         \\
NGC\,5623    &  E7     &  10.7   &    294.2     &       15.4            &        0.32         \\
NGC\,5642    &  E5     &  11.0   &    246.5     &       18.6            &        0.38         \\
NGC\,5684    &  E3     &  10.8   &    192.0     &       17.8            &        0.27         \\
NGC\,5689    &  E6     &  10.9   &    190.0     &       17.0            &        0.75         \\
NGC\,5928    &  E4     &  10.9   &    201.2     &       15.8            &        0.31         \\
NGC\,5966    &  E4     &  10.8   &    201.7     &       18.6            &        0.38         \\
NGC\,6020    &  E4     &  10.7   &    207.3     &       19.0            &        0.30         \\
NGC\,6021    &  E5     &  10.9   &    216.8     &       9.50            &        0.27         \\
NGC\,6125    &  E1     &  11.1   &    253.4     &       21.7            &        0.04         \\
NGC\,6411    &  E4     &  10.9   &    195.1     &       34.0            &        0.35         \\
NGC\,6515    &  E3     &  11.0   &    195.9     &       19.0            &        0.34         \\
NGC\,7550    &  E4     &  11.2   &    242.5     &       24.5            &        0.09         \\
NGC\,7562    &  E4     &  11.2   &    264.9     &       20.9            &        0.32         \\
NGC\,7619    &  E3     &  11.3   &    305.2     &       35.6            &        0.17         \\
UGC\,05771   &  E6     &  11.1   &    253.3     &       12.6            &        0.32         \\
UGC\,10097   &  E5     &  11.3   &    315.0     &       14.6            &        0.17         \\
\hline 
\end{tabular}
\caption{Basic properties of our sample of Es. Morphologies, effective radii and ellipticity ($\epsilon$) are those of \citet{FB17}. Stellar masses were calculated by \citet{mother}, and the effective velocity dispersions at half \re were measured over the same CALIFA cubes used for the stellar populations analysis (see \S~\ref{sec:ana}).}
\label{tab:sample}
\end{center}
\end{table*}

\section{MILES alpha-enhanced models} \label{sec:miles}

To study the stellar population properties in our sample of galaxies, we compared our integral field spectroscopic data to the  [$\alpha$/Fe]-variable MILES stellar population synthesis models \citep{miles,alpha}. These new set of models were built by combining the theoretical response functions of \citet{Coelho05,Coelho07} with the MILES stellar library \citep{Pat06}, taking into account the [Mg/Fe] determination for the individual MILES stars \citep{Milone11}. The stellar spectra were first used to populate, in a consistent way, solar-scaled and [$\alpha$/Fe]-enhanced BaSTi isochrones \citep{Pietrinferni04,Pietrinferni06}. Latter, the resulting single stellar population models were [$\alpha$/Fe]-corrected using the theoretical response functions. 

The [$\alpha$/Fe]-variable MILES models cover a wide range of ages, from 0.03 to 14 Gyr, as well as nominal metallicities from -2.27 to +0.40 dex. In addition, the models allow for [$\alpha$/Fe] variations from 0.0 to +0.4 dex. Given the weak dependence of the selected spectral features on the IMF slope, we assumed a universal Kroupa-like IMF shape \citep{mw,kroupa}.

Note that all $\alpha$ elements are varied simultaneously in the [$\alpha$/Fe]-variable MILES models, and therefore, the interpretation of the measured abundance pattern should be done carefully. Since our analysis is based on Fe and Mg optical features (see \S~\ref{sec:pop}),  we are only sensitive to the [Mg/Fe] ratio. However, if a different approach is followed (e.g. full spectral fitting) the resulting [$\alpha$/Fe] value would be determined by the average effect of all $\alpha$ elements on the analyzed spectrum. Since elemental abundances vary from galaxy to galaxy, the same {\it average} [$\alpha$/Fe] could be measured in two galaxies with a completely different partition of individual $\alpha$-elements. This is particularly important when studying wavelengths bluewards than $\lambda\sim4500$\AA \ \citep{alpha}.  Similarly, when applying the [$\alpha$/Fe]-variable MILES models to optical features sensitive to different elements such as Mg, Ca or Ti, one could obtain different answers for each line/element.

\section{Analysis} \label{sec:ana}
 
In order to make full use of the CALIFA integral field spectroscopy, we derived two-dimensional maps of all measured quantities. Data cubes were spatially binned using the Voronoi tessellation described in \citet{voronoi} to a common signal-to-noise of 60 per \AA.  

To derive the stellar population properties within our Voronoi maps, we first needed to calculate the radial velocity (V) and $\sigma$ of each spectra along the maps. We made use of the pipeline developed by \citet{FB17}, which is fully detailed in their Section 4. In summary, V and $\sigma$ were derived using the Penalized Pixel-Fitting method (pPXF) \citep{ppxf,ppxf2}, without including higher Gauss-Hermite terms. pPXF was fed with stellar templates from the Indo-US Library \citep{valdes}, carefully selected to minimize template mismatch systematics \citep{FB17}. Once the kinematics of each spectra was measured, we corrected them for nebular emission using the Gas AND Absorption Line Fitting algorithm \citep[GANDALF,][]{sarzi}.

\subsection{Stellar populations}  \label{sec:pop}

We measured the stellar population properties using standard line-strength analysis. We used the \citet{Cervantes} definition of the 
H$\beta_\mathrm{o}$ index, combined with the Fe\,4383 \citep{Worthey94}, Fe\,5015 \citep{Worthey94}, Fe\,5270 \citep{burstein}, and Mgb \citep{burstein} standard line indices. Note that the only $\alpha$-element measurable in CALIFA is Mg. Thus, when analyzing and discussing the abundance pattern in our data we explicitly used [Mg/Fe], but not the more general [$\alpha$/Fe]. The [$\alpha$/Fe]-variable MILES models can in principle be used to constrain other $\alpha$-elements.

As described in \S~\ref{sec:data}, the presence of a systematic feature in the data forced us to apply a redshift cut to our sample. In order to maximize the number of galaxies in the final sample, we decided not to include the Fe\,5335 spectral feature (our reddest metallic indicator) in the analysis (see details in Appendix~\ref{sec:resi}). This allowed us to study all galaxies with redshift $z < 0.026$. For larger redshifts, the systematic started to affect the Fe\,5270 index definition. Notice that we also avoided the use of the strong NaD absorption lines, as it depends, not only on metallicity, but strongly on the IMF slope and on the [Na/Fe] abundance \citep{Spiniello12,LB17}

Our stellar population modeling consisted therefore of three free parameters, namely age, metallicity ([M/H]) and [Mg/Fe],  to be constrained with the five line-strength indices listed above. Stellar population properties were measured using the {\it emcee} Bayesian Markov chain Monte Carlo sampler \citep{emcee}. In practice, each walker tried to maximize the following likelihood function

\begin{equation*}
 \ln ({\bf I}  |  \mathrm{age, [M/H], [Mg/Fe]} ) = -\frac{1}{2} \mathlarger{\sum}_n \bigg[ \frac{(\mathrm{I}_n - \mathrm{M}_n)^2}{\sigma_n^2}-\ln \frac{1}{\sigma_n^2}\bigg]
\end{equation*}

\noindent where I$_n$, M$_n$ and $\sigma_n$ are the observed line-strength, model value, and error of our five indices, respectively. We assumed flat priors for the three free parameters. Model predictions M$_n$ were calculated at the measured velocity dispersion of each spectrum. In such a way, we avoid degrading the resolution of the spectra, making full use of their information. We assume a constant resolution for the CALIFA data, with an instrumental FWHM=6 \AA. Line-strength predictions were obtained by convolving MILES SSP models ($\mathrm{FWHM}_\mathrm{MILES}=2.51$\AA, \citealt{FB11b}) with a gaussian kernel whose FWHM depends on the measured velocity dispersion for that particular radial bin ($\mathrm{FWHM}_\mathrm{gal}$, \S~\ref{sec:ana}):

\begin{equation*}
 \mathrm{FHWM}_\mathrm{ker}^2 = \mathrm{FHWM}_\mathrm{CALIFA}^2+\mathrm{FHWM}_\mathrm{gal}^2-\mathrm{FHWM}_\mathrm{MILES}^2
\end{equation*}

\begin{figure}
\begin{center}
\includegraphics[width=8.7cm]{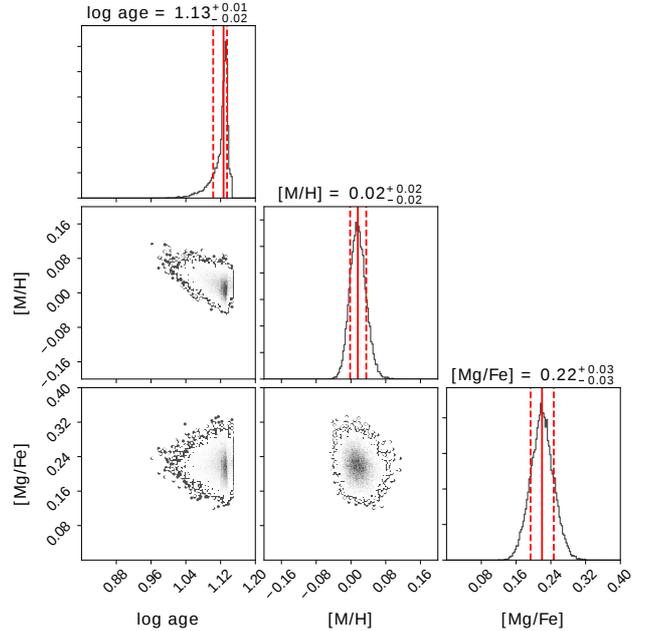}
\end{center}
\caption{Example of a corner plot produced by our stellar populations fitting code, in this case for the galaxy NGC\,7619. The cloud of points indicates the posterior distributions, while red lines mark the best-fitting (solid line) and 1$\sigma$ confidence intervals (dashed lines). We are capable of breaking the degeneracies between the different stellar populations parameters for a typical signal-to-noise of $\sim 60$ per \AA.}
\label{fig:corner}
\end{figure}

A typical corner plot is shown in Fig.~\ref{fig:corner}, where the degeneracies between age, metallicity and [Mg/Fe] are partially broken.  We use the median of the posterior distributions as our best-fitting solutions (red solid lines in Fig.~\ref{fig:corner}), and 1-$\sigma$ uncertainties correspond to the 16th and 84th percentiles.

\section{Results} \label{sec:resu}

After the analysis of the CALIFA data cubes, for each galaxy in our sample we had a map of V, $\sigma$, age, metallicity and [Mg/Fe]. As an example, the case of NGC\,7619 is shown in Fig.~\ref{fig:map}, and the rest of maps can be found in Appendix~\ref{sec:maps}.

\begin{figure*}
\begin{center}
\includegraphics[width=12.5cm]{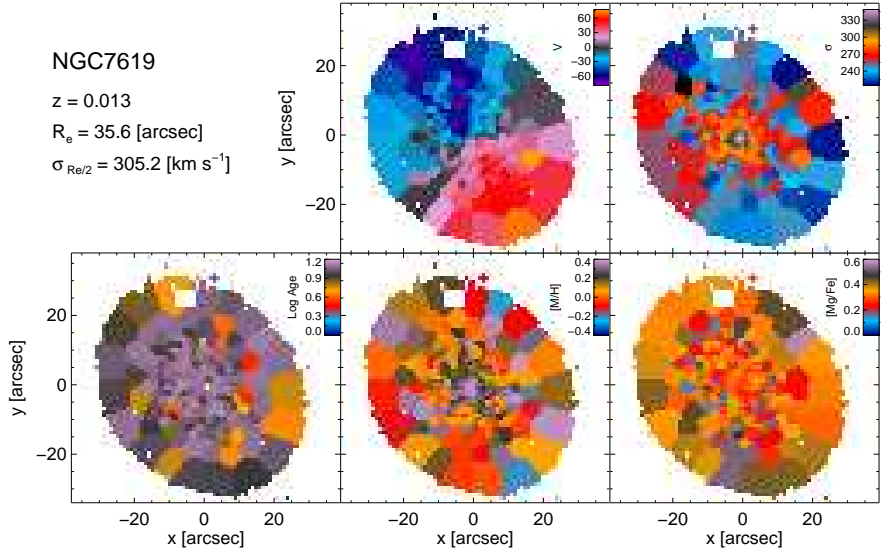}
\end{center}
\caption{Kinematical and stellar population parameters measured for NGC\,7619. From, top to bottom and left to right we show here the velocity, $\sigma$, $\log$ age, [M/H] and [Mg/Fe] maps. The redshift, \re, and effective $\sigma$ are also indicated. There is a clear hint of rotation, as well as a radially declining $\sigma$ profile. Ages are very old through all the galaxy, and the metallicity decreases outwards. The [Mg/Fe] maps shows a mild positive radial gradient.}
\label{fig:map}
\end{figure*}

\begin{figure*}
\begin{center}
\includegraphics[width=13.cm]{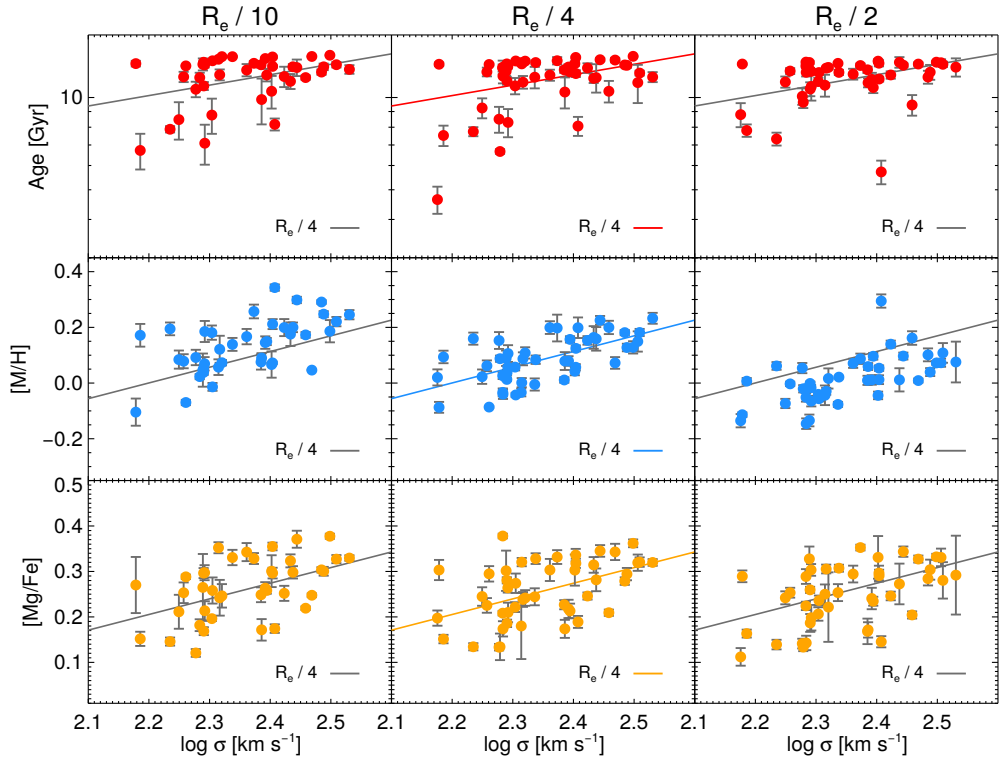}
\end{center}
\caption{Integrated stellar populations trends with velocity dispersion. From left to right, each column correspond to an integrated measurement over one tenth, one fourth and half \re, respectively. As a reference, the solid line indicates the best fitting trend with $\sigma$ as measured at \re/4. Es become more metal rich and more [Mg/Fe]-enhanced with increasing $\sigma$. Our sample of Es exhibit very old ages (red dots), with only a small fraction of younger populations, mostly at low velocity dispersions. The negative [M/H] gradient at all masses is clear from the second row (blue dots). Orange dots indicate the [Mg/Fe] measurements, which do not show any net radial variation.}
\label{fig:global}
\end{figure*}

\subsection{Global scaling relations}\label{sec:sca}

First, we analyzed the relation between integrated stellar population properties and velocity dispersion. In Fig.~\ref{fig:global} we show how age, metallicity and [Mg/Fe] change as a function of the effective $\sigma_{0.5\mathrm{R_e}}$. From left to right, the stellar population properties were measured over an integrated aperture of one tenth, one fourth, and half \re. On each panel, the solid line indicates the best fitting $\sigma$ relation at \re/4. 

As expected, we found old stellar populations at all radii, with the scatter suggesting the presence of small fractions of younger populations. As indicated in the introduction, this apparent {\it downsizing} is robustly supported by numerous observational studies \citep[e.g.][]{Trager00,Thomas05,Pat06,Kuntschner10,McDermid15}, challenging a pure hierarchical assembly of massive ETGs. Residual recent star formation as seen in the inner regions of our sample are needed to explain the scatter of age-sensitive spectral features in the infrared \citep{FB11}, in the optical \citep{Trager00b} and in the ultraviolet \citep{Vazdekis16}.

\citet{Thomas05} suggested that the Mgb-$\sigma$ relation (top panel in Fig.~\ref{fig:sample}) is mainly driven by an increase in metallicity with increasing galaxy $\sigma$. Although the physical origin of this relation is still under debate \citep{Keller16,Bower16}, it is clear from Fig.~\ref{fig:global} that galaxies with higher $\sigma$ exhibit higher metallicities. The distribution of [M/H] measurements clearly moves towards lower values as we probe the most external regions of our Es, indicating a negative metallicity gradients in our sample. This was already noticed by the early studies of \citet{Mehlert03,SB07} and \citet{Spolaor09}, and it is usually understood as an imprint of the early gas collapse, later shaped by mergers. 

Finally, the [Mg/Fe] trends as a function of $\sigma$ and galactocentric distance are also shown in Fig.~\ref{fig:global}. We did find the expected enhancement in [Mg/Fe] with increasing sigma $\sigma$. Interestingly, and contrary to the behavior of metallicity, there is not a clear trend with radius. The [Mg/Fe] enhancement of the inner regions is also observed beyond half \re. This is in agreement with previous studies \citep{Peletier99,Mehlert03,SB07,Kuntschner10}, and suggests a fundamental difference between the development of metallicity and abundance pattern gradients within ETGs.

\subsection{Characteristic gradients}\label{sec:mgrad}

\begin{figure}
\begin{center}
\includegraphics[width=7.5cm]{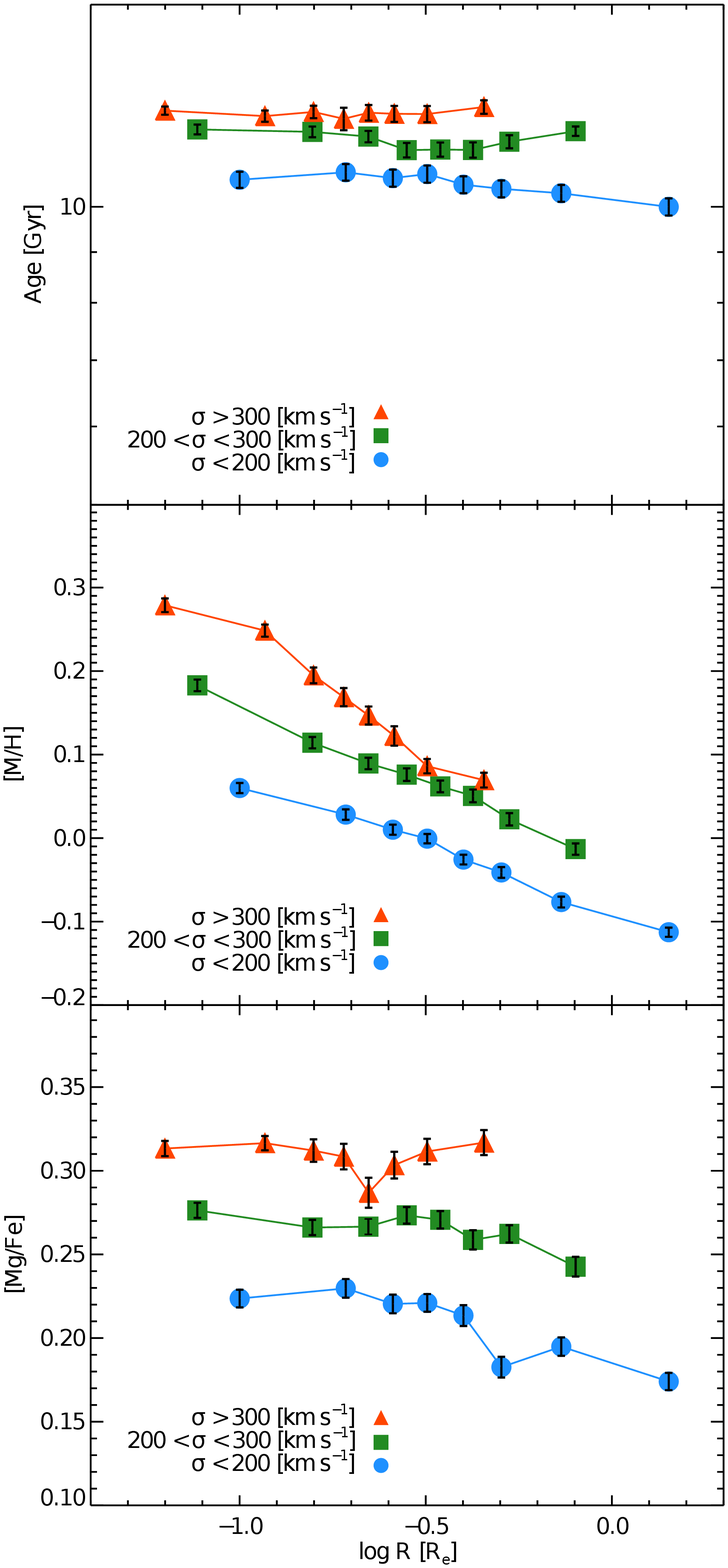}
\end{center}
\caption{Average stellar population gradients as a function of galaxy velocity dispersion. Each radial profile consists of 8 bins, each of them containing the same number of {\it individual} radial measurements. Error bars indicate the uncertainty on the mean value. The average profile of galaxies with  effctive velocity dispersions below 200 \kms are indicated with blue filled circles, green squares correspond to galaxies with 200 < $\sigma$ < 300 \kms, and finally the gradients of the most massive galaxies ($\sigma$>300 \kms) are shown with red filled triangles. From, top to bottom, we show the age, metallicity and [Mg/Fe] average radial gradients. Age and [Mg/Fe] behave similarly, as galaxies become older and more [Mg/Fe]-enhanced with increasing $\sigma$, with rather flat profiles. Galaxies with higher $\sigma$ are also more metal rich, and the slope of the gradient significantly steepens for galaxies with larger velocity dispersions.}
\label{fig:grad}
\end{figure}

Alternatively, we also analyzed the stellar populations properties by looking at the average radial age, metallicity, and [Mg/Fe] gradients in different $\sigma$ bins.  Following Fig.~\ref{fig:sample}, we separated our sample into three bins  according to the central velocity dispersion, and combined all the measurements to obtain the average radial profiles.  These characteristic radial profiles, for a given central velocity dispersion, are shown in Fig.~\ref{fig:grad}.  To quantify the differences among them, in Table~\ref{tab:gradi} we report the ($\chi^2$) best-fitting linear gradients. Note that the fits are based on all individual radial measurements, i.e., using the 2D stellar populations maps normalized to the effective radius (see Fig.~\ref{fig:map}), and not on the average profiles shown in  Fig.~\ref{fig:grad}. 

{\bf -\,Age.} The downsizing mentioned in \S~\ref{sec:sca} is also clear from Fig.~\ref{fig:grad}, where low-$\sigma$ galaxies are younger at all radii than high-$\sigma$ objects. Interestingly, the small fraction of young populations seen in Fig.~\ref{fig:global} is not longer visible in the mean age profile. This shows that the bulk of the stellar mass is dominated by old stars (age $> 10$ Gyr). The slope of the age gradients, although rather small, becomes steeper (i.e., more negative) with decreasing velocity dispersion, suggesting an increasing fraction of younger stellar populations in the outer parts of low-$\sigma$ Es.

{\bf -\,Metallicity.} We found that the characteristic metallicity gradients are negative in the velocity dispersion range probed by our sample, with more massive galaxies being more metal rich. However, in contrast to \citet{SB07} and \citet{Kuntschner10} \citep[but see also][]{Gorgas90,Gonzalez95,Francesco12}, we found that the average metallicity gradient steepens with increasing galaxy stellar velocity dispersion. Hence, galaxies with higher $\sigma$, are not only more metal rich, but also exhibit more pronounced metallicity gradients. This change in the slope of the metallicity gradient is mostly related to the central regions ($\log \mathrm{R}/$\re$<-0.5$). Excluding later morphological types (E6 and E7) does not affect the steepening of the metallicity gradients and thus, confusion between Es and S0s is not likely to drive the $\nabla_\mathrm{[M/H]}$ variation.

\begin{figure}
\begin{center}
\includegraphics[width=8 cm]{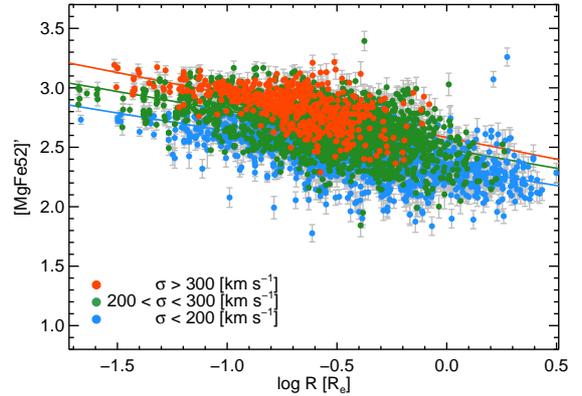}
\end{center}
\caption{[MgFe52]' index gradients color-coded a function of galaxy central velocity dispersion. As in Fig.~\ref{fig:grad}. red, green and blue filled symbols correspond to individual measurements within galaxies with $\sigma$>300 \kms, 200 < $\sigma$ < 300 \kms, and $\sigma$<200 \kms, respectively. The best-fitting relation for each cloud of points is shown as a solid colored line. Being independent of any model assumption, the radial trends of this total metallicity-sensitive line-strength support the steepening of the metallicity gradients with increasing velocity dispersion shown in Fig.~\ref{fig:grad}.}
\label{fig:rindex}
\end{figure}

To assess whether the steepening of the metallicity gradients is an actual feature, or whether it is driven by systematics in the stellar populations analysis, we compared in Fig.~\ref{fig:rindex} the radial variation of the combined index [MgFe52]', as defined in \citet{Kuntschner10}, of our three velocity dispersion bins. This index is almost a pure metallicity indicator, weakly depending on [Mg/Fe], and therefore it is a rather model-independent proxy for metallicity variations. Mimicking the behavior of metallicity gradients shown in Fig~\ref{fig:grad}, the radial variation in the [MgFe52]' index gets steeper with increasing velocity dispersion. Using the same linear fitting scheme as above and considering only the central regions ($\log \mathrm{R}/$\re$<-0.5$), the measured [MgFe52]' gradients equal to $\nabla_\mathrm{[MgFe52]'}=-0.21{\pm0.03}$, $-0.30{\pm0.02}$ and $-0.37{\pm0.03}$ for the low-$\sigma$, intermediate-$\sigma$ and high-$\sigma$ bins, respectively, further supporting that galaxies with higher stellar velocity dispersions exhibit steeper metallicity gradients.

{\bf -\,[Mg/Fe].} The rather flat character of the [Mg/Fe] profiles is clearly shown in Fig.~\ref{fig:grad}, and confirmed by the best-fitting gradients listed in Table~\ref{tab:gradi}, which suggest a mild trend with $\sigma$. Galaxies with higher velocity dispersions have, on average, slightly positive [Mg/Fe] radial gradients, but this mean gradient tends to become more negative for galaxies with lower velocity dispersions.  Notice that, at all radii, high-$\sigma$ galaxies are more [Mg/Fe]-enhanced than low-$\sigma$ objects.

\begin{table*}
\centering
\begin{center}\begin{tabular}{c c c | c c | c c}

\hline
 & Age$_{Re}$  &  $\nabla_\mathrm{age}$  &  [M/H]$_{Re}$  &  $\nabla_\mathrm{[M/H]}$   &  [Mg/Fe]$_{Re}$  & $\nabla_\mathrm{[Mg/Fe]}$ \\
  & [$\log$ Gyr] &  [$\log\mathrm{Gyr}/\log R$ ]  &  [dex] &   [dex$/\log R$ ]  & [dex] &   [dex$/\log R$ ]\\
\hline 
$\sigma$ > 300 \kms &  1.113$^{\pm0.008}$   & 0.013$^{\pm0.006}$     &  -0.029$^{\pm0.010}$    &  -0.269$^{\pm0.012}$    &        0.312$^{\pm0.007}$   & 0.001$^{\pm0.009}$          \\
200 < $\sigma$ < 300 \kms & 1.087$^{\pm0.004}$   &  -0.005$^{\pm0.003}$     &  -0.031$^{\pm0.005}$    &    -0.187$^{\pm0.007}$      &       0.252$^{\pm0.004}$             &        -0.024$^{\pm0.006}$          \\
$\sigma$ < 200 \kms & 1.020$^{\pm0.005}$   &  -0.024$^{\pm0.008}$     &  -0.089$^{\pm0.003}$   &    -0.158$^{\pm0.006}$      &       0.186$^{\pm0.003}$             &        -0.047$^{\pm0.006}$          \\
\hline 
\end{tabular}
\caption{Best-fitting stellar population gradients for our three velocity dispersion bins. The age, metallicity and [Mg/Fe] radial gradients were fit to a $\alpha + \beta \log R/R_e$ linear relation. We list here the best-fitting coefficients, namely, the stellar populations value at the \re \ ($\alpha$), and the logarithmic gradient slope ($\beta$).}
\label{tab:gradi}
\end{center}
\end{table*}

\section{Discussion} \label{sec:blabla}

We have shown that both the normalization and the slope of the stellar populations gradients in Es depend on the central velocity dispersion. Regarding the age of the stellar populations within our sample, the bulk of stars are old, typically formed $\gtrsim 10$ Gyr ago. On top of the dominant old stellar component, younger populations can form later due to the either metal-rich, recycled gas expelled by supernovae and evolved stars, or due to accretion of more pristine gas from the intergalactic medium. Although these new generations of stars marginally contribute to the mass budget, their high luminosity explains the diversity of age values shown in Fig.~\ref{fig:global}. This so-called {\it frosting} populations were first proposed by \citet{Trager00b}, and have also be invoked to explain the ultraviolet properties of nearby ETGs \citep{Vazdekis16}.

The averaged age gradients (top panel in Fig.~\ref{fig:grad}) probe the overall formation of Es, minimizing the stochastic frosting effect. The transition from a slightly positive age gradient in massive Es towards slightly negative for lighter objects (see Table~\ref{tab:gradi}) is expected according to closed-box ETG formation scenarios \citep{Pipino04,Pipino06}. Due to the continuous gas infall at the formation epoch, star formation lasted longer in the central regions of massive galaxies, leading to the observed positive age gradient in the highest $\sigma$ bin. As stars and dark matter become less concentrated with decreasing galaxy mass \citep[e.g.][]{Bertin02,Ferrarese06,Kormendy09,vu13,Tortora14}, the efficiency to drive and retain the gas in the center also decreases, hence, creating flatter or even negative age profiles. This transition in the age gradients, from high- to low-$\sigma$ galaxies, is in agreement with the gradual dominance of rotationally-supported systems among lower-mass (M$\sim10^{11}$\msun) ETGs \citep{Emsellem07,Emsellem11,Cappellari13}, as well as with implementations of semi-analytical models \citep{Porter14,Lacey16,Tonini17}.

As expected from the standard interpretation of the [Mg/Fe] as a star formation timer, we found that average age and [Mg/Fe] radial variations seem to be, at least, partially correlated (see Fig.~\ref{fig:grad}). The marginally positive [Mg/Fe] gradient of our highest $\sigma$ bin suggests that the star formation lasted longer in the center than in the outskirts, which is consistent with the age profile. On the other hand, the low [Mg/Fe] values in the outer regions of low-$\sigma$ galaxies might indicate a more extended star formation, which is consistent with the observed age gradient. 

Our measurements favor a gradual steepening of the metallicity gradients with increasing galaxy velocity dispersion. Although studies over wide ranges of masses and morphological types support similar trends \citep{GD15}, detailed stellar populations analysis of ETGs tend to favor the opposite, i.e., galaxies with higher velocity dispersions exhibiting flatter metallicity gradients \citep{SB07,Kuntschner10,Spolaor09,Spolaor10}. Theoretically, a monolithic-like formation path for ETGs should result in a clear trend between galaxy mass and metallicity gradient slope, as suggested by our measurements, with most massive galaxies hosting steeper metallicity gradients \citep{Chiosi02,Kawata03}. To what extent this applies to real ETGs, in particular due to the effect of successive mergers \citep{Cook16}, will be discussed in the next section. 

\subsection{Cosmological evolution of the stellar population gradients} \label{sec:relics}

The short formation time-scales \citep[e.g.][]{Thomas05}, old ages \citep[e.g.][]{Trager00}, light profile concentration \citep[e.g.][]{Graham01} and recurrent appearance of old galaxies at high redshifts \citep{Lonoce15,Kriek16,LC17} ought to indicate that ETGs were formed very rapidly, in gas rich, high star formation rate events beyond $z\sim2$, or even earlier \citep{Wellons}. This would imply that the bulk of stars within ETGs were formed {\it in situ}, in a monolithic-like fashion, leading to the observed scaling relations between stellar population properties and galaxy mass.

\begin{figure}
\begin{center}
\includegraphics[width=7.2cm]{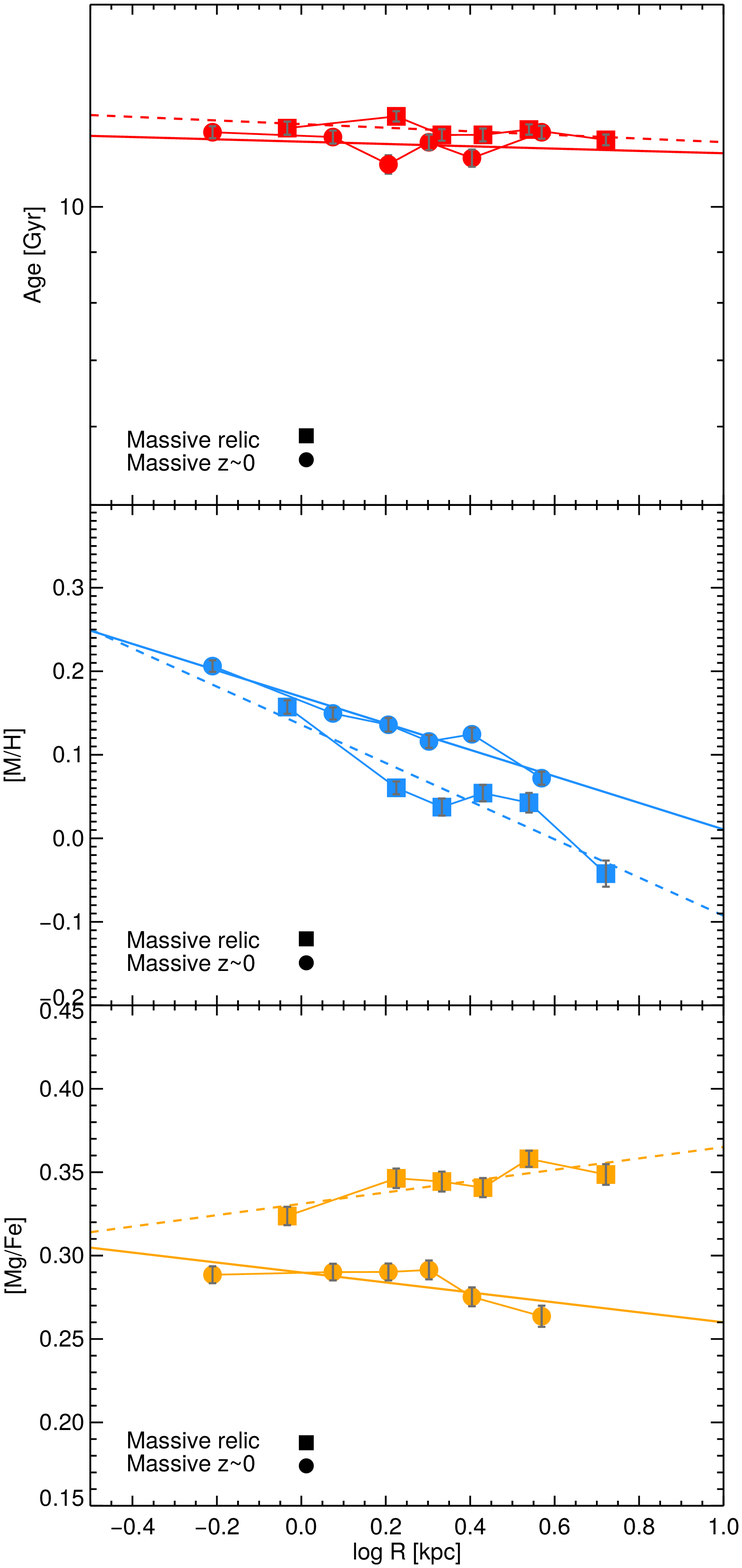}
\end{center}
\caption{Stellar populations gradients of massive relic galaxies. Same as in Fig.~\ref{fig:grad}, but now comparing a sample of nearby relic galaxies (squares) and standard Es of similar masses (circles). Red, blue and yellow correspond to the age, metallicity and [Mg/Fe] measurements. In order to account for the size evolution of massive ETGs since $z\sim2-3$, we plot the stellar populations properties as a function of radial distances in units of kpc, rather than \re. Age profiles are equally flat, but {\it red nuggets} descendants are on average slightly older. On the other hand, the metallicity gradients strongly differ, being flatter in standard nearby Es. The steep metallicity gradient and the increasing [Mg/Fe] with radius are compelling evidence supporting that massive relic galaxies are direct descendants of the high-$z$ population of massive galaxies ({\it blue/red nuggets}).}
\label{fig:reli}
\end{figure}

However, in a $\Lambda$-CDM Universe, galaxies and their dark matter halos do not evolve in isolation. The role of mergers in shaping the evolution of galaxies is particularly relevant for ETGs, as suggested by the morphology-density relation \citep{Dressler80,Postman84,Cappellari11}, and by the lack of net angular momentum, particularly in high-mass objects \citep{Emsellem11,Cappellari13,FB17}. Furthermore, massive ETGs have grown in size a factor of $\sim4$ since $z\sim2$ \citep[e.g.][]{Trujillo04,Trujillo07,Daddi05,Franx08,vdk10b,Arjel14}, presumably through the accretion of less massive satellites \citep{Naab09,Hopkins10,Hilz12}.

To reconcile monolithic and merger-driven properties of ETGs, a mixed formation path has been proposed and developed over the last years, both theoretically \citep{Oser,Oser12,navarro,Shankar13} and observationally \citep{Pastorello14,MN15b,Foster16}. This so-called two-phase formation scenario can be summarized as follows:

(1) The gas rich environment at redshift \mbox{$z\sim2-3$} favored the formation of massive, compact and highly star-forming disks, either via violent instabilities \citep{Gammie01,Dekel09,Krumholz10} or through gas-rich mergers \citep{Barnes91,Mihos96,Hopkins06}. Note that, given their dissipative character, these two mechanisms of gas fueling towards the center of galaxies are virtually indistinguishable after they take place. The end products of this first stage are massive, compact and rotating {\it blue nuggets}, that ultimately quench into {\it red nuggets} \citep[e.g.][]{Barro13,Dekel14,Barro17}. The stellar populations properties of these {\it red nuggets} should be those expected in a classical monolithic collapse, i.e., steep and negative radial gradients in metallicity, but a radially increasing [Mg/Fe] \citep{Chiosi02,Kawata03,Pipino04,Pipino06}. The scaling relations between stellar population properties and galaxy mass get established during this stage.

(2) After the star formation within {\it red nuggets} is quenched, their later evolution is driven by mergers. In the case of ETGs, these mergers have to be dry, otherwise they would host young stellar populations at $z\sim0$. However, it is also possible that at least a fraction of {\it red nuggets} have accreted a significant amount of gas since they were formed, leading to bulge-dominated spiral galaxies \citep{Graham13,Graham15,dlr16}. Massive ETGs sometimes exhibit complex internal kinematics, signatures of major (dry) mergers \citep{Bois11,Emsellem11,Naab14}. However, since the pristine scaling relations imprinted in phase (1) are conserved, major mergers can not be responsible for the size evolution of ETGs. Thus, dry accretion of low-mass satellites is thought to be the main evolutionary channel for ETGs during this second stage \citep{Naab09,Hopkins10,Hilz12}, although additional mechanisms might me required\citep{akin17}. This size and mass growth is mostly driven by the accretion of relatively massive satellites, with masses typically between 1:4 and 1:10 that of the central \citep{Carlos12,Bluck12,Ferreras14}. As these minor satellites fall towards the center of the more massive central ETG, dynamical friction strips apart the stars of the satellites and their stellar mass remains in the outskirts \citep{Lackner12}. Such dry growth preserves the monolithic-like properties of massive ETGs, but makes them evolve in size.

Therefore, the stellar populations properties observed in present day massive ETGs are expected to be the combination of the {\it in situ} formation of the core, plus the {\it ex situ} (accreted) material from low-mass satellites deposited in the outer regions \cite[e.g.][]{Oser,Oser12,Ciotti07,navarro,Hirschmann14,RG16}. Hence, if one wants to safely understand and interpret the stellar population properties at $z\sim0$, {\it in situ} and {\it ex situ} processes have to be decoupled. Fortunately, since galaxy mergers are an stochastic process, it is possible that a fraction of {\it red nuggets} have passively evolved since $z\sim2$ without accreting any {\it ex situ}-formed material \citep{Quilis13,Stringer15}. Just recently, a sample of galaxies fulfilling all the characteristics to be considered nearby {\it red nuggets} (or massive relic compact galaxies) has been identified in the local Universe \citep{Remco12,Trujillo14,MN15b,Anna17,akin17}. These objects offer the unique possibility of studying the baryonic processes that took place in phase (1), minimizing the influence of the accreted material. Moreover, by comparing these massive relic galaxies to those measured in standard $z\sim0$ Es, we isolate the effects of phase (2) on the stellar populations properties.

In Fig.~\ref{fig:reli} we show the mean age, metallicity and [Mg/Fe] gradients of Es from the present work, compared to those measured for the massive relic galaxy sample of \citet{akin17} with similar stellar velocity dispersion ($\sigma_e \sim 270$ \kms). To avoid systematics in the stellar populations analysis, we repeated the fitting process described in \S~\ref{sec:pop} for the massive relic galaxies, instead of using the values published in \citet{akin17}. 

The mean age gradients for standard nearby Es and for compact relic galaxies are both rather flat. The main difference among them is that nearby Es are slightly ($\sim$ 1 Gyr) younger, which is consistent with a more extended star formation. Although mergers in phase (2) have to be predominantly dry, ETGs are not completely gas-free \citep{Young14,Lagos14}. Thus, a small fraction of merger-induced star formation is expected, in agreement with the age gradients shown Fig.~\ref{fig:reli}. 

On the other hand, radial metallicity gradients in compact relic galaxies are significantly steeper than in standard Es, in remarkable agreement with theoretical predictions \citep{Taylor17}. These differences can not be explained by late star formation, as this would also radially affect the age profile. Alternatively, the flattening of the metallicity gradient is naturally expected if minor mergers actually drive the size growth of ETGs. Basic dynamical arguments indicate that accreted satellites populate the outskirts of the central galaxy, at a distance determined by their mass ratio \citep{Tal11}. The more massive a satellite is, the further inside the central galaxy it is deposited. This, combined with the aforementioned mass-dependent contribution to the size growth \citep{Carlos12}, implies a differential radial accretion, with the mass growth taking place predominantly in the outskirts of the central galaxy. Note also that the mass--metallicity relation would imply a radial stratification in the metallicity of the accreted material: stars from (massive) metal rich satellites will populate the inner regions of the central ETGs, which are already more metal rich than the outskirts. The combination of these two processes, differential growth plus metallicity stratification, can explain the observed flattening in the metallicity profiles of nearby Es. This scenario could also explain our finding of a steepening in the metallicity gradients with increasing $\sigma$, as well as the differences with respect to previous results. Our gradients, in particular for the high-$\sigma$ bin, are restricted to the central regions (R$\lesssim$0.5\re), where the signature of a possible monolithic-like collapse is expected to be less affected by mergers, and star formation is also expected to last longer.

Interesting differences are also found in the [Mg/Fe] gradients. In relic compact galaxies, [Mg/Fe] increases with radius, in perfect agreement with monolithic-like formation scenarios \citep{Pipino06}, further supporting that this kind of objects are the nearby descendant of the high-$z$ {\it red nuggets}. Over cosmic time, the [Mg/Fe] gradients have experienced a strong variation ($\delta \nabla_\mathrm{[Mg/Fe]} / \nabla_\mathrm{[Mg/Fe]} \sim -1.8$), even more important than the change in metallicity ($\delta \nabla_\mathrm{[M/H]} / \nabla_\mathrm{[M/H]} \sim -0.5$). Although naively one would expect a similar variation in both quantities, metallicity and [Mg/Fe] values measured in a galaxy are set in different time-scales. Whereas the overall metallicity is rapidly determined by core-collapse supernovae, in scales of a few hundred Myr, [Mg/Fe] probes much larger time-scales ($\sim$ Gyrs), driven by successive Type Ia supernovae explosions \citep[e.g.][]{Vazdekis96}. Hence, metallicity hardly evolves with time, but [Mg/Fe] does \citep{Feltzing01,Bensby14,Bergemann14}. As the size growth of massive ETGs occurs, the stellar populations accreted in their outskirts are more evolved, and presumably they would be less [Mg/Fe]-enhanced. We therefore speculate that the strong variation in the [Mg/Fe] gradient of ETGs, since they were formed as {\it red nuggets} to what we observe at $z\sim0$, is due to the late accretion of low-mass satellites with depleted [Mg/Fe]: the later the accretion, the later the satellite quenching, the lower the [Mg/Fe] value of the accreted stars. Metallicity, on the other hand, is rather independent of the satellite accretion time, and therefore its radial gradient shows a weaker variation with redshift. 

In Appendix~\ref{sec:inde}, we prove that the differences in the radial [Mg/Fe] trends of compact and massive Es are clearly seen at the level of line-strengths (i.e. without relying on stellar population models), hence proving the robustness of our results.

It is worth mentioning that our sample selection, excluding S0 galaxies, provides a cleaner test for the cosmological evolution of the stellar population gradients, as the influence of mergers is likely to be negligible compared to Es \citep[e.g.][]{Cappellari16}. Moreover, given the high masses of relic compact galaxies, we expect that their $z\sim0$ descendant will be mostly Es, as the number density of S0 galaxies decreases with increasing galaxy mass. However, an absolute measurements of the change in the slope of the stellar populations properties needs a more complete sample of both nearby Es and massive relic galaxies.

\subsection{On the origin of the [Mg/Fe] variations}

\begin{figure}
\begin{center}
\includegraphics[width=8cm]{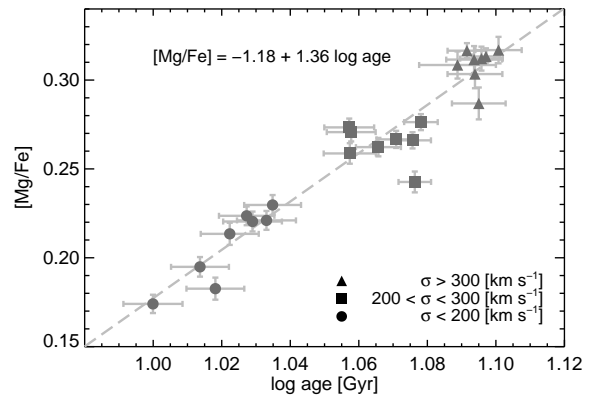}
\end{center}
\caption{Age-[Mg/Fe] relation for the CALIFA sample. Combination of top and bottom panels of Fig.~\ref{fig:grad}. Filled triangles, squares and circles correspond to our high-, intermediate- and low-$\sigma$ bins. The oldest stellar populations found in the most massive galaxies are more [Mg/Fe]-enhanced than the young populations of low-mass objects. Since both dwarfs and massive galaxies started forming stars at similar times, the age-[Mg/Fe] relation for Es should be interpreted as a consequence of the different formation time-scales: whereas massive galaxies formed their stellar population very rapidly, low-mass objects have experienced more extended star formation histories, leading to lower [Mg/Fe] values and effectively younger ages.}
\label{fig:age}
\end{figure}

As described in \S~\ref{sec:intro}, the question of whether the [Mg/Fe] is entirely defined by the formation time-scale of a galaxy or whether it also depends on a variable IMF is still open. In Fig.~\ref{fig:age} we show the mean age--[Mg/Fe] relation for the stellar populations in our sample, as derived from the radial measurements shown in Fig.~\ref{fig:grad}. The trend represented in Fig.~\ref{fig:age} could be interpreted as the combination of two different phenomena: (1) low-$\sigma$ galaxies formed their stellar populations later and (2) over more extended periods of time, being therefore younger and less [Mg/Fe]-enhanced. Both properties (young stars plus low [Mg/Fe] ratios) could, in principle, be completely independent. However, very old stars, with ages $\gtrsim12$ Gyr, constitute a significant fraction of the stellar mass for all ETG masses, from dwarfs \citep{Koleva09,aga15} to giants \citep{McDermid15,dlr}. Therefore, the  age--[Mg/Fe] relation shown in Fig.~\ref{fig:age} is seems to be mainly driven by the star formation time-scale, in agreement with the standard interpretation of the [Mg/Fe] ratio. Low-$\sigma$ galaxies started forming stars at the same time as high-$\sigma$ ones, but since star formation took longer, they are less [Mg/Fe] enhanced, and look effectively younger.

However, the analysis of our observations suggests that other mechanisms, in addition to the star formation history, might be needed to explain the abundance pattern of Es. This claim is based on two main results. First of all, age and [Mg/Fe] gradients in massive relic galaxies are anti-correlated (Fig.~\ref{fig:reli}). If the star formation duration was to be the only mechanism driving the abundance pattern in massive ETGs, massive relic galaxies should exhibit a positive radial age variation of $\sim 1.5$ Gyr, to be consistent with the [Mg/Fe] gradient (Fig.~\ref{fig:age}). \citet{Molaeinezhad17} have reported a similar decoupling between age and [Mg/Fe] in the bulges of nearby barred galaxies.

\begin{figure}
\begin{center}
\includegraphics[width=8cm]{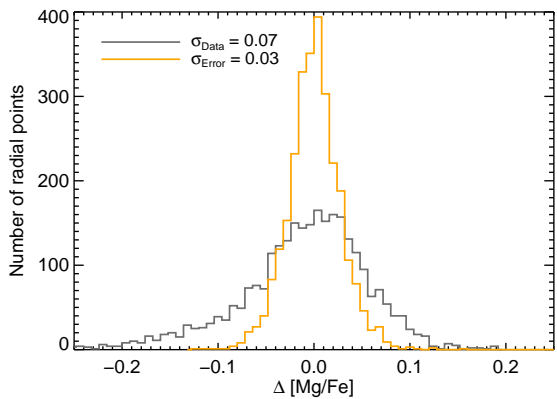}
\end{center}
\caption{$\Delta \mathrm{[Mg/Fe]}$ distribution within our sample of CALIFA Es (grey). $\Delta \mathrm{[Mg/Fe]}$ describes the scatter in the [Mg/Fe] values of stellar populations with a given age. Therefore, in the absence of additional mechanisms other than the duration of the star formation histories, it should be consistent with the observed errors ( orange). The notable differences between the two distributions, both in width and in shape, suggest that [Mg/Fe] is not only driven by the star formation time-scale.}
\label{fig:scatter}
\end{figure}

The effect of complementary mechanisms can be explored by analyzing the [Mg/Fe] distribution at fixed age. For every spatial bin in our stellar populations maps (Fig.~\ref{fig:map}), we defined the quantity $\Delta$[Mg/Fe] as

\begin{equation*}
 \Delta \mathrm{[Mg/Fe]} = \mathrm{[Mg/Fe]}_{(x,y)} - \langle \mathrm{[Mg/Fe]_{\mathrm{age}(x,y)}} \rangle
\end{equation*}

\noindent
where $\mathrm{[Mg/Fe]}_{(x,y)}$ is the abundance pattern measurement at the $(x,y)$ position, and $\langle \mathrm{[Mg/Fe]_{\mathrm{age}(x,y)}} \rangle$ is the median [Mg/Fe] value of all the points in that galaxy with the same age. Given our limited age resolution\footnote{For old stellar populations, the age step in BaSTI-based MILES is $\Delta_\mathrm{age}=0.5$ Gyr.}, this latter median [Mg/Fe] was evaluated in steps of $\delta \mathrm{age} = 0.5$ Gyr. In practice, $\Delta \mathrm{[Mg/Fe]}$ defines the scatter in the [Mg/Fe] value for a given age, and its distribution is shown in Fig.~\ref{fig:scatter} combining all galaxies in our sample.

\begin{figure}
\begin{center}
\includegraphics[width=8cm]{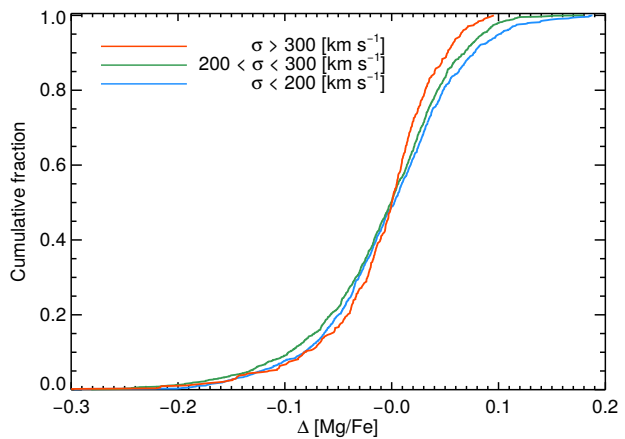}
\end{center}
\caption{Cumulative distributions of $\Delta \mathrm{[Mg/Fe]}$ within our three velocity dispersion bins. Galaxies with higher velocity dispersion have a more asymmetrical distribution, with a larger fraction of negative $\Delta \mathrm{[Mg/Fe]}$ but lower regions where $\Delta \mathrm{[Mg/Fe]}$ is above the average. This change in the skewness, from $\gamma_1=-1.31$ for the high-$\sigma$ bin to $\gamma_1=-0.42$ for the low-$\sigma$ one, is compatible with the expected effect of a variable IMF as a function of galaxy mass.}
\label{fig:cum}
\end{figure}

If the only driver of the [Mg/Fe] variations is the duration of the star formation, the distribution of $\Delta \mathrm{[Mg/Fe]}$ (gray) should be entirely explained by our observational uncertainties (orange). However, it is clear that uncertainties alone do not explain the observed distribution, further suggesting additional mechanisms regulating the [Mg/Fe] enhancement. As discussed in \S~\ref{sec:intro}, IMF variations have been invoked to explain the abundance pattern variations in ETGs. A non universal IMF is particularly appealing to explain Fig.~\ref{fig:scatter}, given the pronounced tail of the $\Delta \mathrm{[Mg/Fe]}$ distribution towards negative values. The assumption of a bottom-heavier IMF in massive ETGs would reduce the number of massive stars that will end up as core-collapse supernovae, leading to lower [Mg/Fe] values than expected assuming a standard IMF slope \citep[][]{MN16b}. Given the IMF slope--$\sigma$ relation \citep[e.g.][]{Treu,Cappellari,labarbera}, if the IMF is responsible for the asymmetry observed in the $\mathrm{[Mg/Fe]}$ distribution, this should be more prominent in our high-$\sigma$. In other words, the skewness ($\gamma_1$) of the $\Delta \mathrm{[Mg/Fe]}$ distribution should gradually become more negative when moving from the low-$\sigma$ to the high-$\sigma$ bin. The analysis of our data are fully consistent with such a scenario, as $\gamma_1$ equals to $-0.42$, $-0.80$, and $-1.31$ in our low-$\sigma$, intermediate, and high-$\sigma$ bins, respectively. The cumulative distribution for the three $\sigma$ bins are shown in Fig.~\ref{fig:cum}.

Alternatively, it may be possible to explain the $\Delta \mathrm{[Mg/Fe]}$ tail towards negative values as a consequence of more extended star formation histories. Since our age bins are not infinitively narrow ($\delta \mathrm{age} = 0.5$), there is room for slight differences in star formation time-scales. We have investigated this possibility by comparing the star formation histories in the high and low $\Delta \mathrm{[Mg/Fe]}$ ends, but not differences were found. In addition, there is no correlation between $\Delta \mathrm{[Mg/Fe]}$ and age. These results seem to suggest that subtle differences in the star formation history are not likely responsible for the shape of the $\Delta \mathrm{[Mg/Fe]}$, at least within the level of precision allowed by the data. Whether weaker variations in the star formation at fixed age are important in shaping $\Delta \mathrm{[Mg/Fe]}$ is yet to be investigated.

It is worth noticing two facts before further interpreting these results. First of all, we have not measured the IMF in our sample of galaxies. Thus, although the quantitative behavior of $\Delta \mathrm{[Mg/Fe]}$ agrees with a variable IMF scenario, no conclusive statements can be made. Second, the observed trend between IMF slope and galaxy velocity dispersion only applies to the low-mass end of the IMF, i.e. below M$\sim1$\msun, but not necessarily to the massive stars driving the [Mg/Fe] ratio. Although there is observational evidence of a variable high-mass end IMF slope \citep{Hoversten08,Meurer09,Gunawardhana11,Nanayakkara17}, as also supported by the chemical composition of massive ETGs \citep{Vazdekis96,Vazdekis97,weidner:13,Ferreras15}, results are less conclusive than those focused on the low-mass end.

\section{Summary and conclusions} \label{sec:fin}

We have analyzed a sample of 45 Es drawn from the CALIFA sample, measuring spatially-resolved age, metallicity and [Mg/Fe] maps. We have compared these measurements with a sample of massive relic galaxies. Our findings can be summarized as follows:

\begin{enumerate}

 \item Age and [Mg/Fe] gradients in nearby Es are rather flat, with no strong dependence on galaxy velocity dispersion. However, in contrast with previous studies, we find that the central (R$\lesssim$0.5\re) metallicity gradients become steeper with increasing $\sigma$, which could be understood as the imprint of an early monolithic-like formation, preserved in the innermost regions of Es. However, when compared to nearby {\it red nuggets}, present-day massive Es exhibit flatter metallicity and [Mg/Fe] gradients. We interpret this transition as a combination of the initial monolithic-like, {\it in situ} formation at high redshift ($z\gtrsim2$), plus the later dry accretion of satellites in the outskirts.
 
  \item The [Mg/Fe]--age relation observed in all our velocity dispersion bins further supports the classical interpretation of the [Mg/Fe] value as a proxy for the formation time-scales of ETGs. However, additional mechanisms other than the duration of the star formation seem to be required, as age and [Mg/Fe] appear partially anti-correlated at $z\sim0$. Moreover, there is a significant $\sigma$-dependent scatter in [Mg/Fe] at fixed age, not accounted by the observational uncertainties. Variations in the IMF are consistent with our measurements.

Understanding baryonic processes within massive ETGs is particularly challenging due to their old ages and their merger-driven evolutionary paths. However, we have shown here that detailed stellar populations analysis and the comparison to nearby massive relic galaxies can be used to reconstruct the star formation processes up to $z\sim2$. A more complete analysis, considering IMF and elemental abundance variations, as well as deeper observations will bring additional constraints in the future. Moreover, detailed observations of massive quiescence galaxies at higher redshifts will shed light on our results based on nearby objects. These observations will become a reality with the advent of the James Webb Space Telescope, combined with ground-based infrared spectrographs such as EMIR (GTC) or MOSFIRE (Keck).
 
\end{enumerate}

\section*{Acknowledgements}

We would like to thank the referee for a detailed and helpful review of our manuscript. We acknowledge support from grants AYA2016-77237-C3-1-P and AYA2014-56795-P from the Spanish Ministry of Economy and Competitiveness (MINECO), and from SFB 881 The Milky Way System (subproject A7 and A8) funded by the German Research Foundation. IMN acknowledges support from the EU Marie Curie Global Fellowships. GvdV acknowledges funding from the European Research Council (ERC) under the European Union's Horizon 2020 research and innovation program under grant agreement No 724857 (Consolidator Grant ArcheoDyn).



\bibliographystyle{mnras}
\bibliography{calfa} 

\newpage

\appendix

\section{Data systematic} \label{sec:resi}

\begin{figure}
\begin{center}
\includegraphics[width=7.5cm]{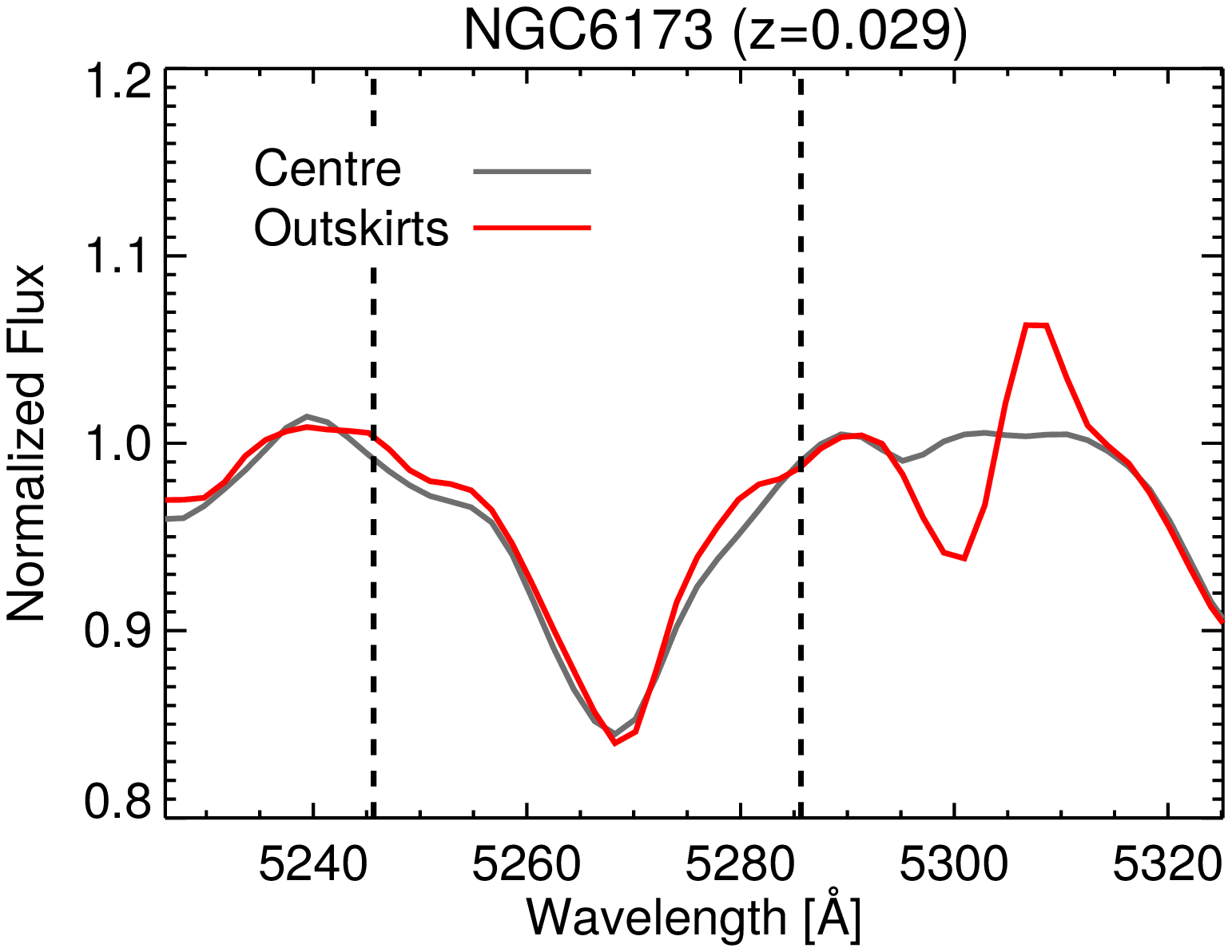}
\includegraphics[width=7.5cm]{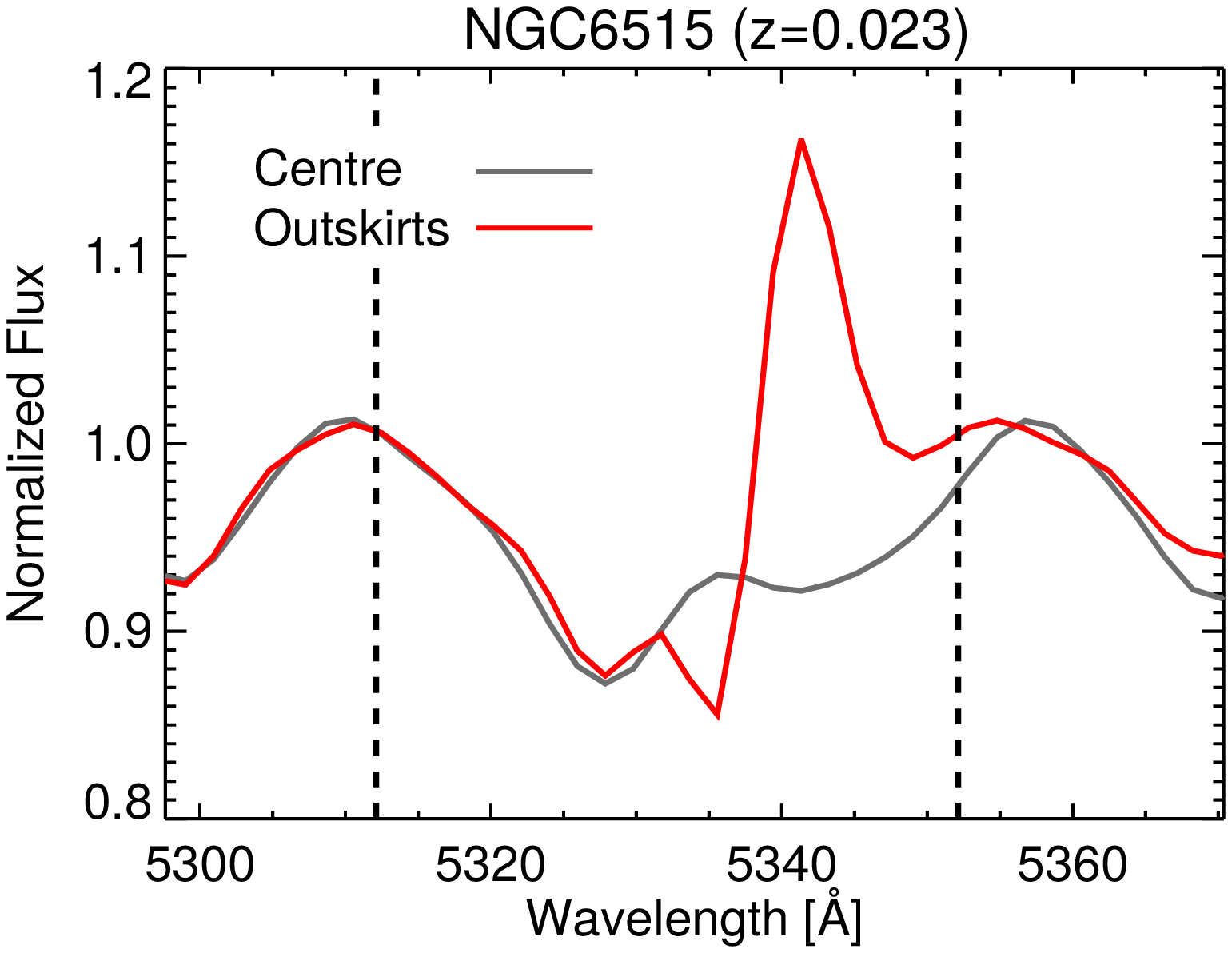}
\end{center}
\caption{Systematic residual at rest-frame $\lambda \sim 5640$ \AA. Grey line shows the central spectrum of the galaxy, whereas the red line correspond to a radial aperture in the outskirts. The top panel shows the spectral region around the Fe\,5270 absorption feature (vertical dashed lines). For NGC\,6173, at redshift $z=0.029$, the systematic affects the red pseudo-continuum of the Fe\,5270 line index. The bottom panel shows Fe\,5335 spectral feature for NGC\,6515 ($z=0.023$). In this case, the systematic lies over the index central band-pass (vertical dashed lines). To avoid contamination from this systematic, we imposed a hard redshift cut to our sample of galaxies (See details in \S~\ref{sec:pop}).}
\label{fig:resi}
\end{figure}

A clear systematic residual appears at a rest-frame wavelength $\lambda \sim 5640$ \AA, as shown in Fig.~\ref{fig:resi}. Although it is negligible in the central parts of the CALIFA field of view, this systematic made impossible a reliable measurement of the affected line-strength indices. Hence, we applied a redshift cut, so we only studied galaxies with $z<0.026$, which allowed a clean and homogeneous stellar population analysis using the Fe\,4383, H$\beta_\mathrm{o}$, Fe\,5015, Mgb, and Fe\,5270 indices.

\section{Model-independent assessment of the [Mg/Fe] gradients} \label{sec:inde}
\begin{figure}
\begin{center}
\includegraphics[width=8.0cm]{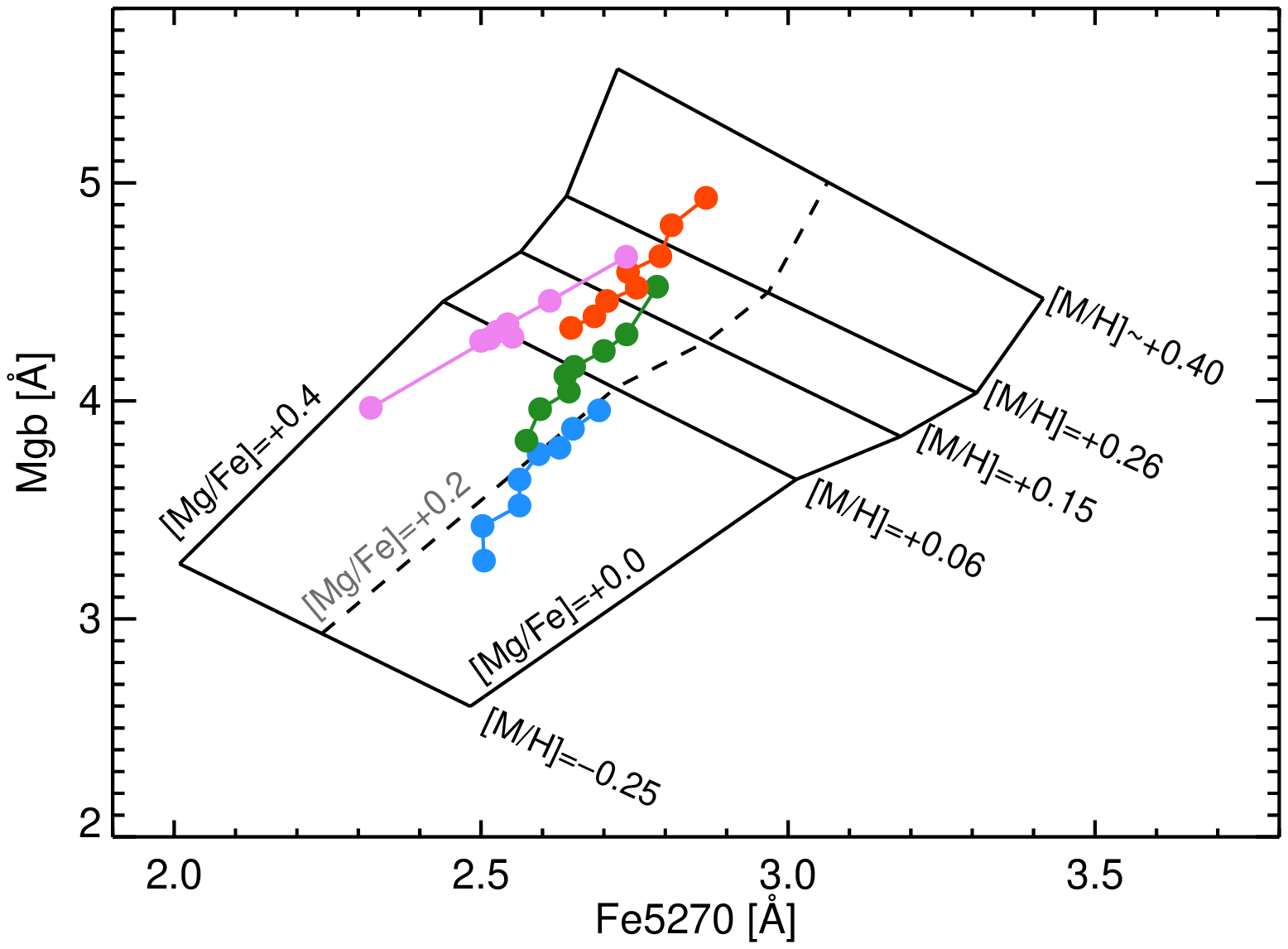}
\end{center}
\caption{Mgb-Fe\,5270 index-index diagram. Red, green and blue filled circles correspond to galaxies with $\sigma$>300 \kms, 200 < $\sigma$ < 300 \kms, and $\sigma$<200 \kms, respectively. For comparison, purple filled circles correspond to the sample of massive relic galaxies. Each symbol indicates a different radial bin, from the center (highest Mgb value) to the outer parts (lowest Mgb value), and are shown at a common velocity dispersion of $\sigma=200$ \kms. Radial distances are the same as in Figs.~\ref{fig:grad},~\ref{fig:reli}. The slope of the Mgb-Fe\,5270 tracks are clearly different when comparing nearby standard Es and high-$z$ relics, pointing towards a much more pronounced (and positive) [Mg/Fe] gradient for the compact relic galaxies.}
\label{fig:grindex}
\end{figure}

As described above, the change in the slope of the [Mg/Fe] gradients from high-$z$ relics to nearby Es is also supported by the Mgb-Fe\,5270 index-index diagram shown in Fig.~\ref{fig:grindex}, where all the data points have been corrected to a common velocity dispersion of $\sigma=200$\kms. Explicitly, the corrected index value is given by

\begin{equation}
 I_\mathrm{cor} = I_\mathrm{obs} \frac{M_{x,\sigma=200}}{M_{x,\sigma_\mathrm{obs}}} 
\end{equation}

\noindent where $I_\mathrm{obs}$ is the observed index value. $M_{x,\sigma=200}$ and $M_{x,\sigma_\mathrm{obs}}$ are the line-strength predictions given the best-fitting stellar population parameters $x=\{$age,[M/H],[Mg/Fe]$\}$, at the reference ($\sigma=200$\kms) and observed ($\sigma_\mathrm{obs}$) resolutions, respectively.

The difference response of Mgb and Fe\,5270 features to total metallicity and [Mg/Fe] allows a model-independent assessment of [M/H] and [Mg/Fe] variations. Radial measurements in nearby Es roughly scatter along constant [Mg/Fe] lines. In contrast, the slope Mgb-Fe\,5270 track of massive relic galaxies clearly evolve from metal-rich and moderately [Mg/Fe] enhanced in the center towards metal-poor (sub-solar) and highly [Mg/Fe] enhancement stellar populations in their outskirts. Although our fitting process included more line-strength indices (see \S~\ref{sec:pop}), and therefore it should be interpreted in a relative way, differences in the radial [Mg/Fe] variation appear clearly even at the level of line-strengths values. 

\newpage

\section{Kinematics and stellar populations maps} \label{sec:maps}

\begin{figure*}
\begin{center}
\includegraphics[width=12cm]{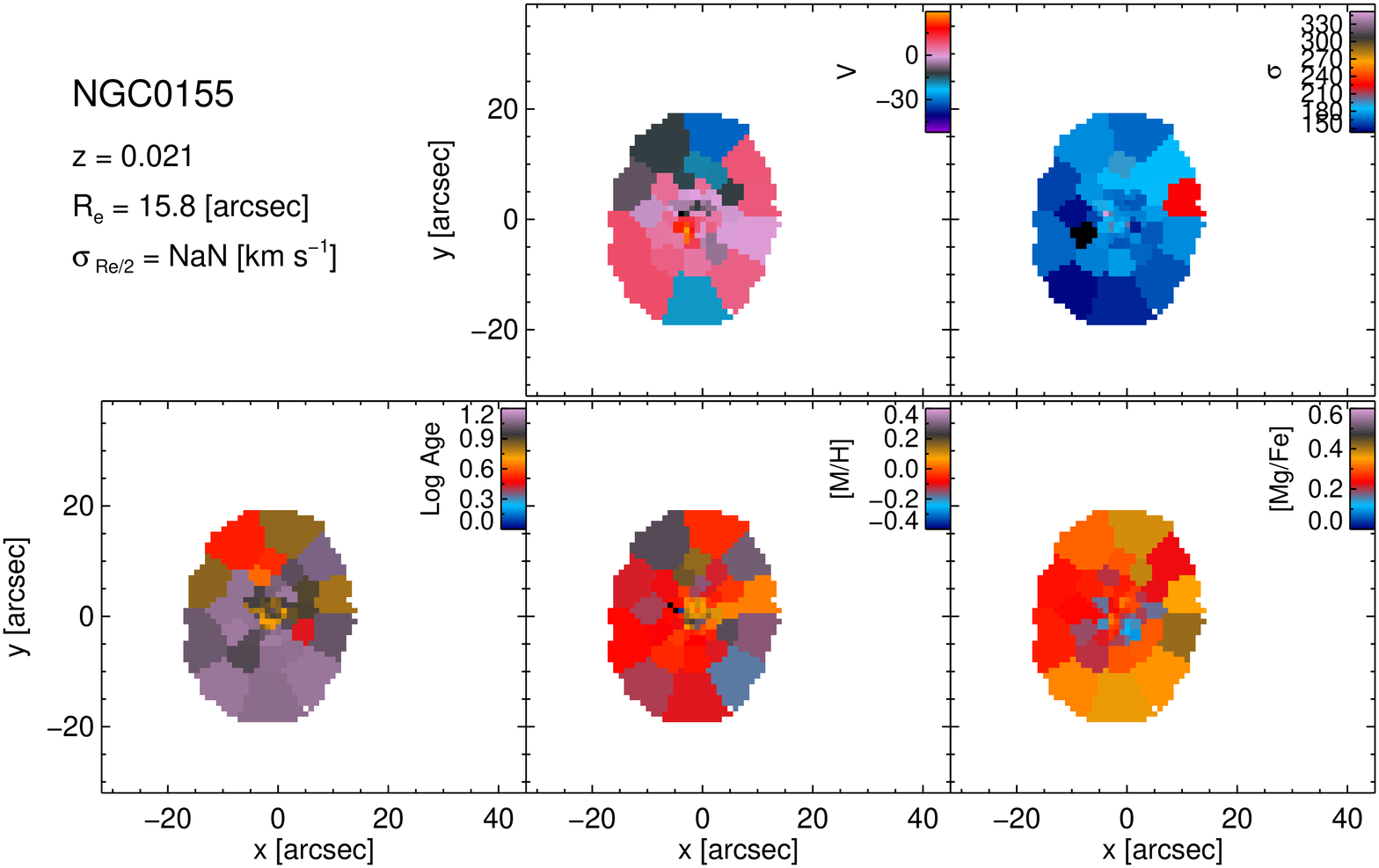}
\includegraphics[width=12cm]{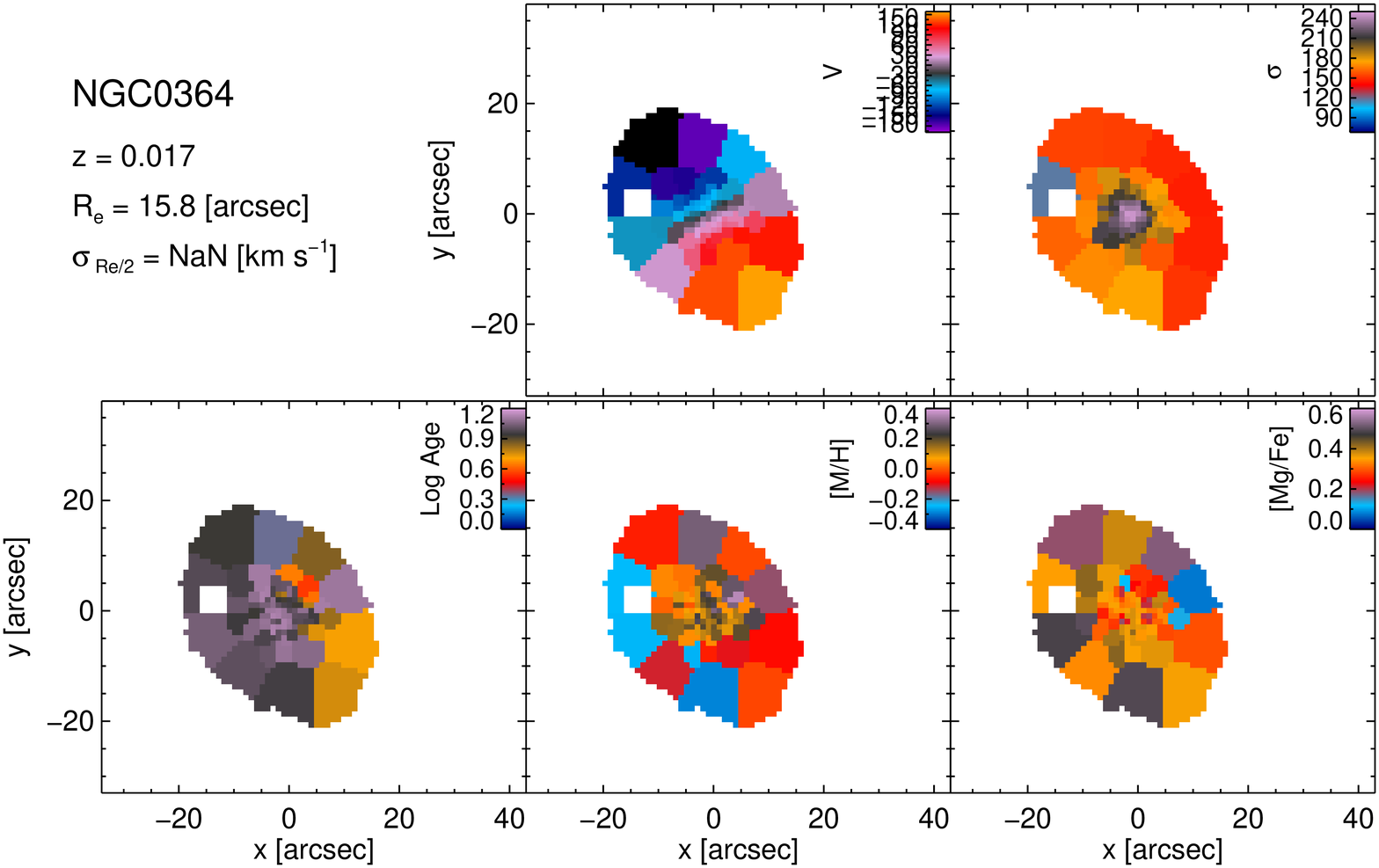}
\includegraphics[width=12cm]{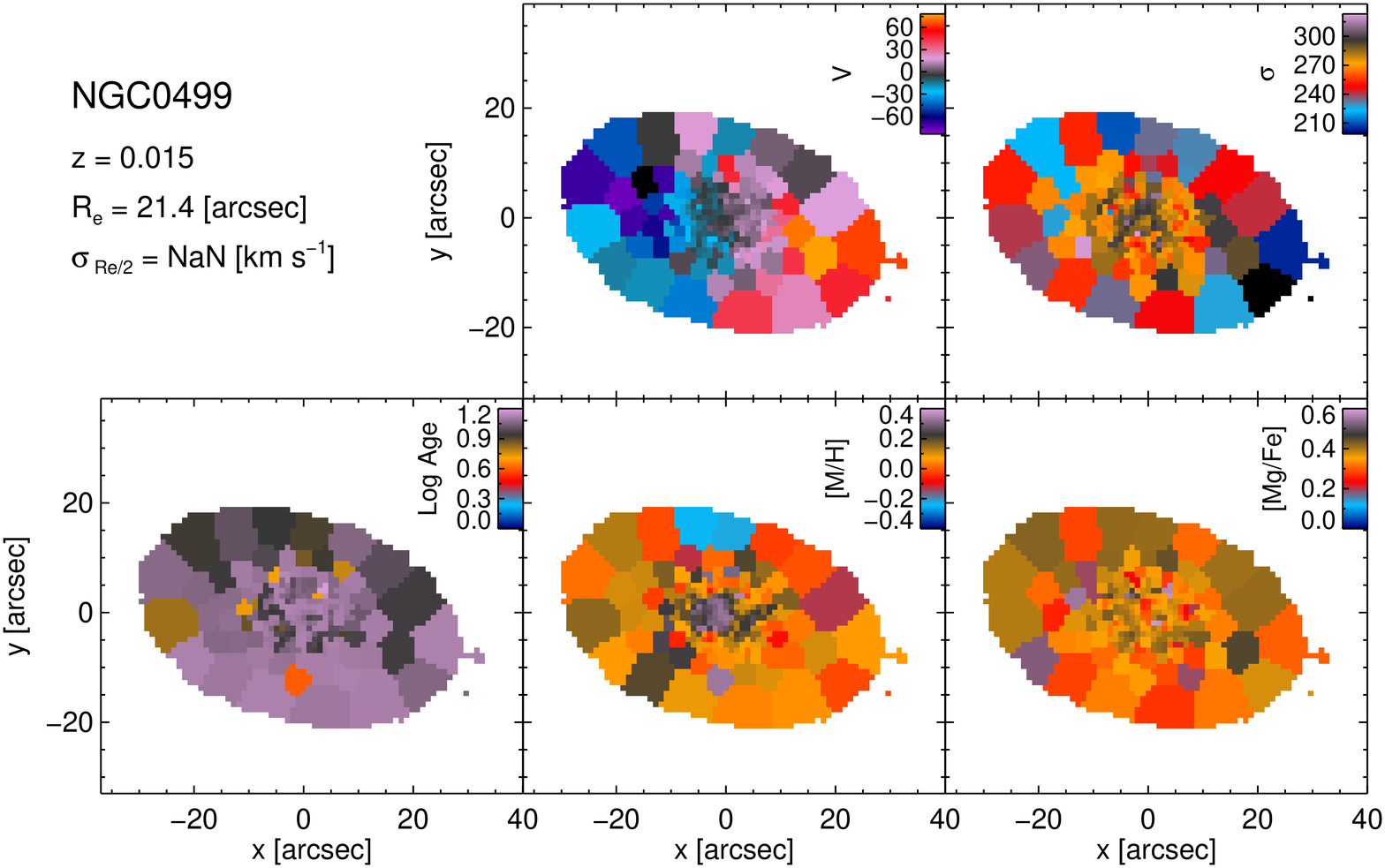}
\end{center}
\end{figure*}

\begin{figure*}
\begin{center}
\includegraphics[width=12cm]{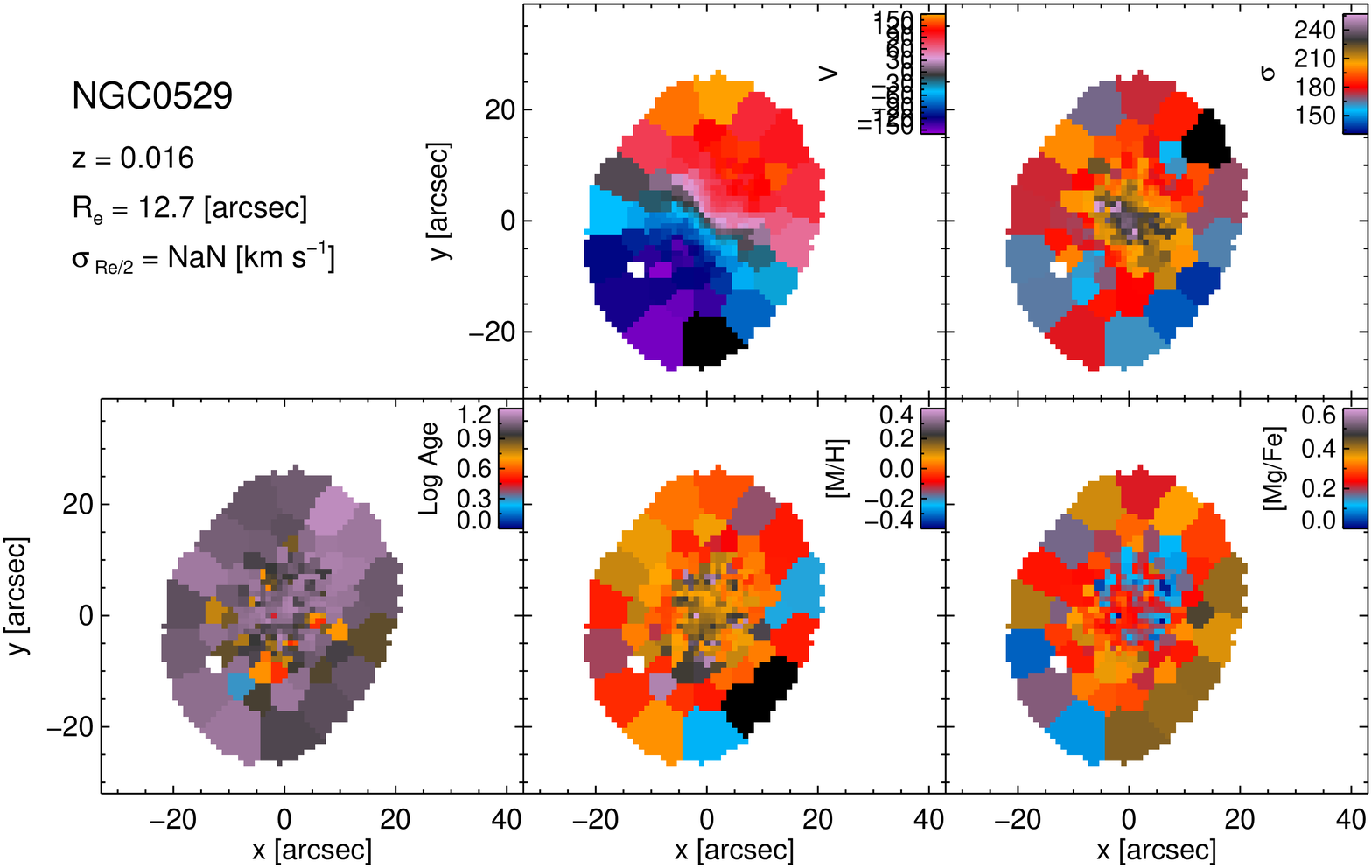}
\includegraphics[width=12cm]{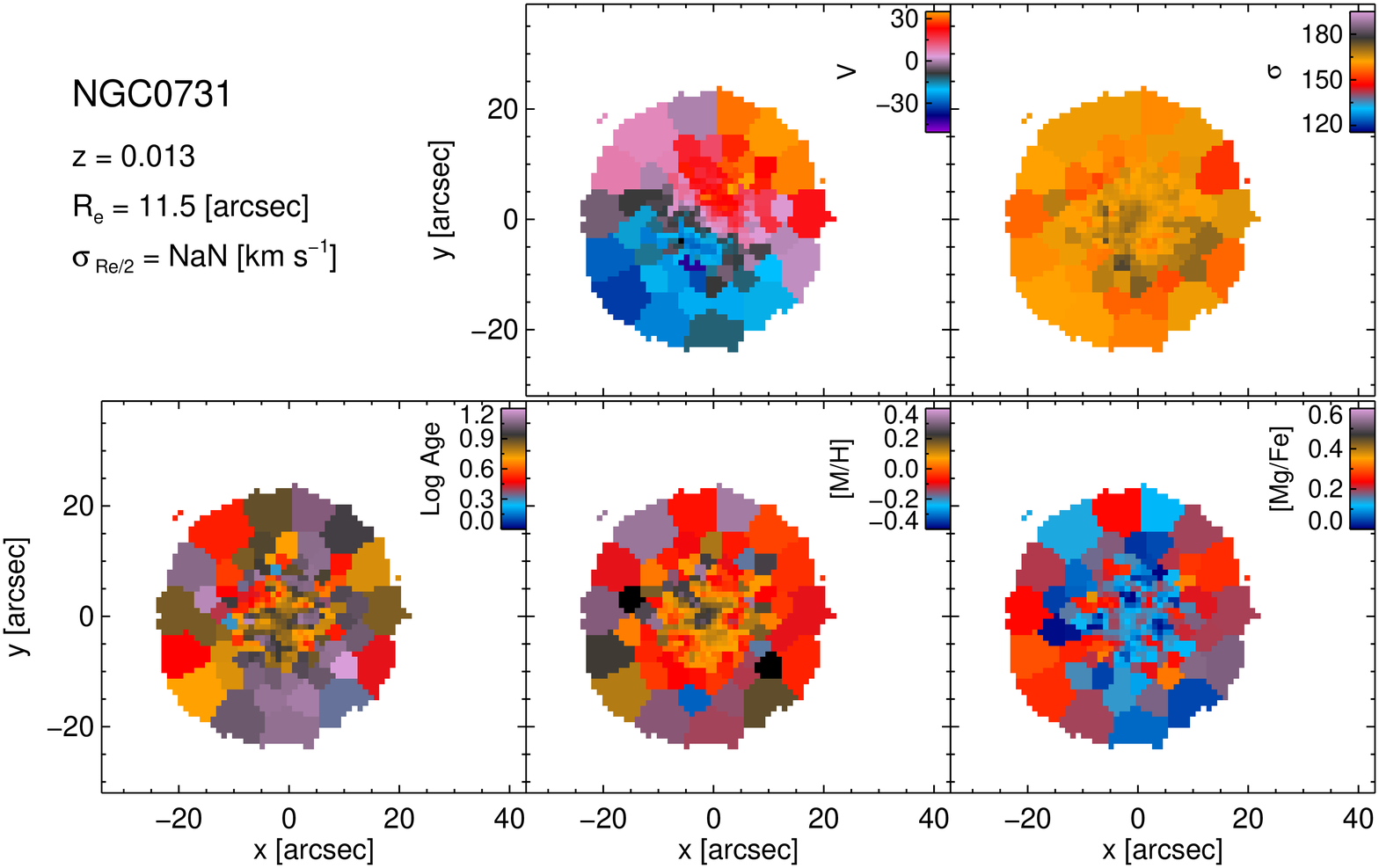}
\includegraphics[width=12cm]{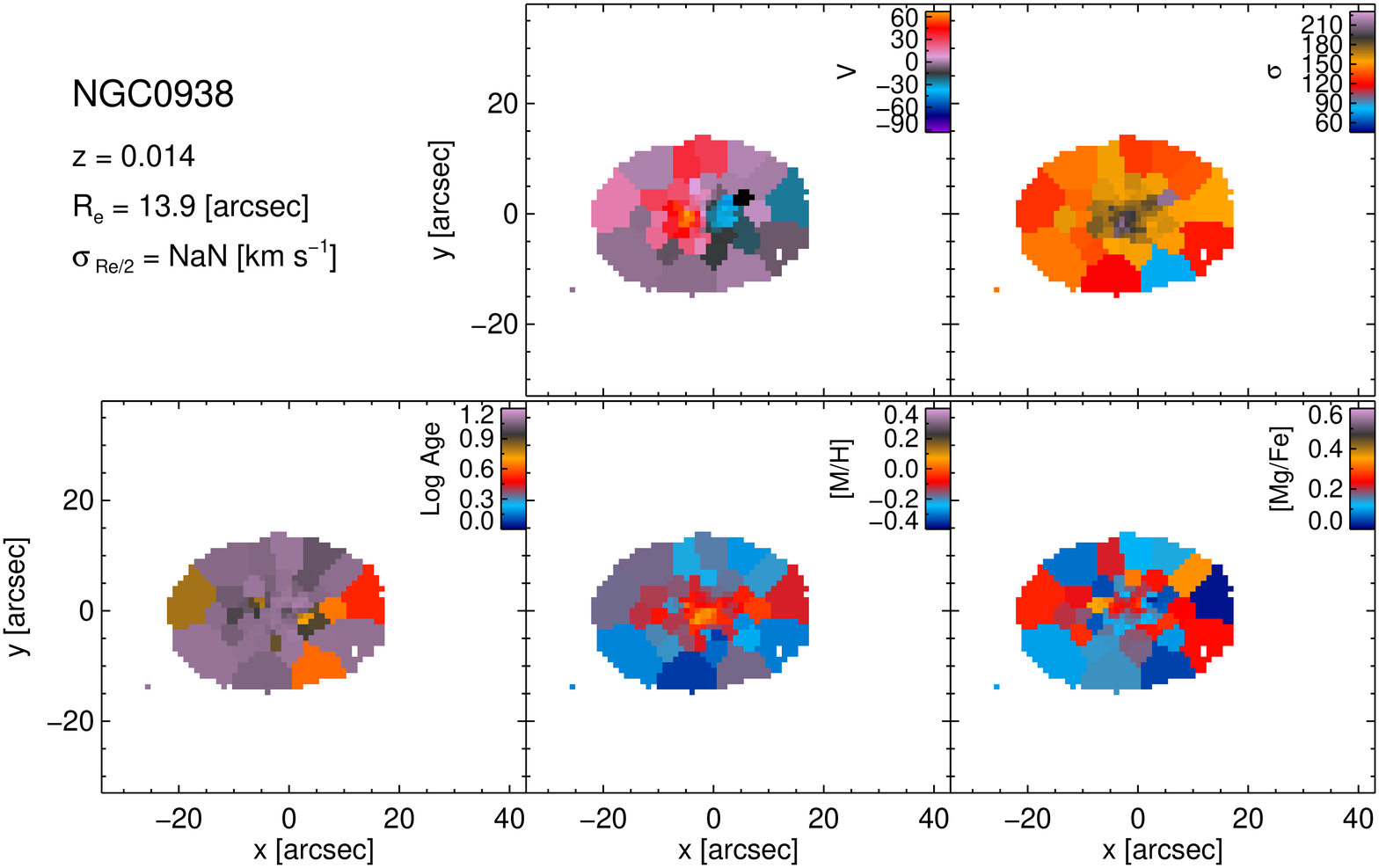}
\end{center}
\end{figure*}

\begin{figure*}
\begin{center}
\includegraphics[width=12cm]{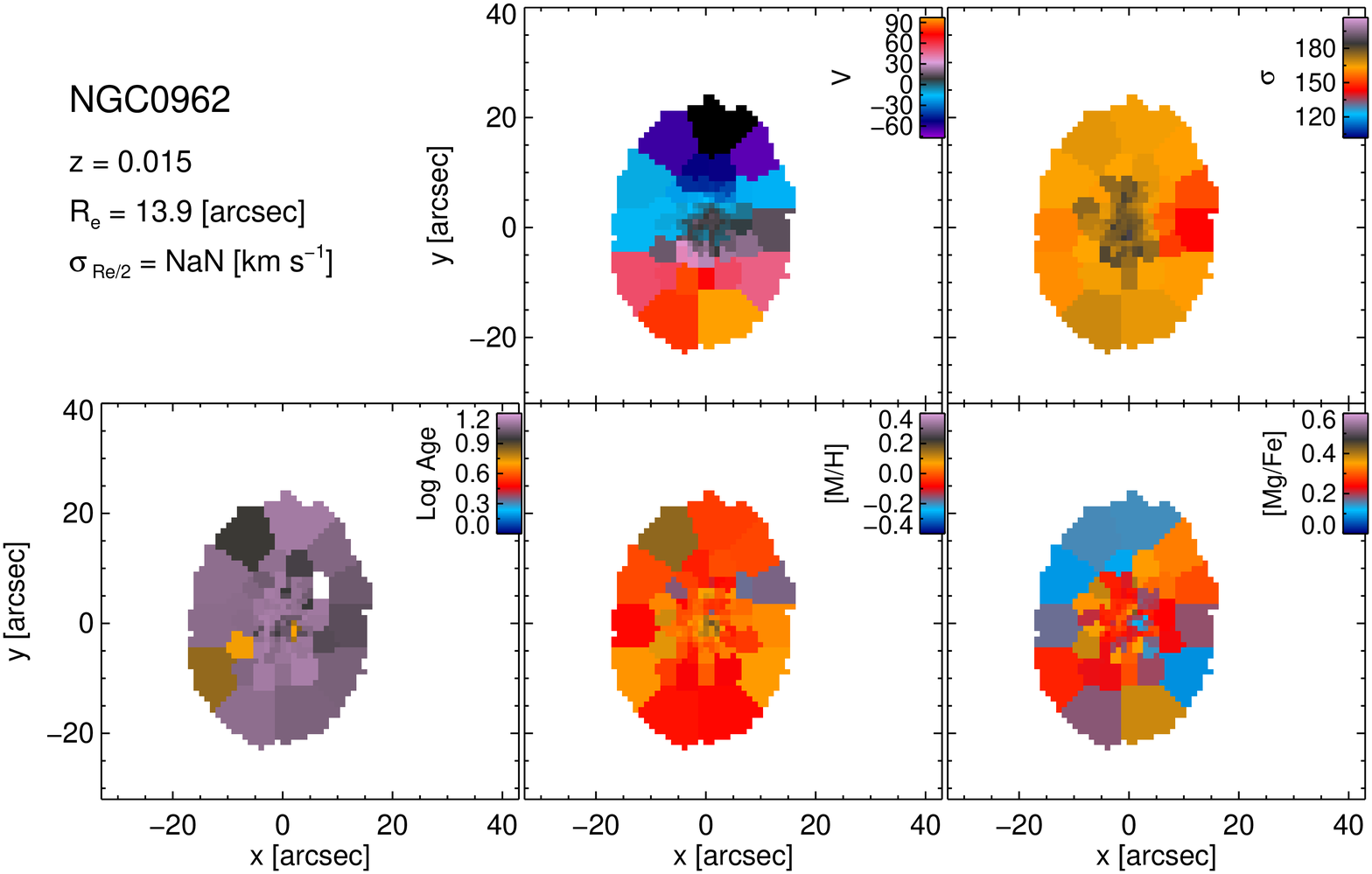}
\includegraphics[width=12cm]{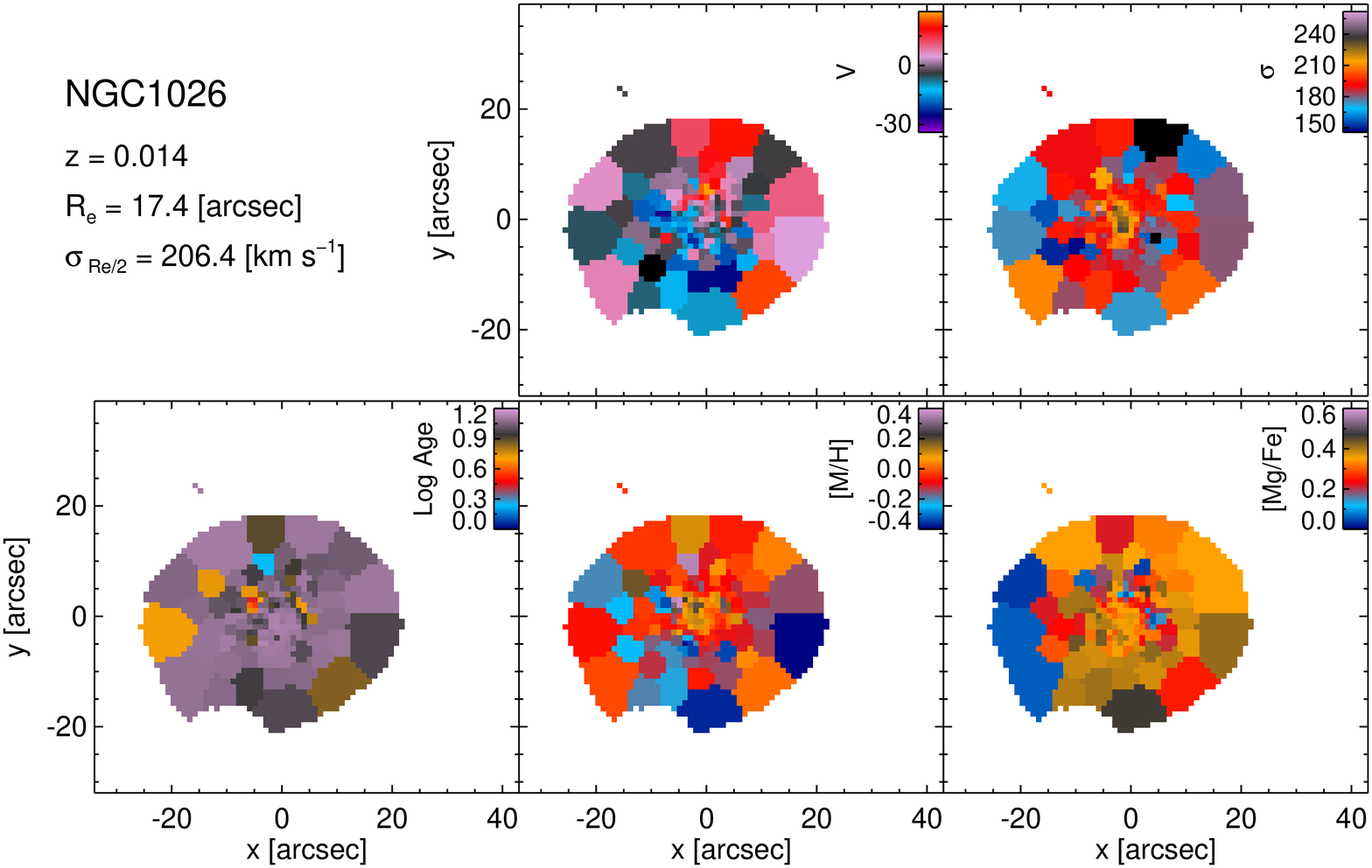}
\includegraphics[width=12cm]{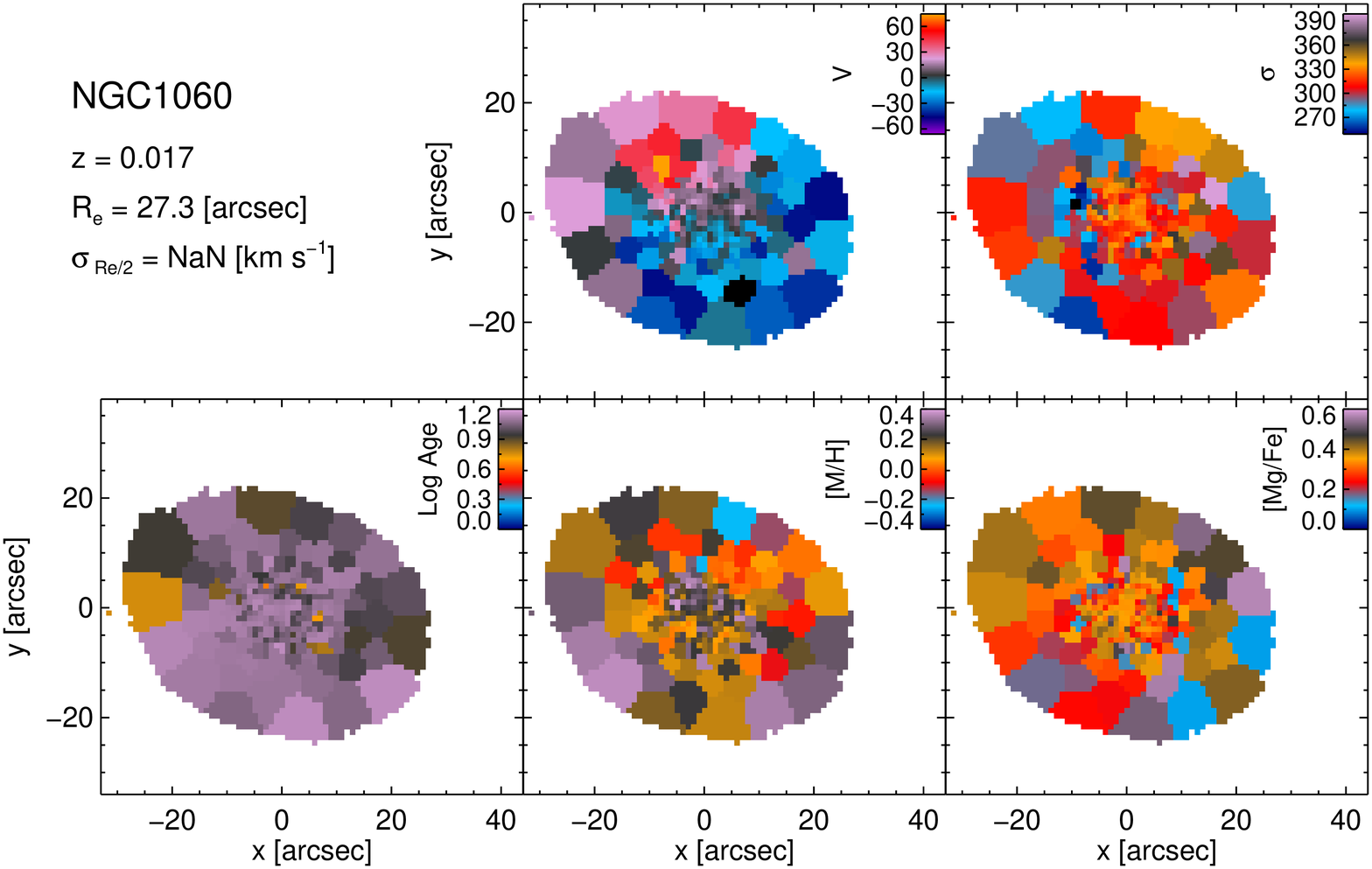}
\end{center}
\end{figure*}

\begin{figure*}
\begin{center}
\includegraphics[width=12cm]{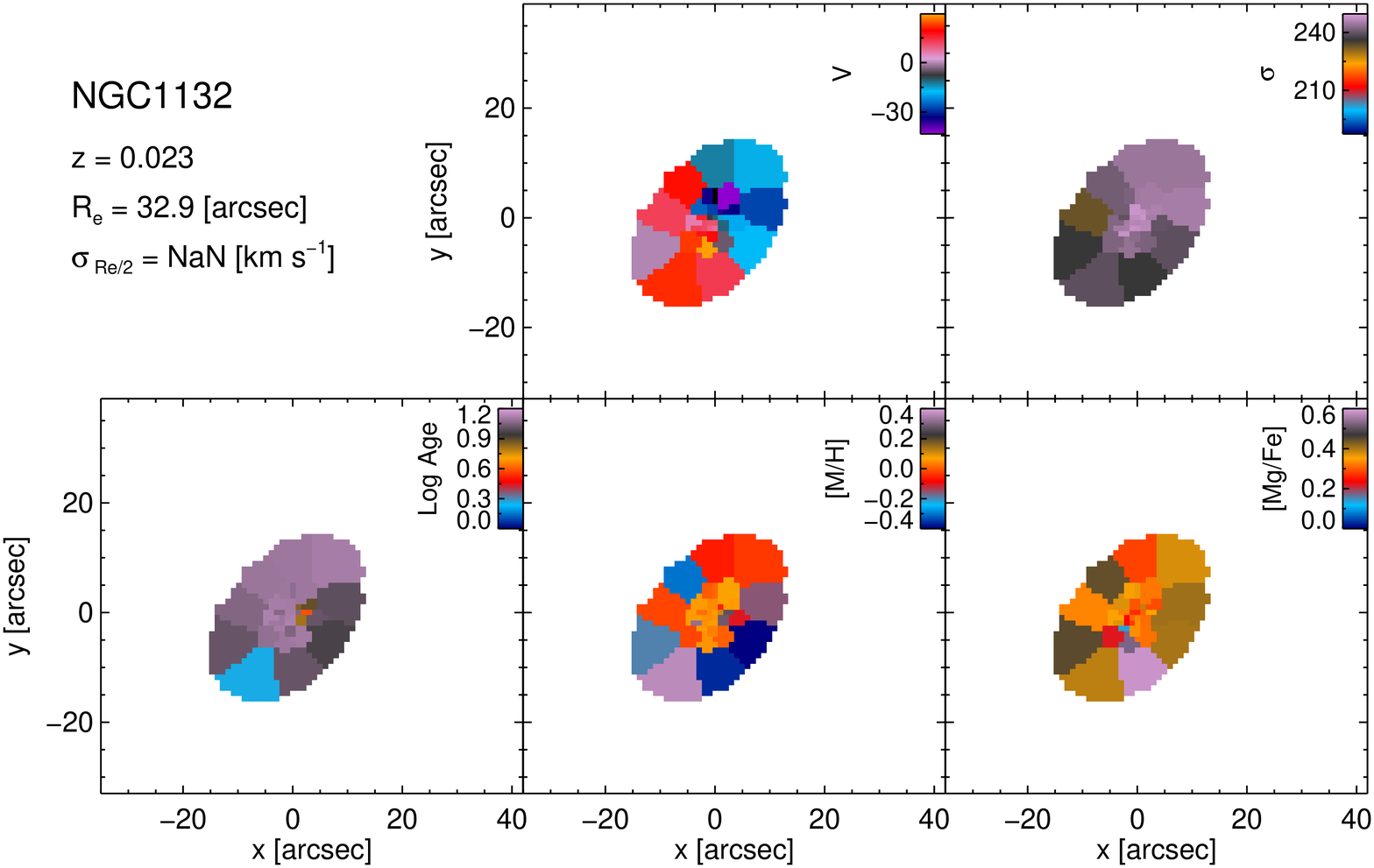}
\includegraphics[width=12cm]{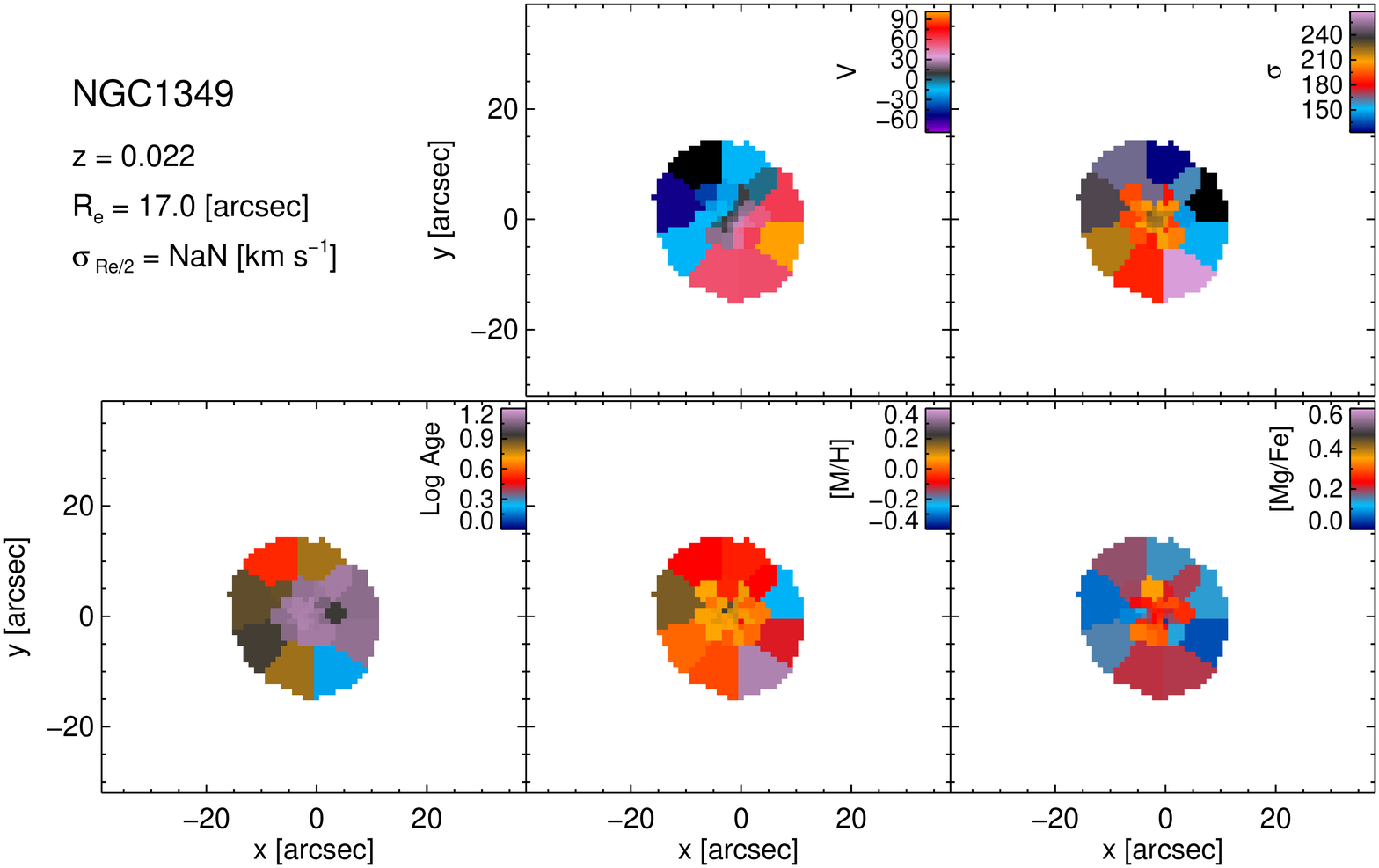}
\includegraphics[width=12cm]{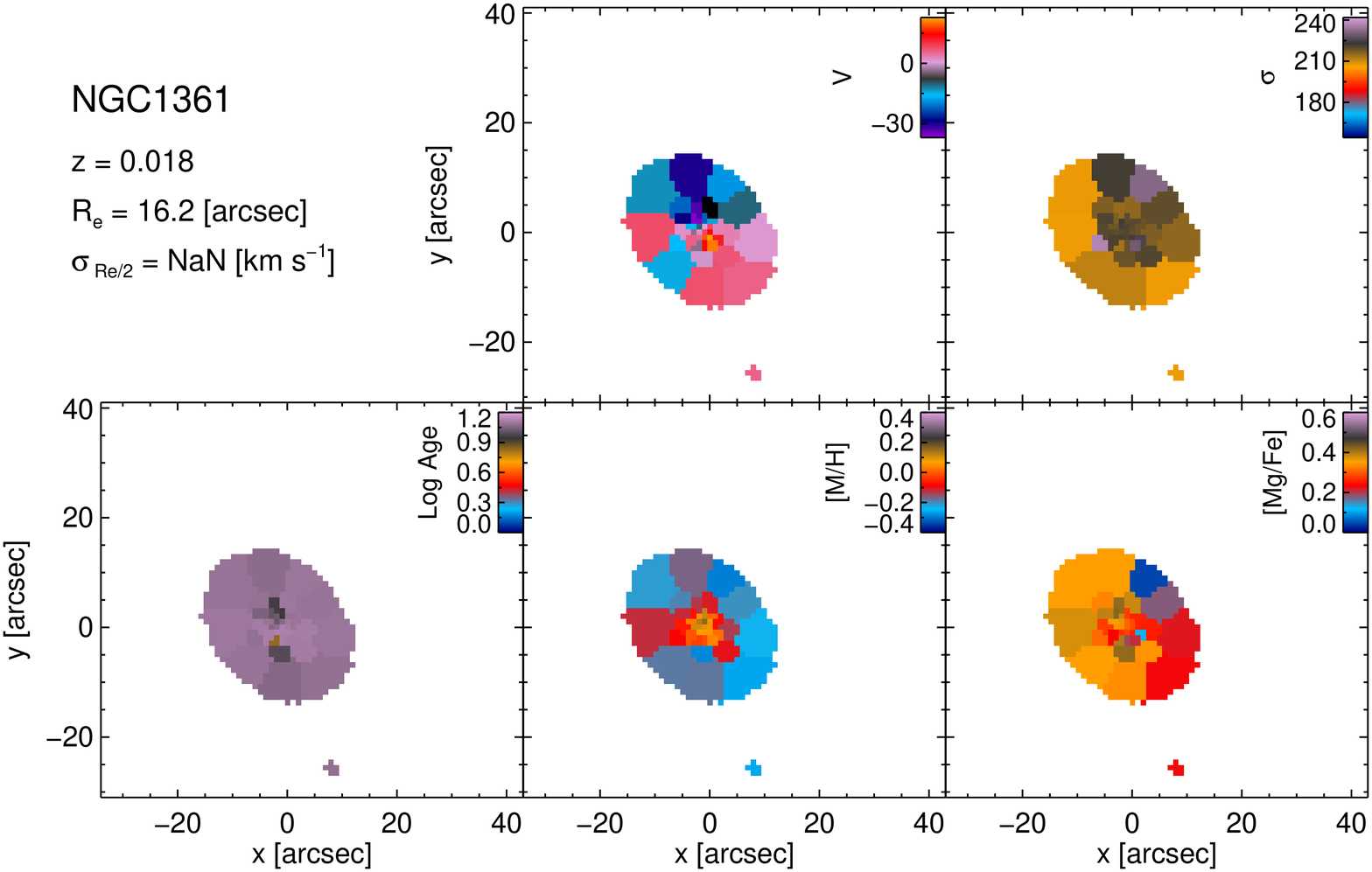}
\end{center}
\end{figure*}

\begin{figure*}
\begin{center}
\includegraphics[width=12cm]{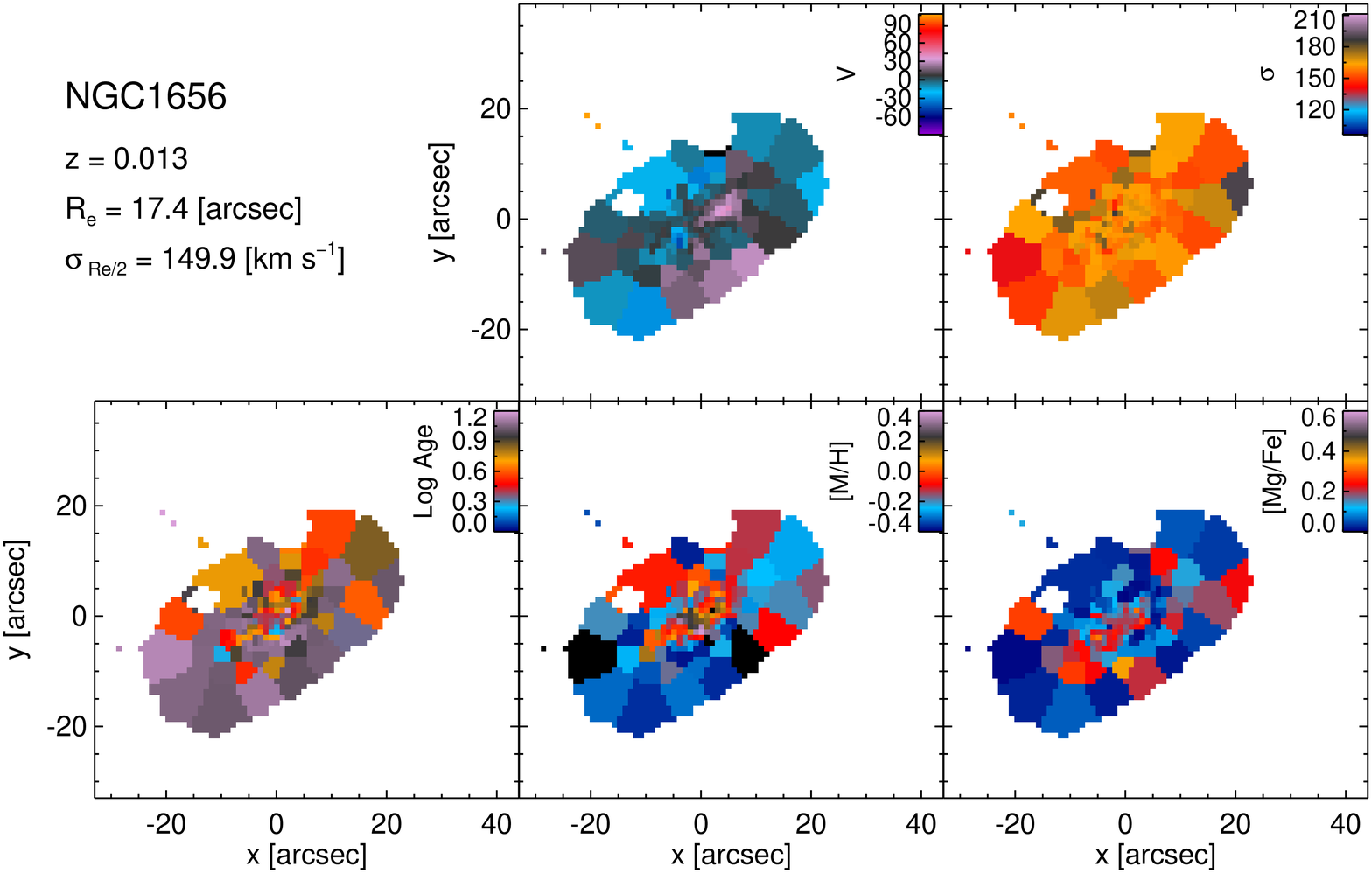}
\includegraphics[width=12cm]{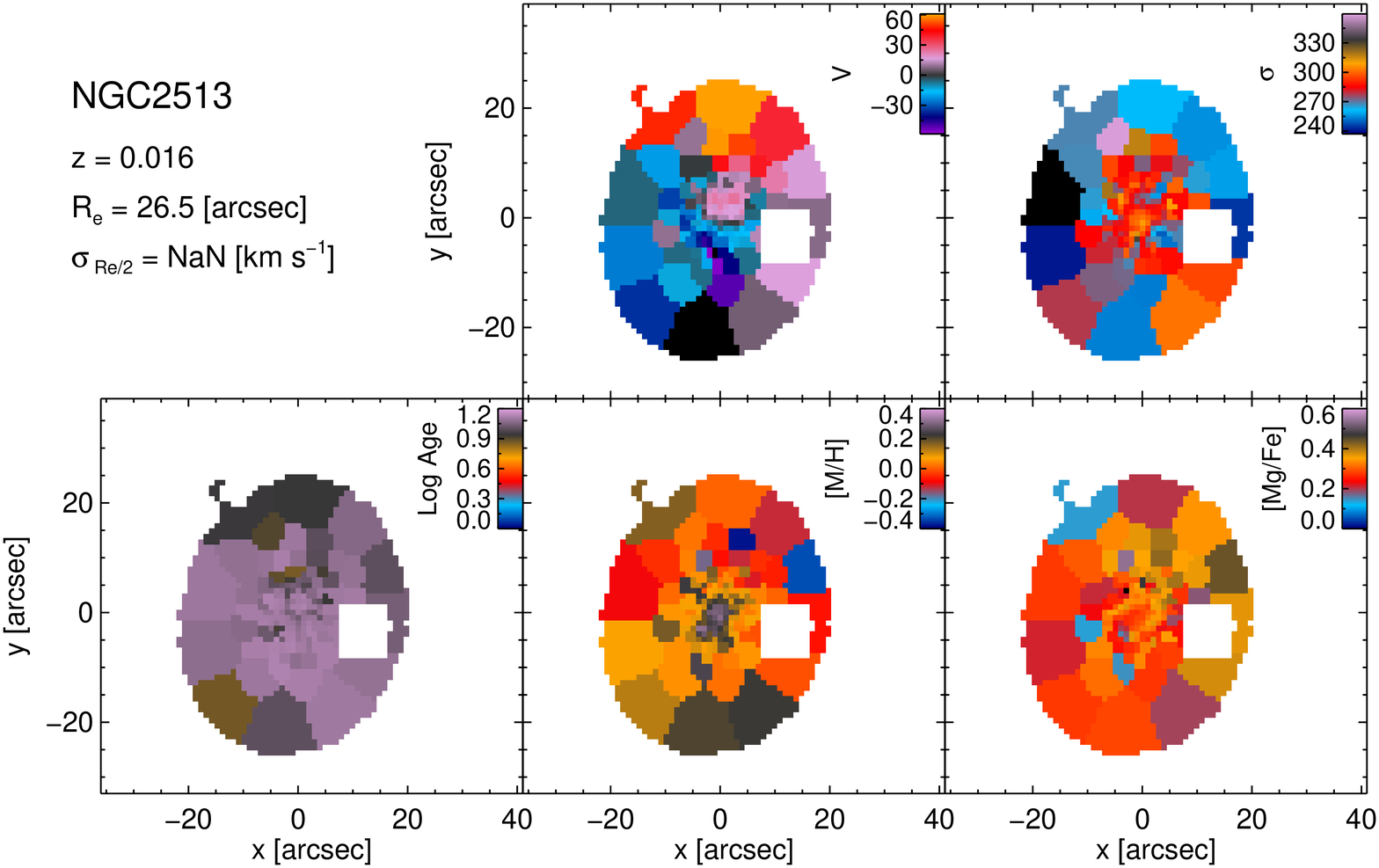}
\includegraphics[width=12cm]{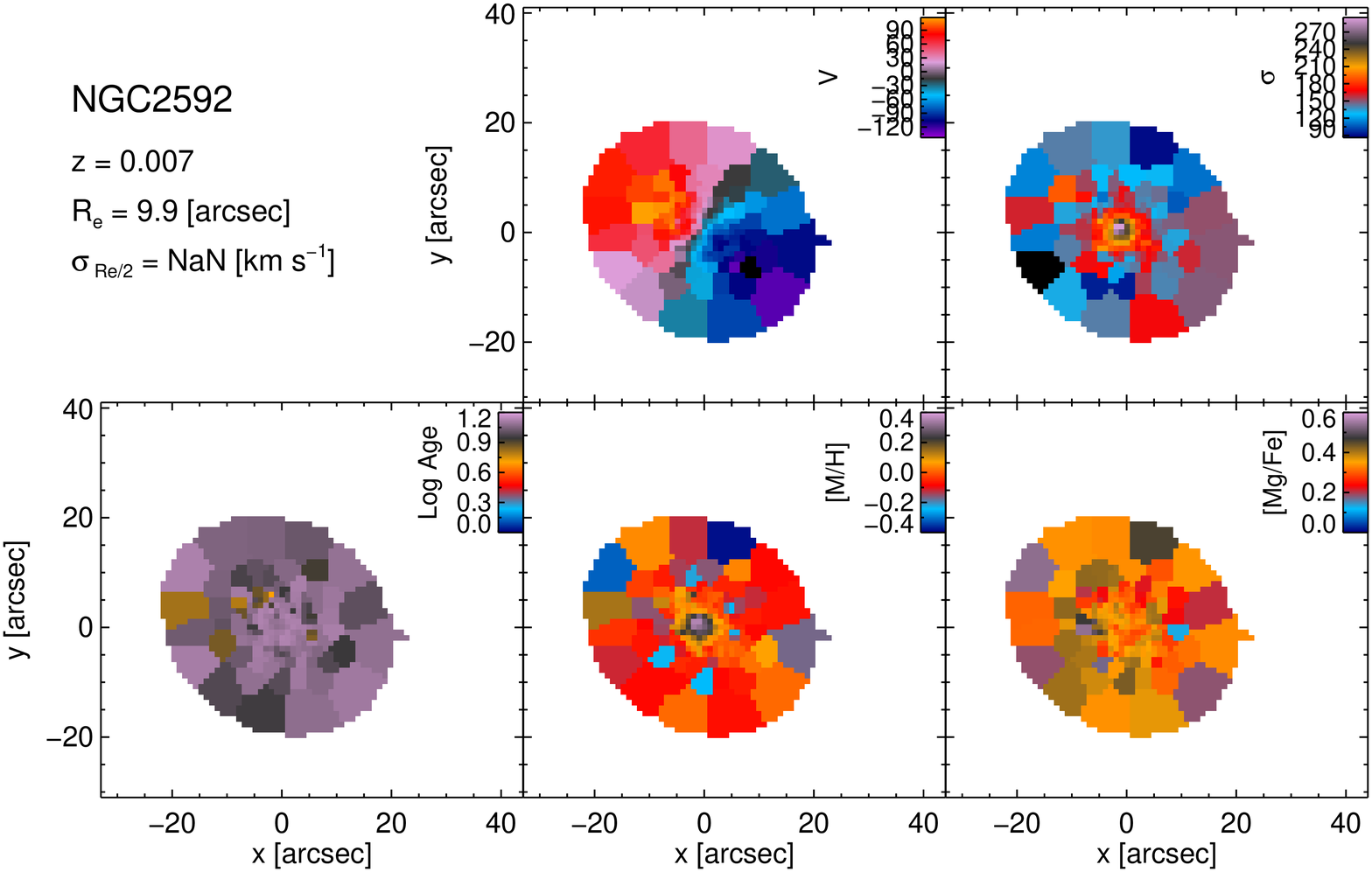}
\end{center}
\end{figure*}

\begin{figure*}
\begin{center}
\includegraphics[width=12cm]{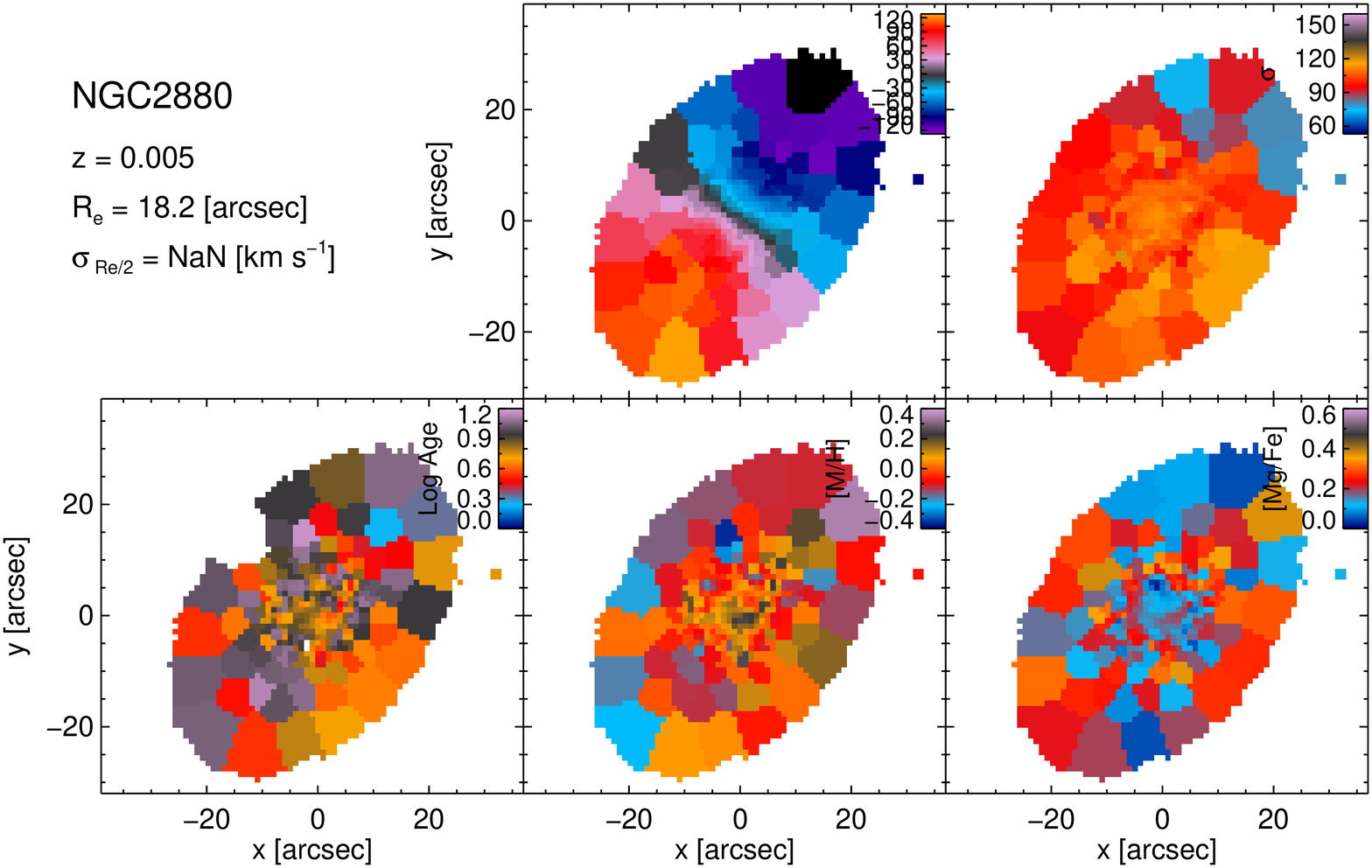}
\includegraphics[width=12cm]{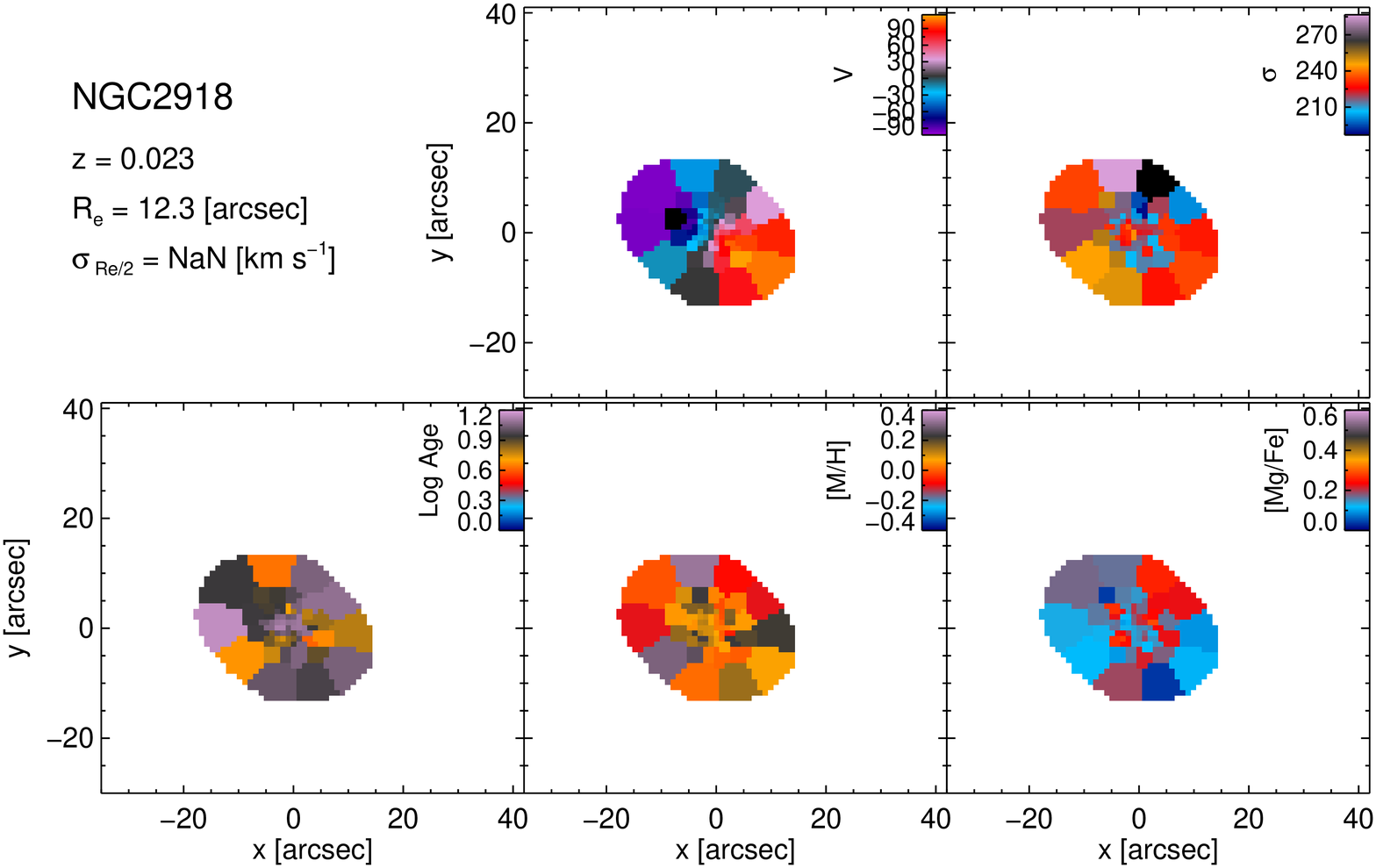}
\includegraphics[width=12cm]{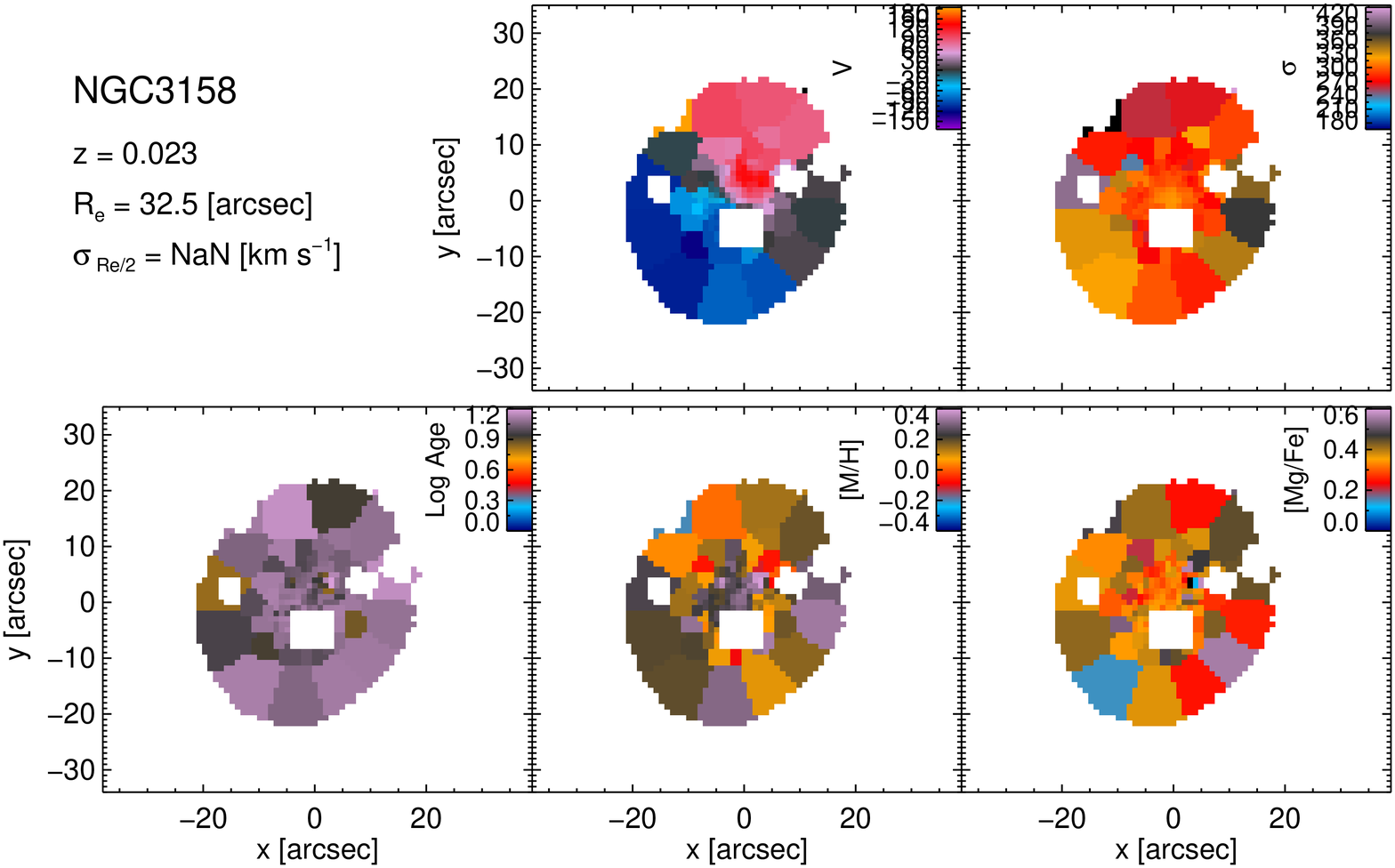}
\end{center}
\end{figure*}

\begin{figure*}
\begin{center}
\includegraphics[width=12cm]{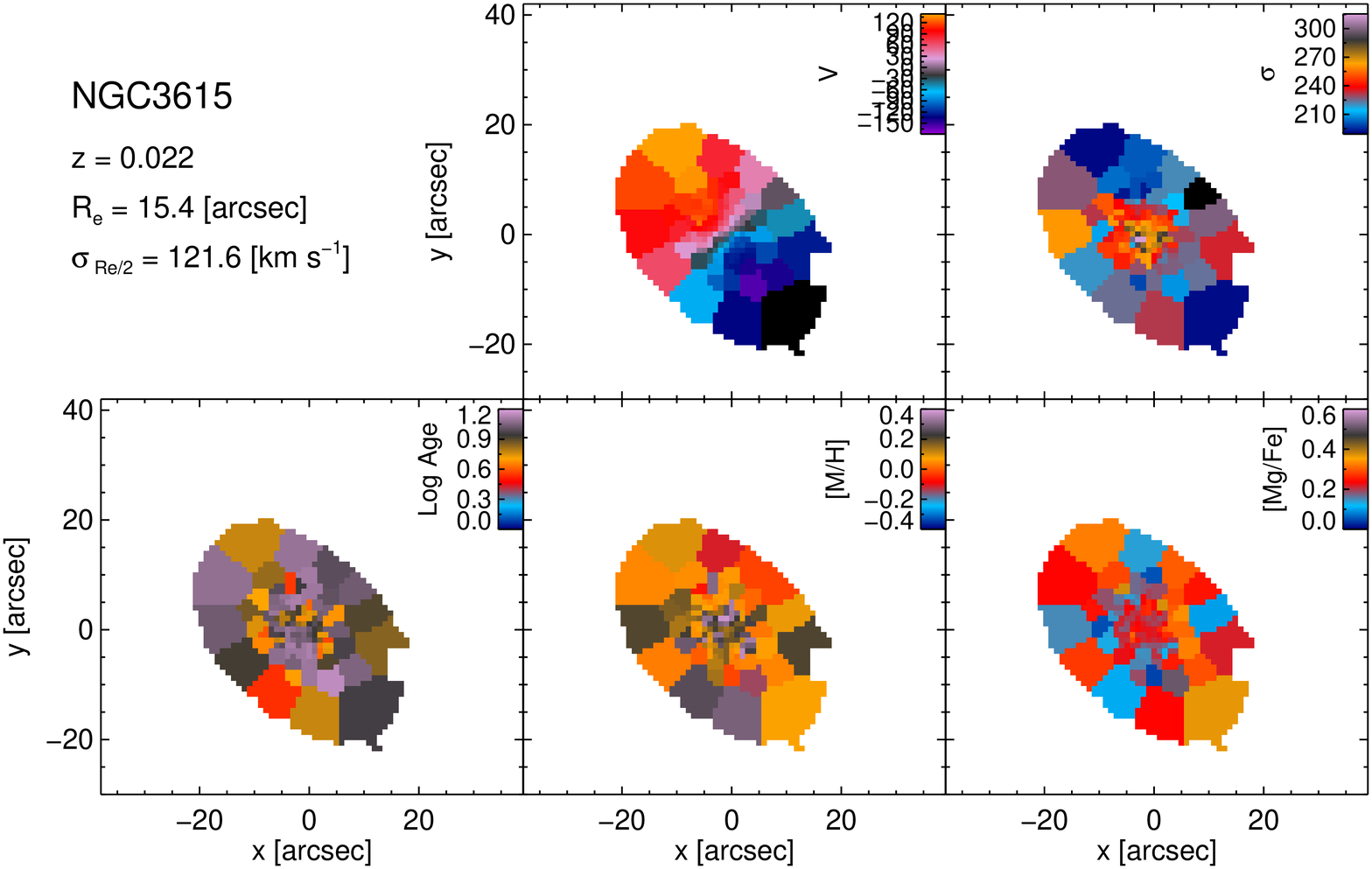}
\includegraphics[width=12cm]{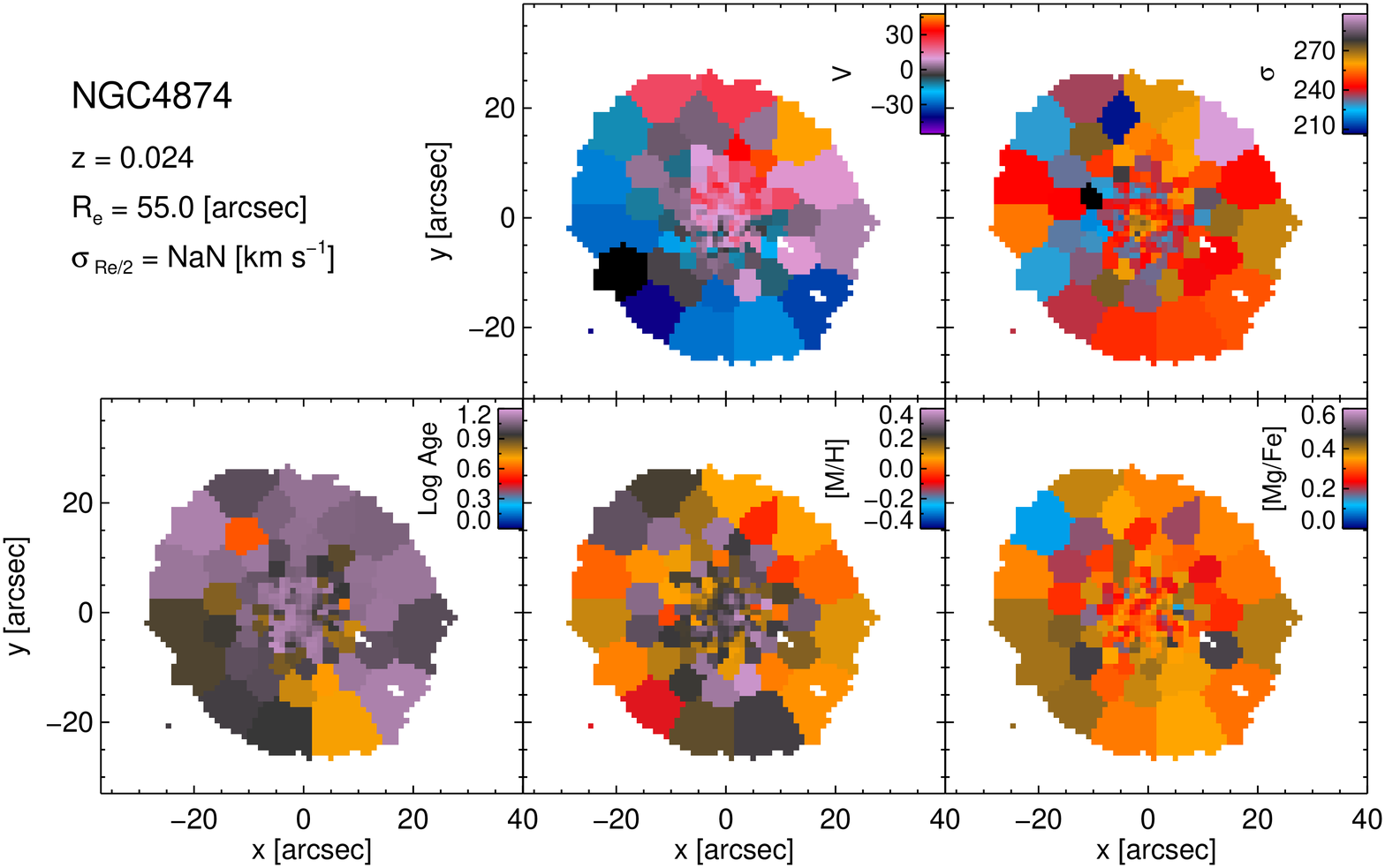}
\includegraphics[width=12cm]{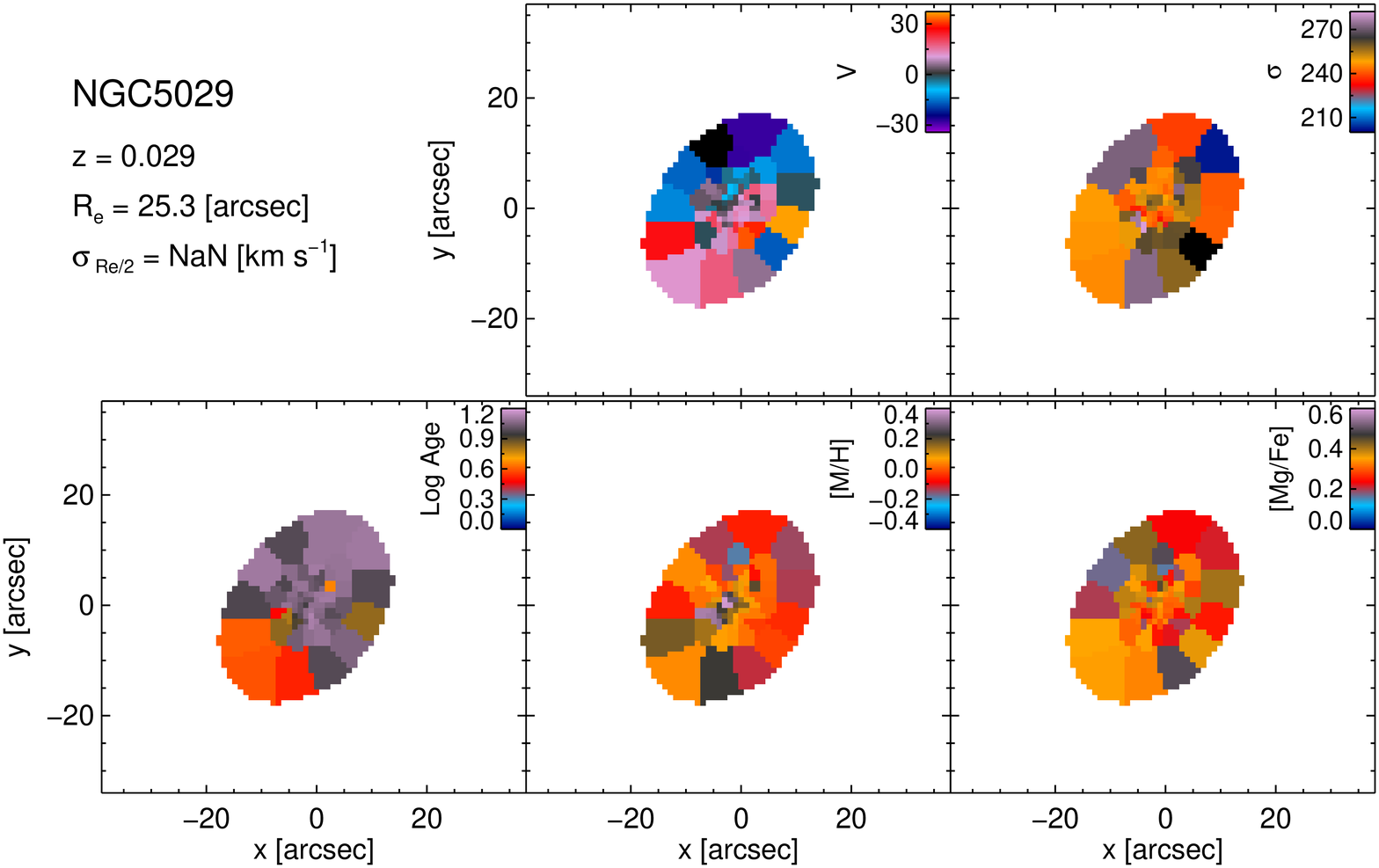}
\end{center}
\end{figure*}

\begin{figure*}
\begin{center}
\includegraphics[width=12cm]{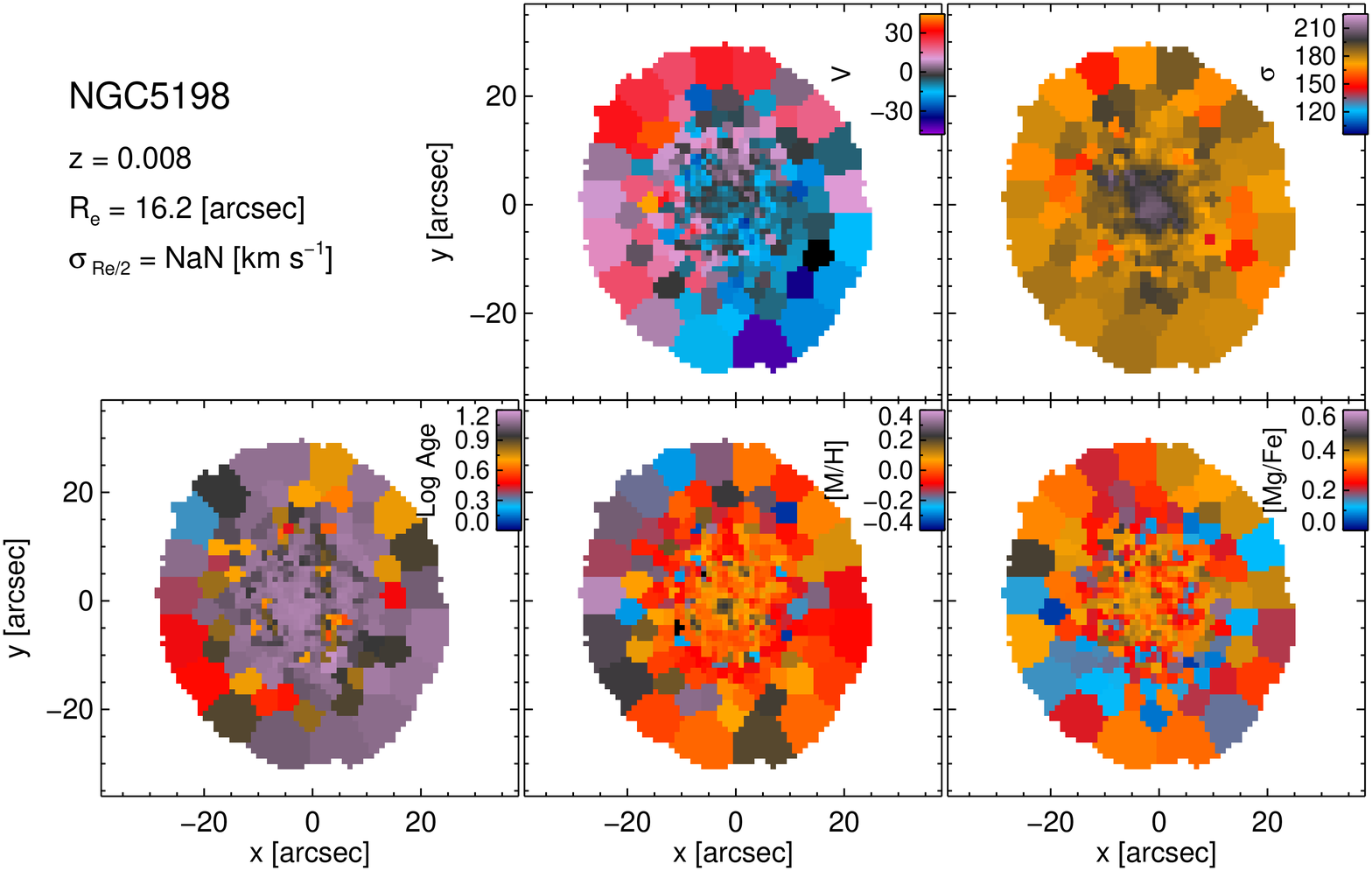}
\includegraphics[width=12cm]{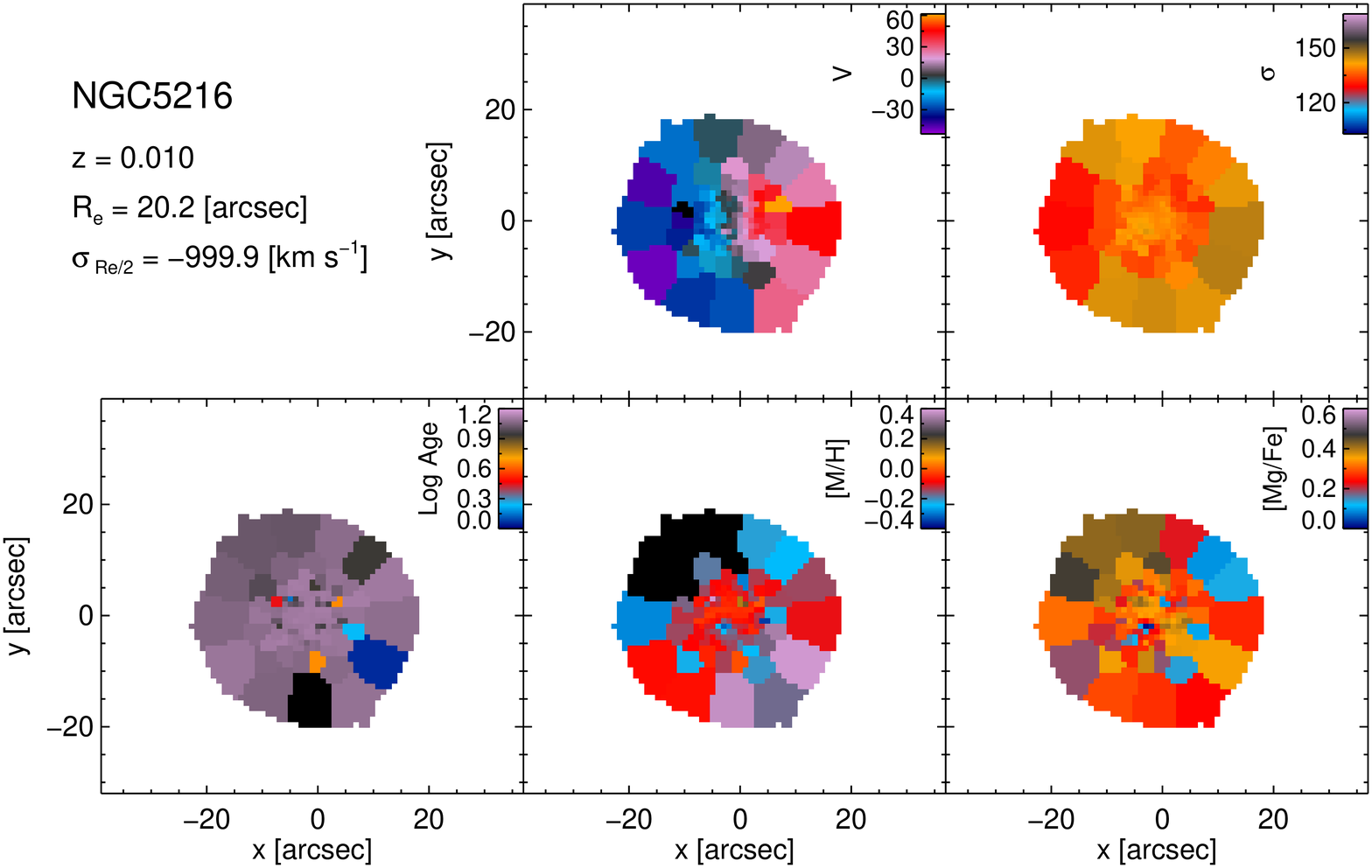}
\includegraphics[width=12cm]{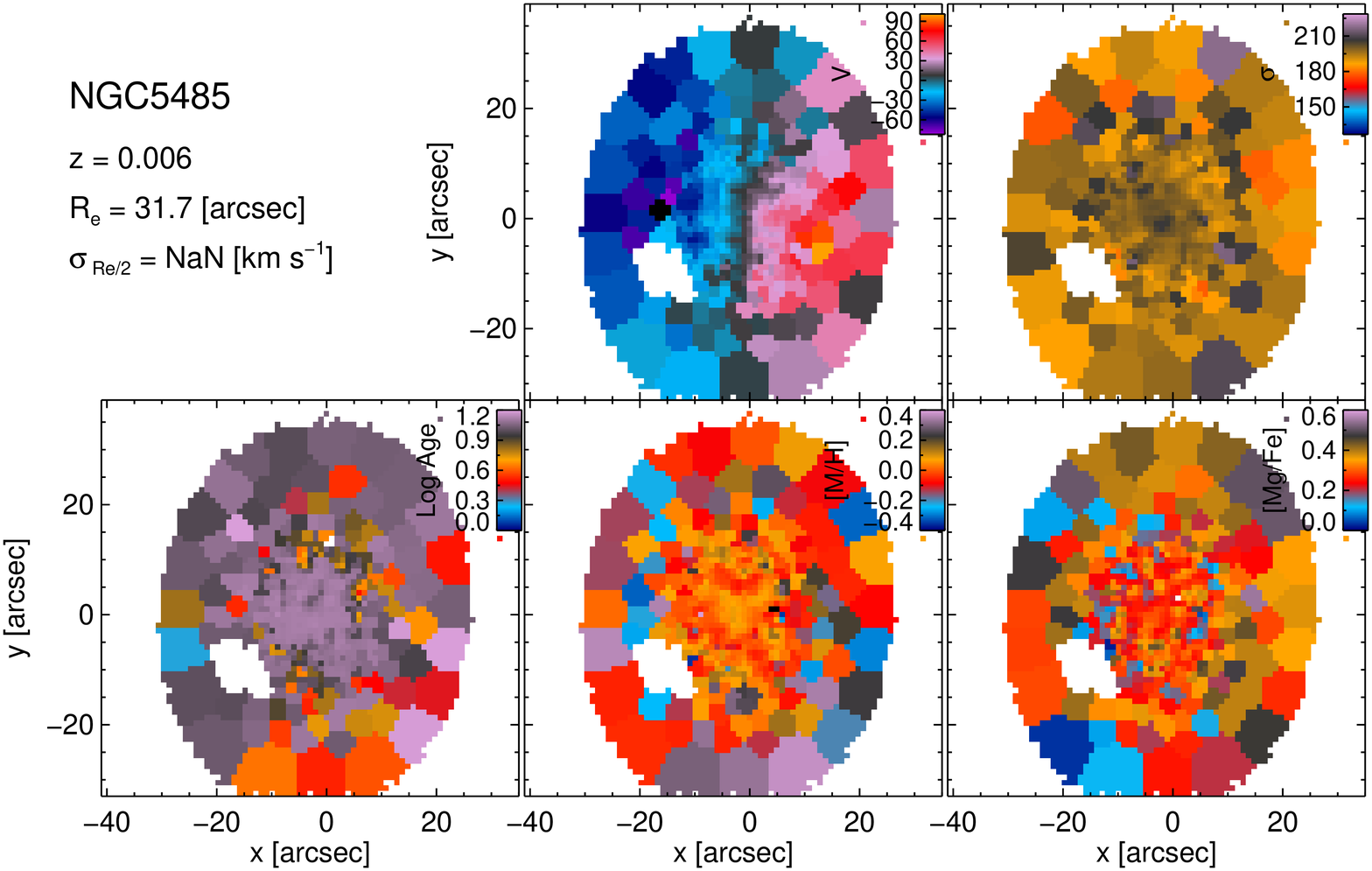}
\end{center}
\end{figure*}

\begin{figure*}
\begin{center}
\includegraphics[width=12cm]{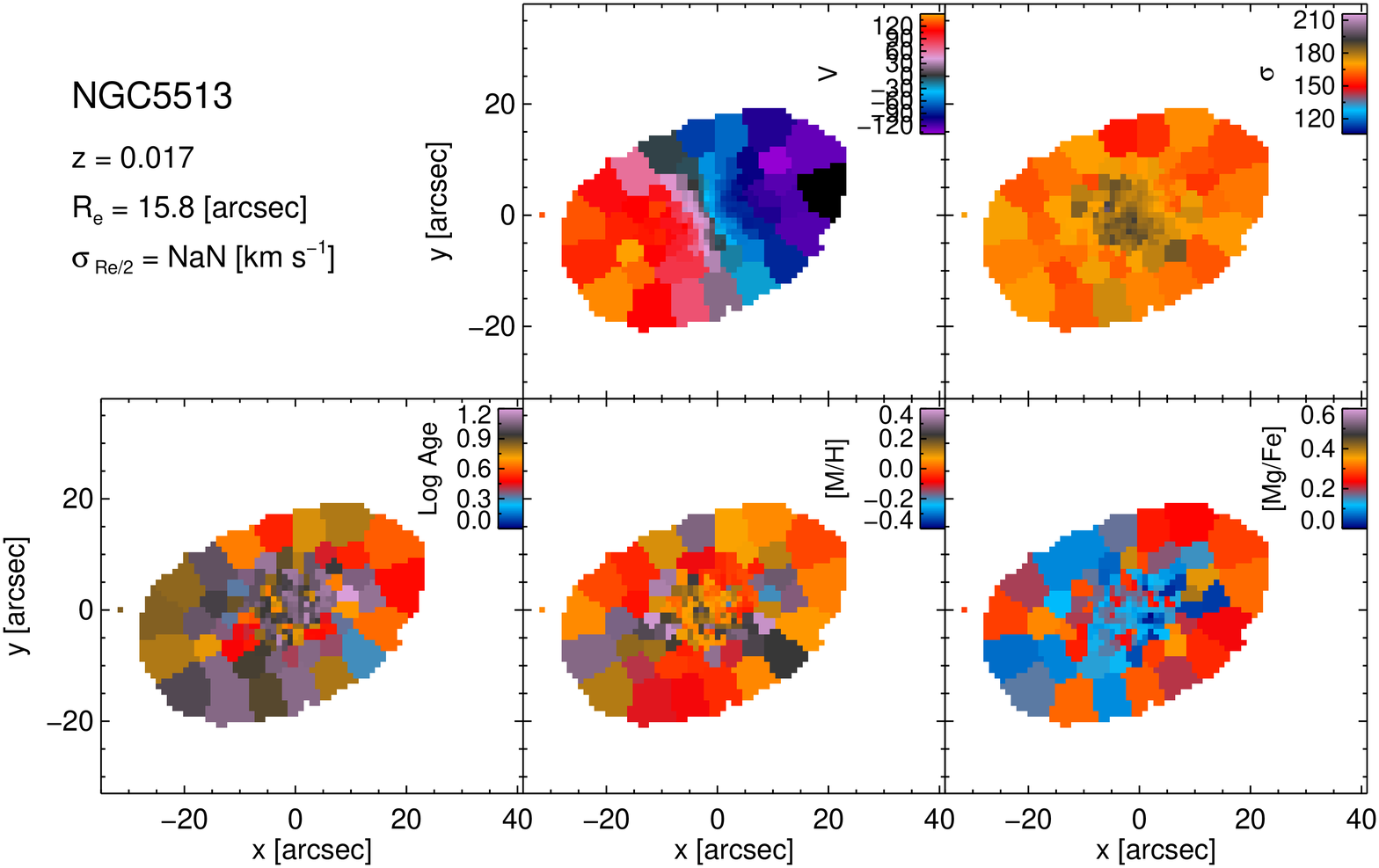}
\includegraphics[width=12cm]{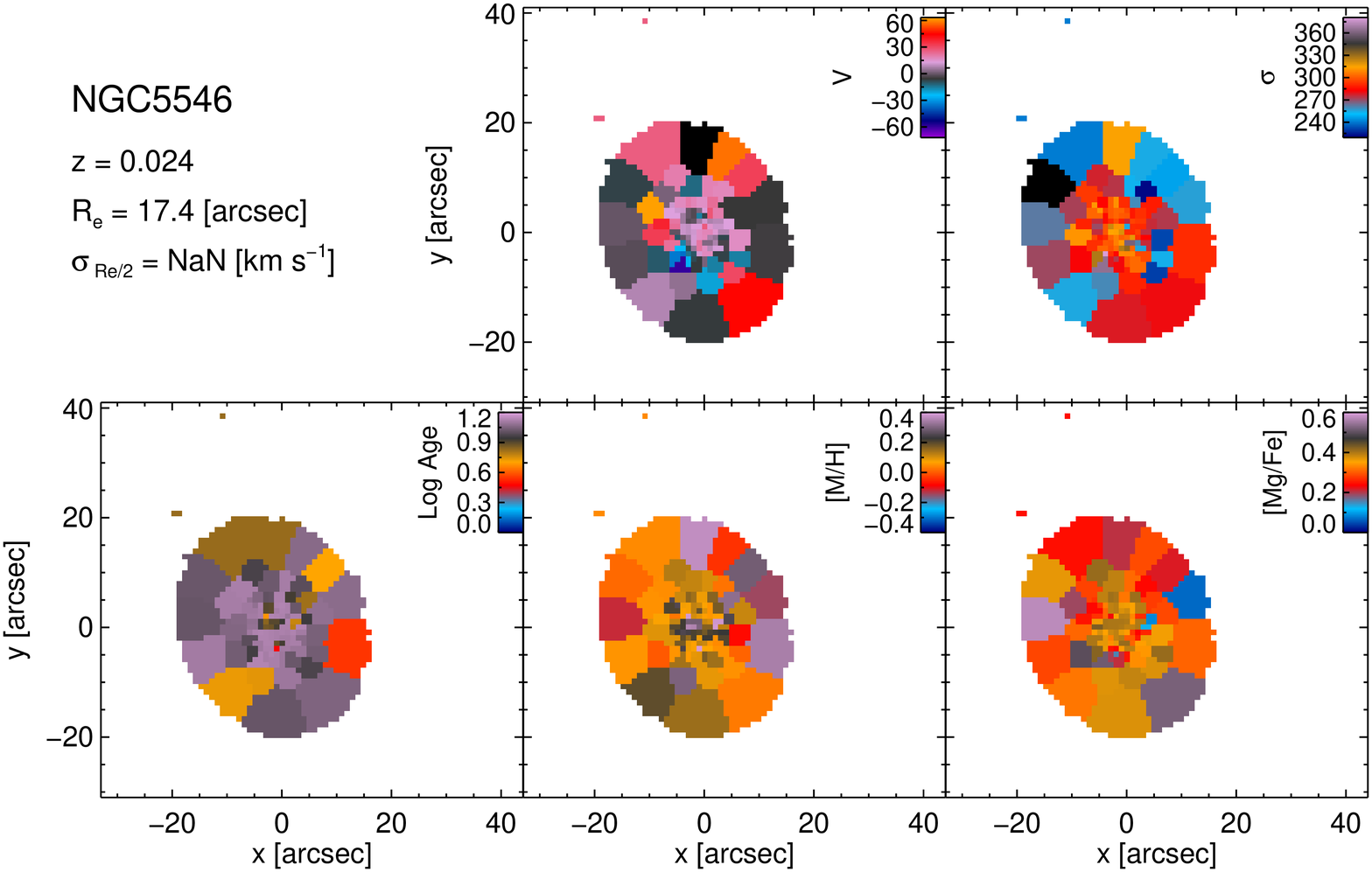}
\includegraphics[width=12cm]{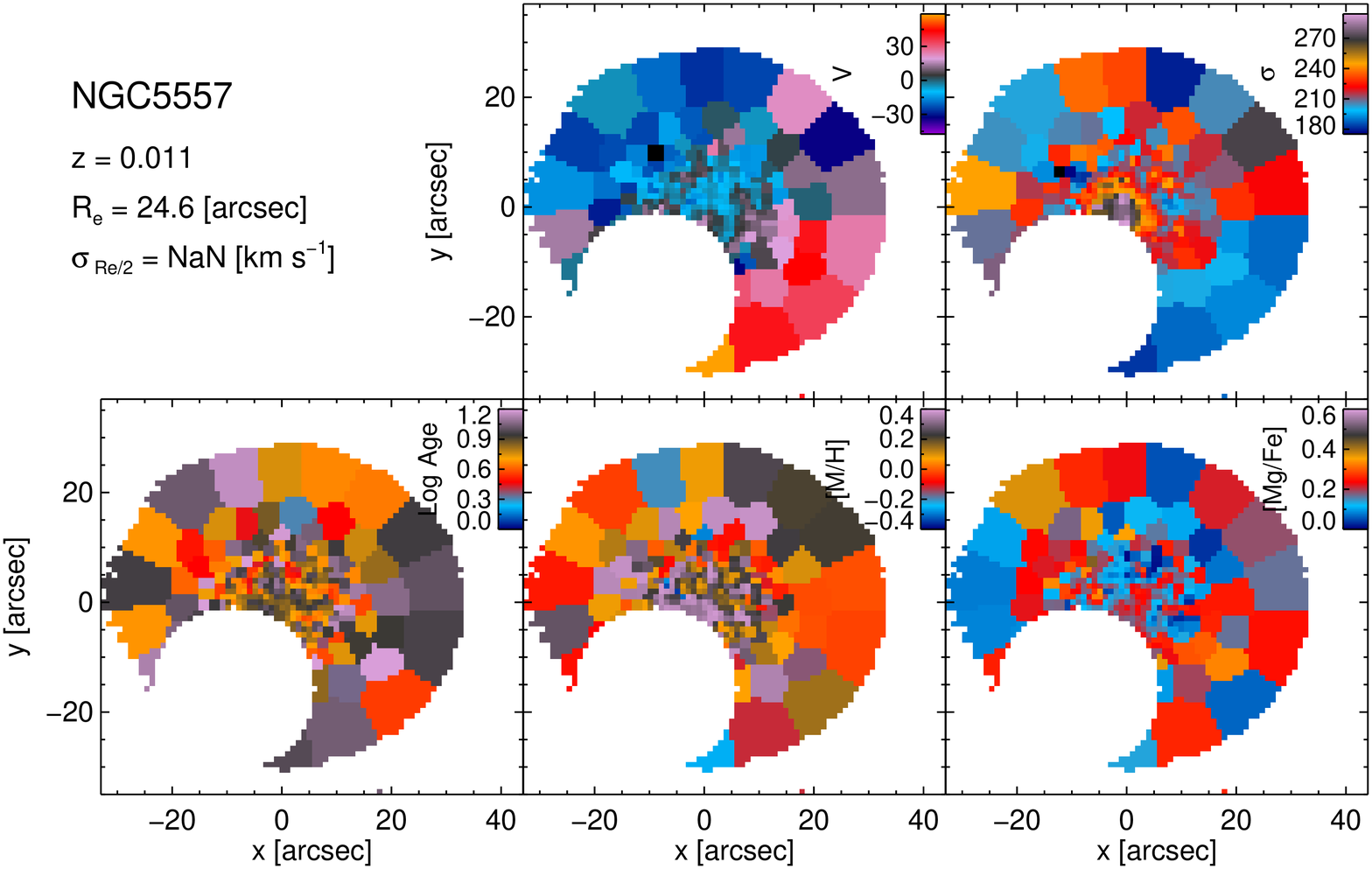}
\end{center}
\end{figure*}

\begin{figure*}
\begin{center}
\includegraphics[width=12cm]{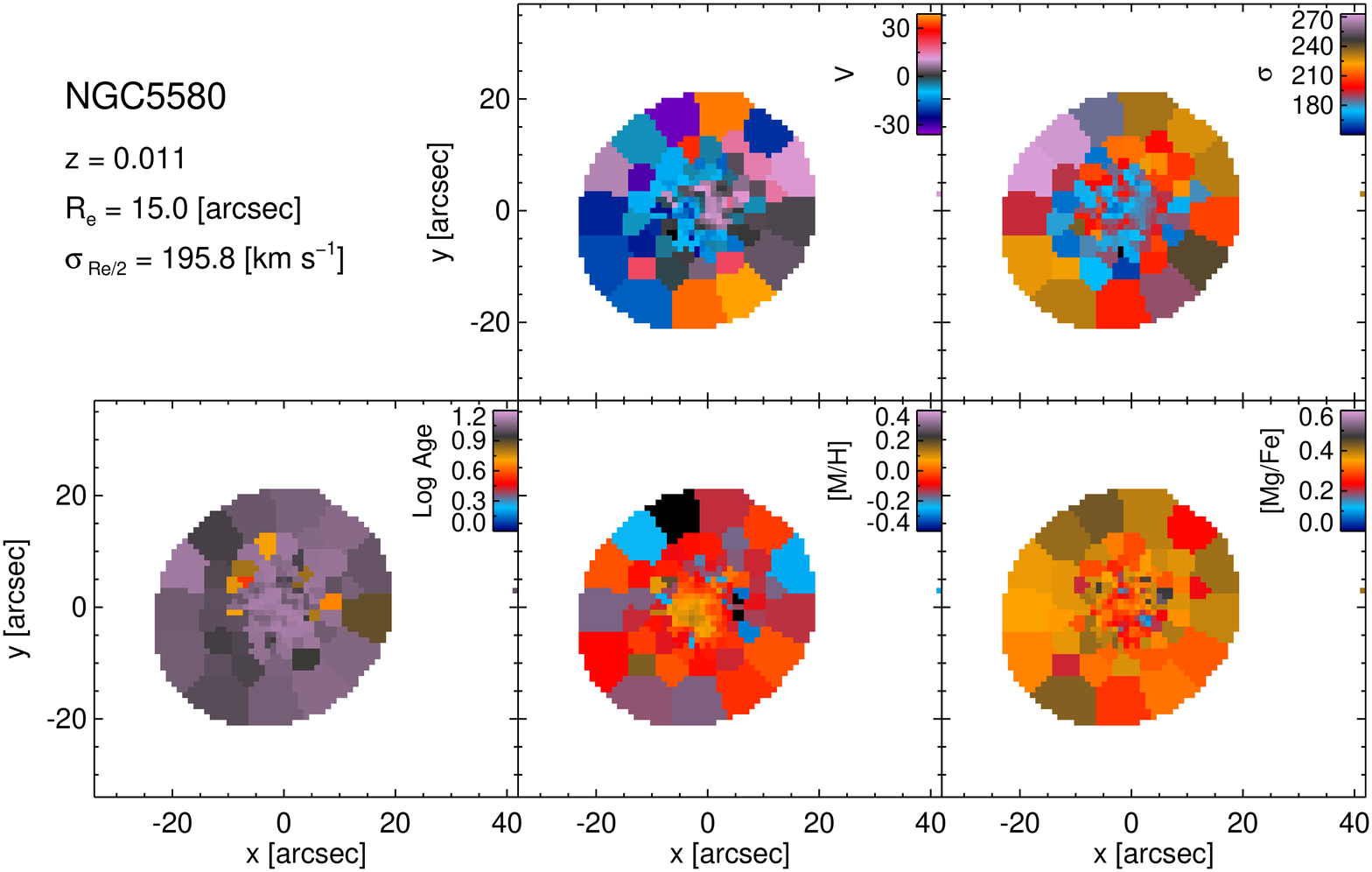}
\includegraphics[width=12cm]{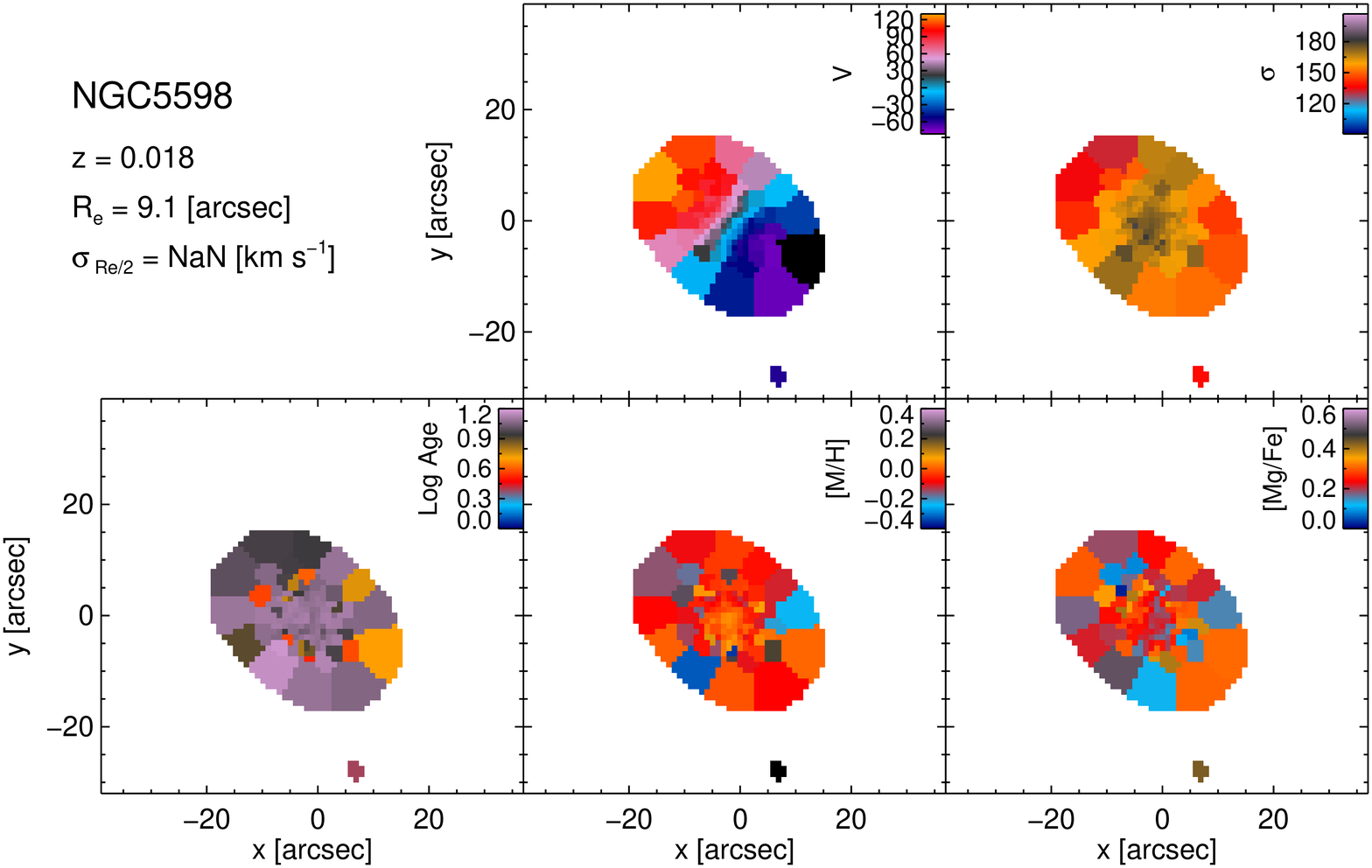}
\includegraphics[width=12cm]{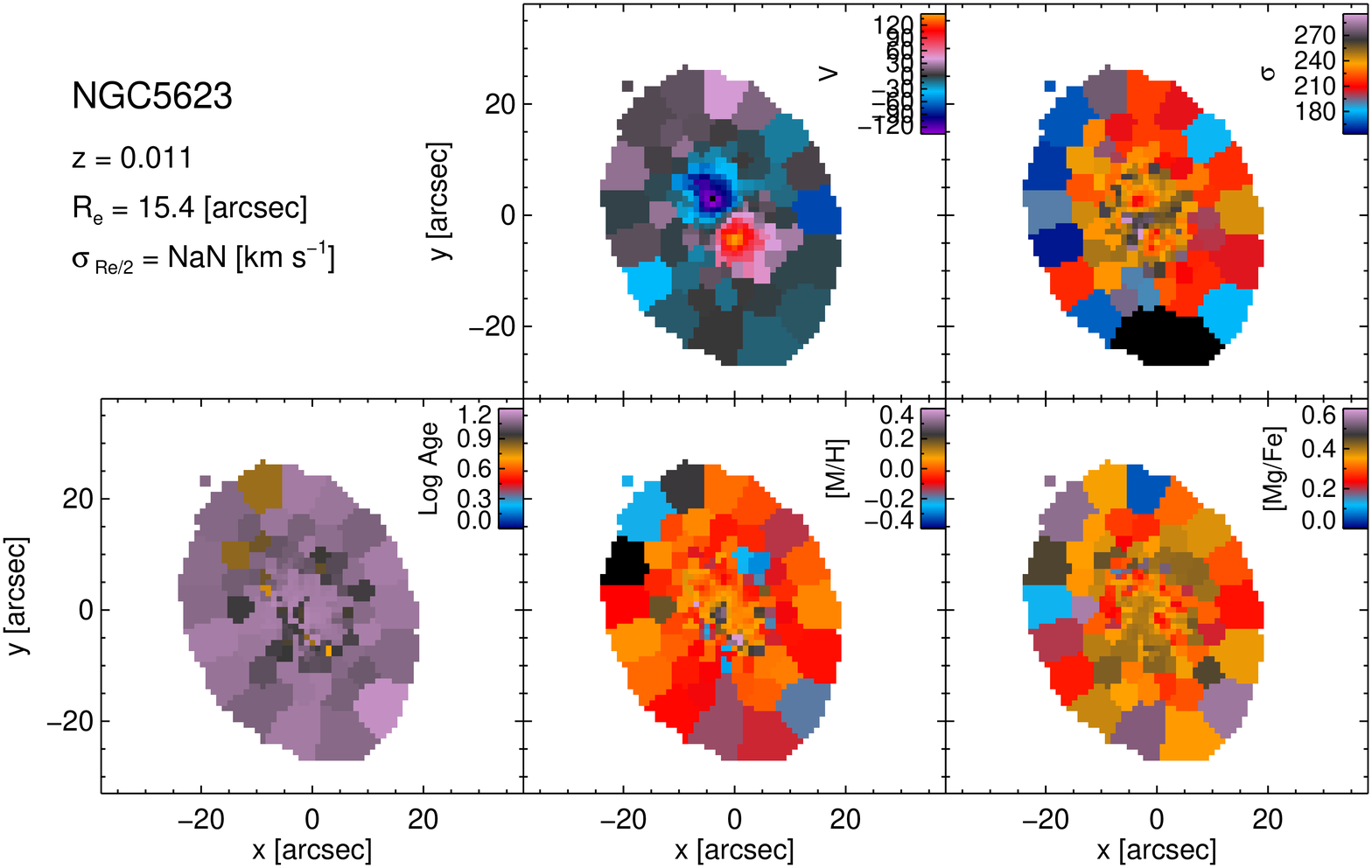}
\end{center}
\end{figure*}

\begin{figure*}
\begin{center}
\includegraphics[width=12cm]{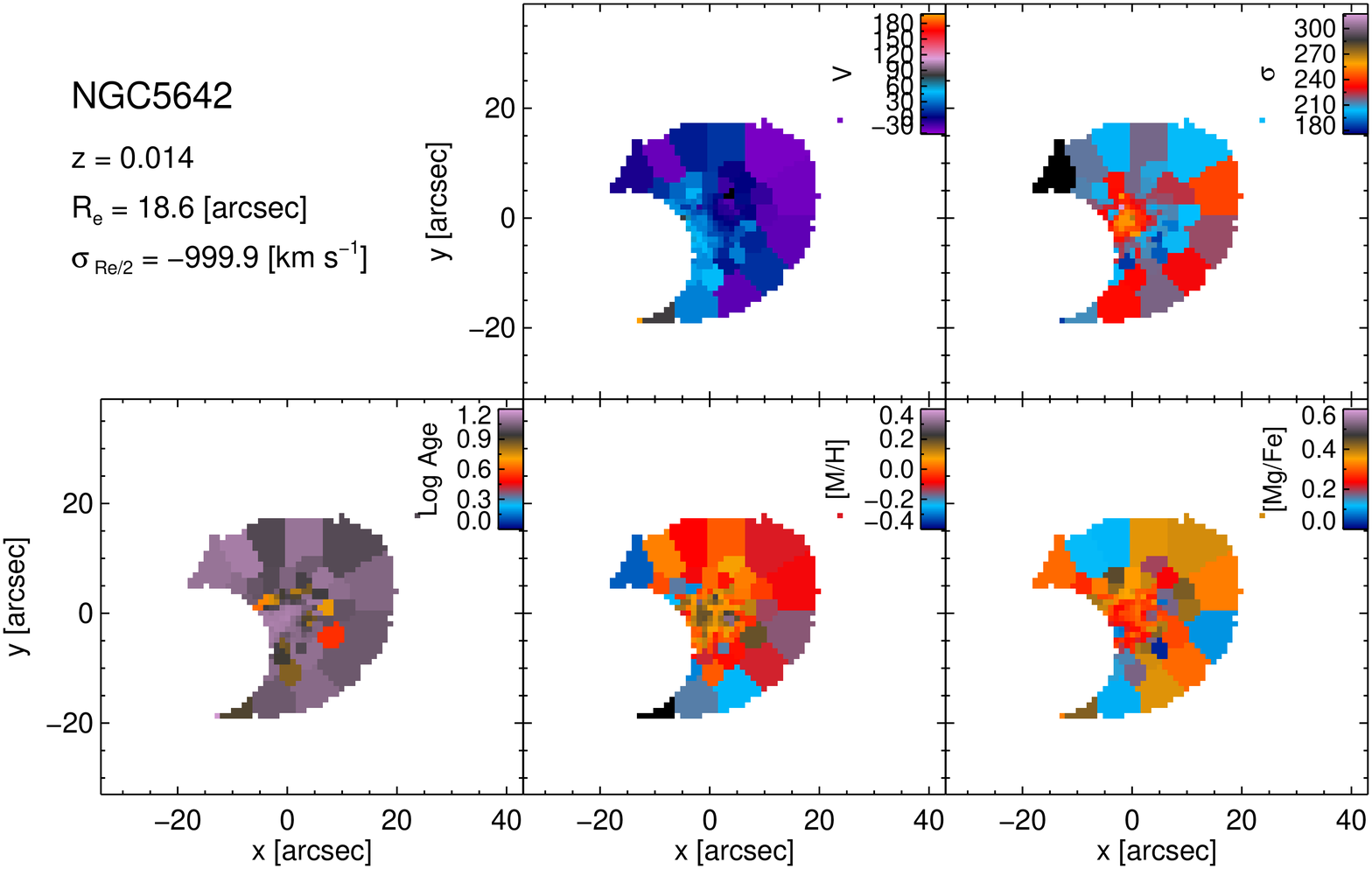}
\includegraphics[width=12cm]{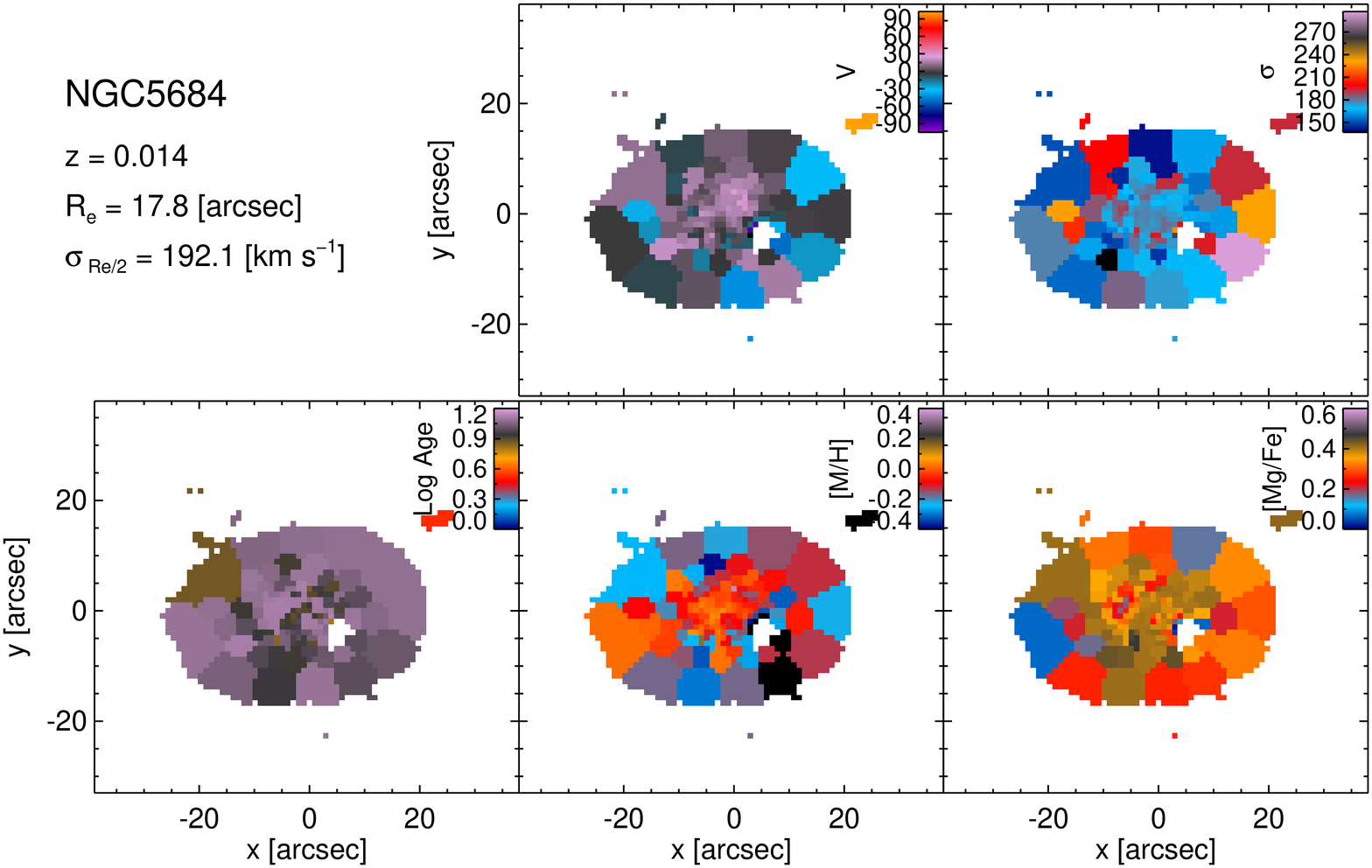}
\includegraphics[width=12cm]{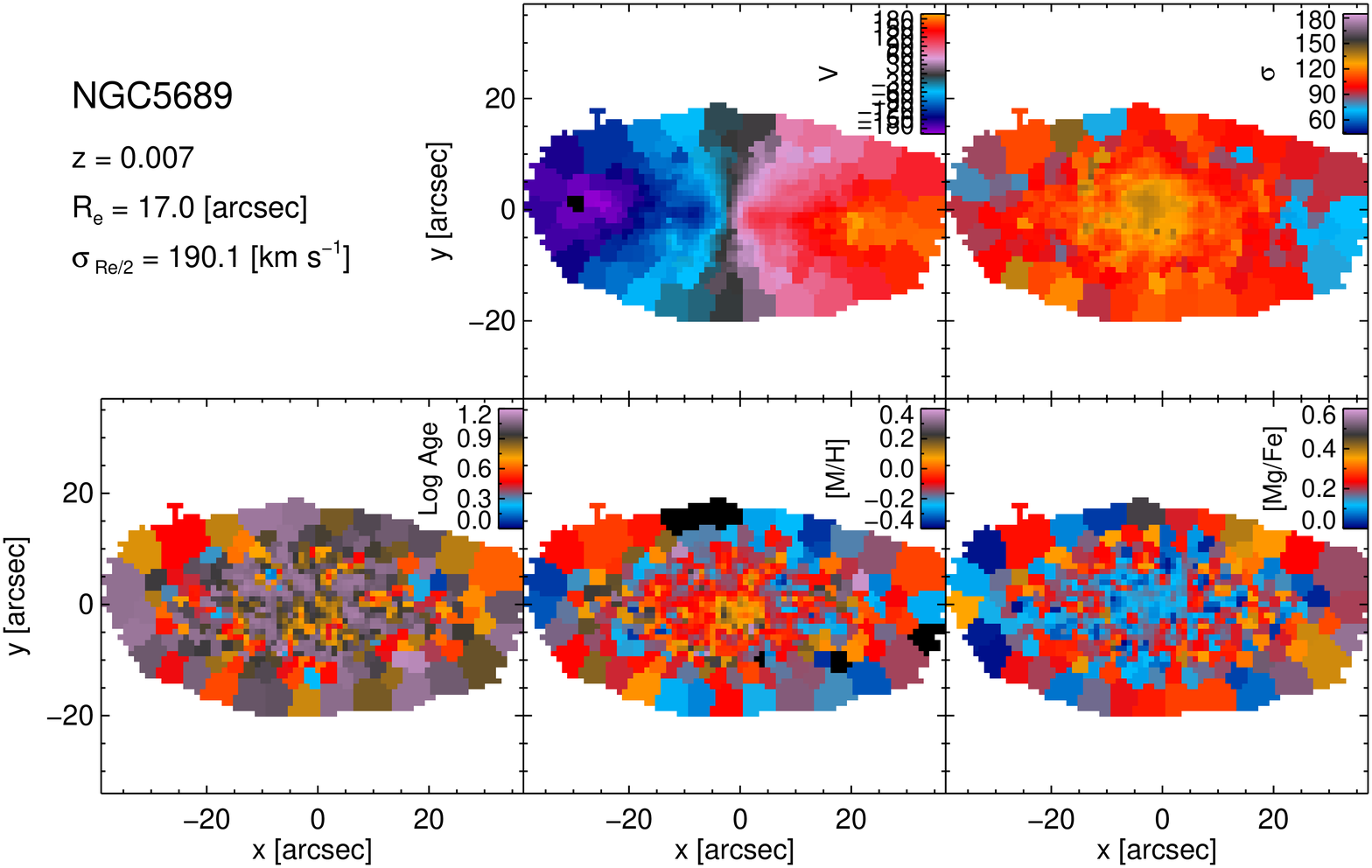}
\end{center}
\end{figure*}

\begin{figure*}
\begin{center}
\includegraphics[width=12cm]{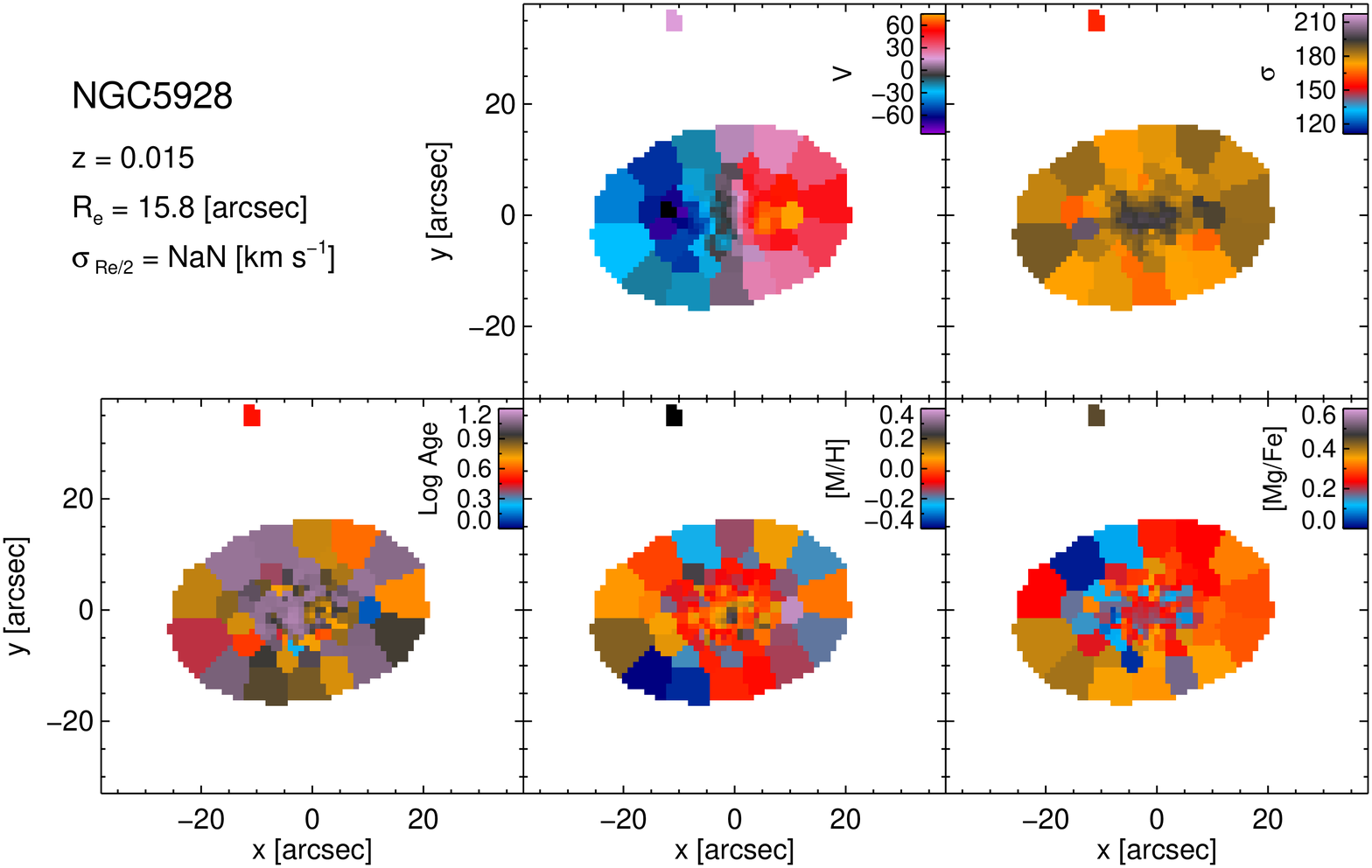}
\includegraphics[width=12cm]{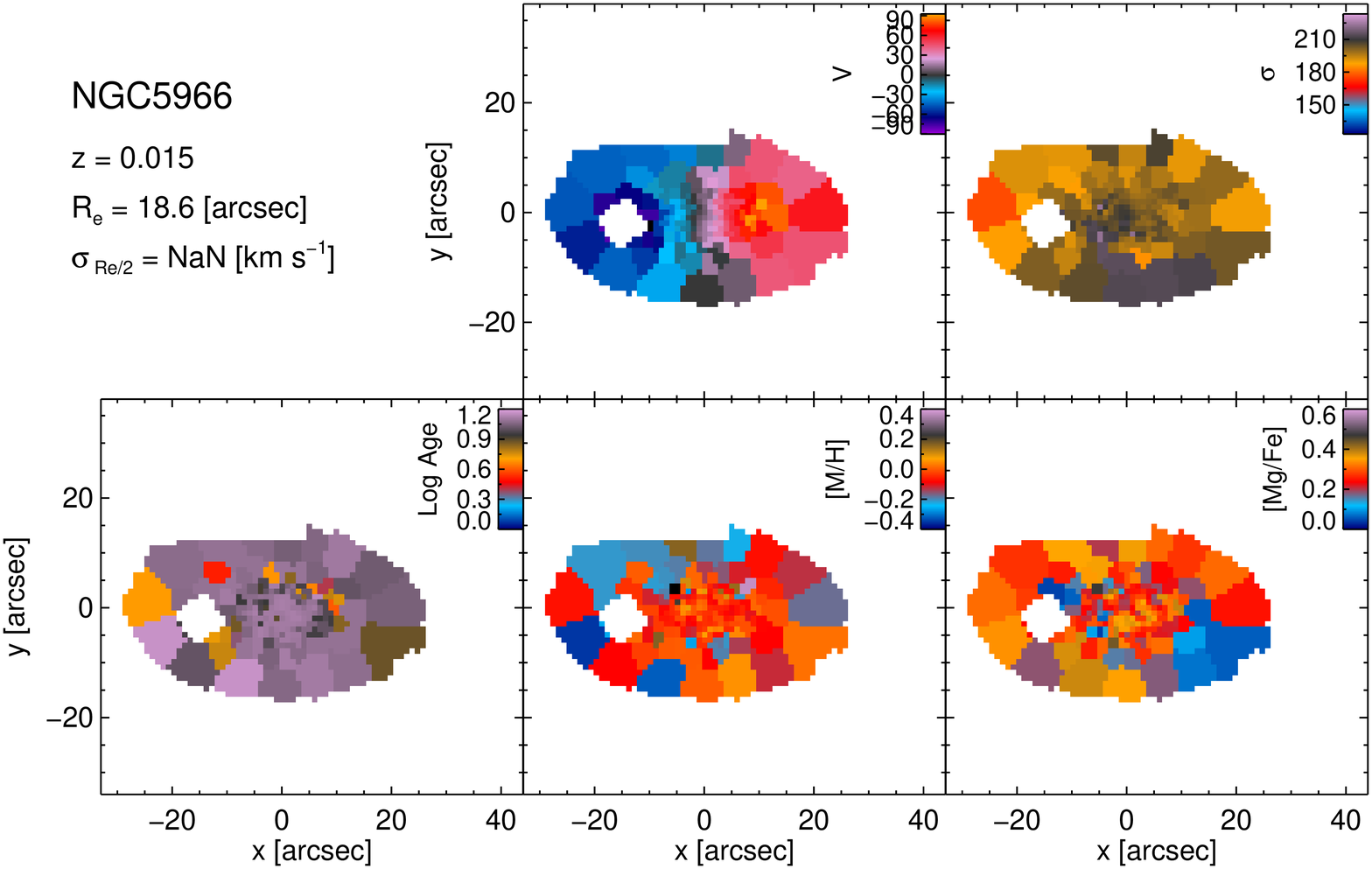}
\includegraphics[width=12cm]{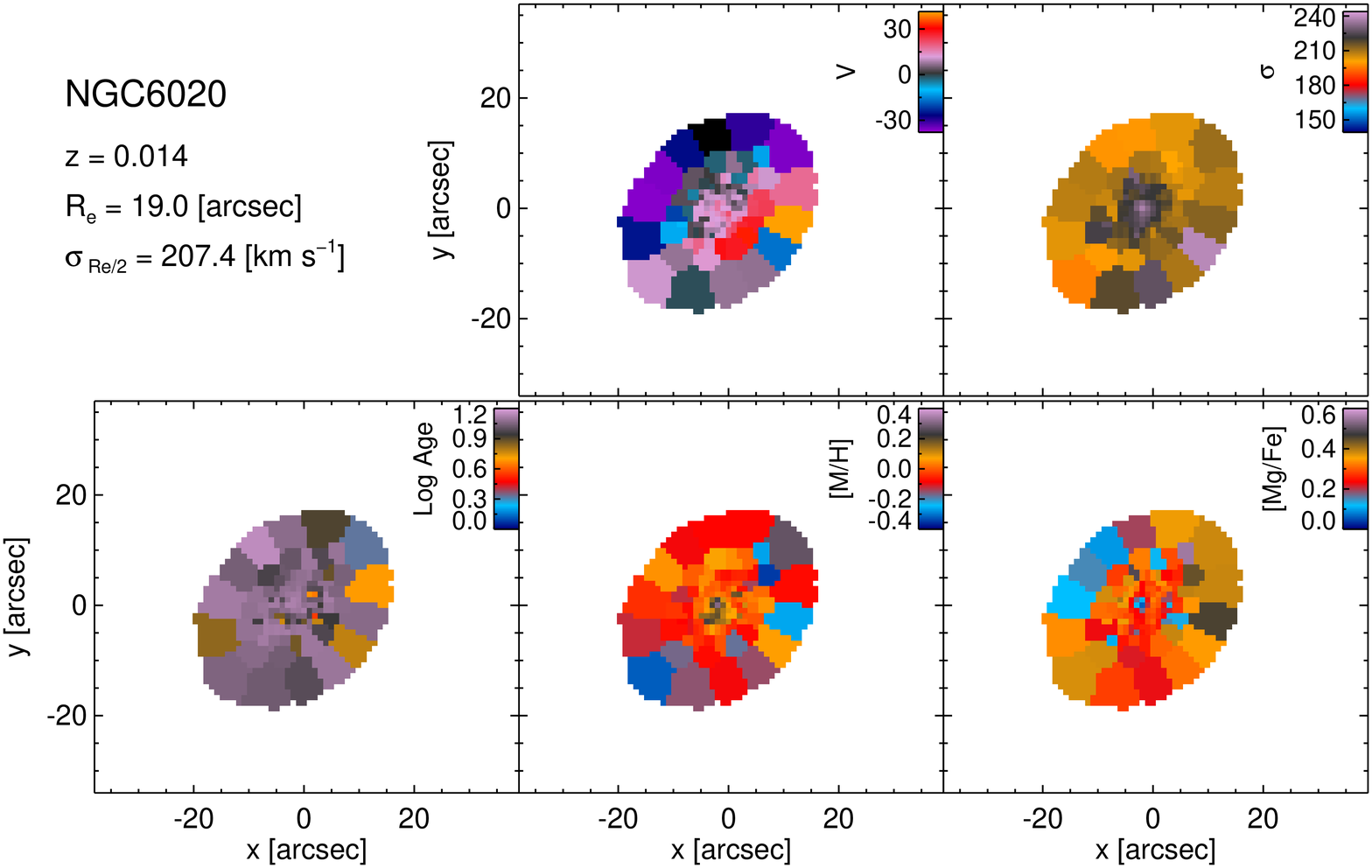}
\end{center}
\end{figure*}

\begin{figure*}
\begin{center}
\includegraphics[width=12cm]{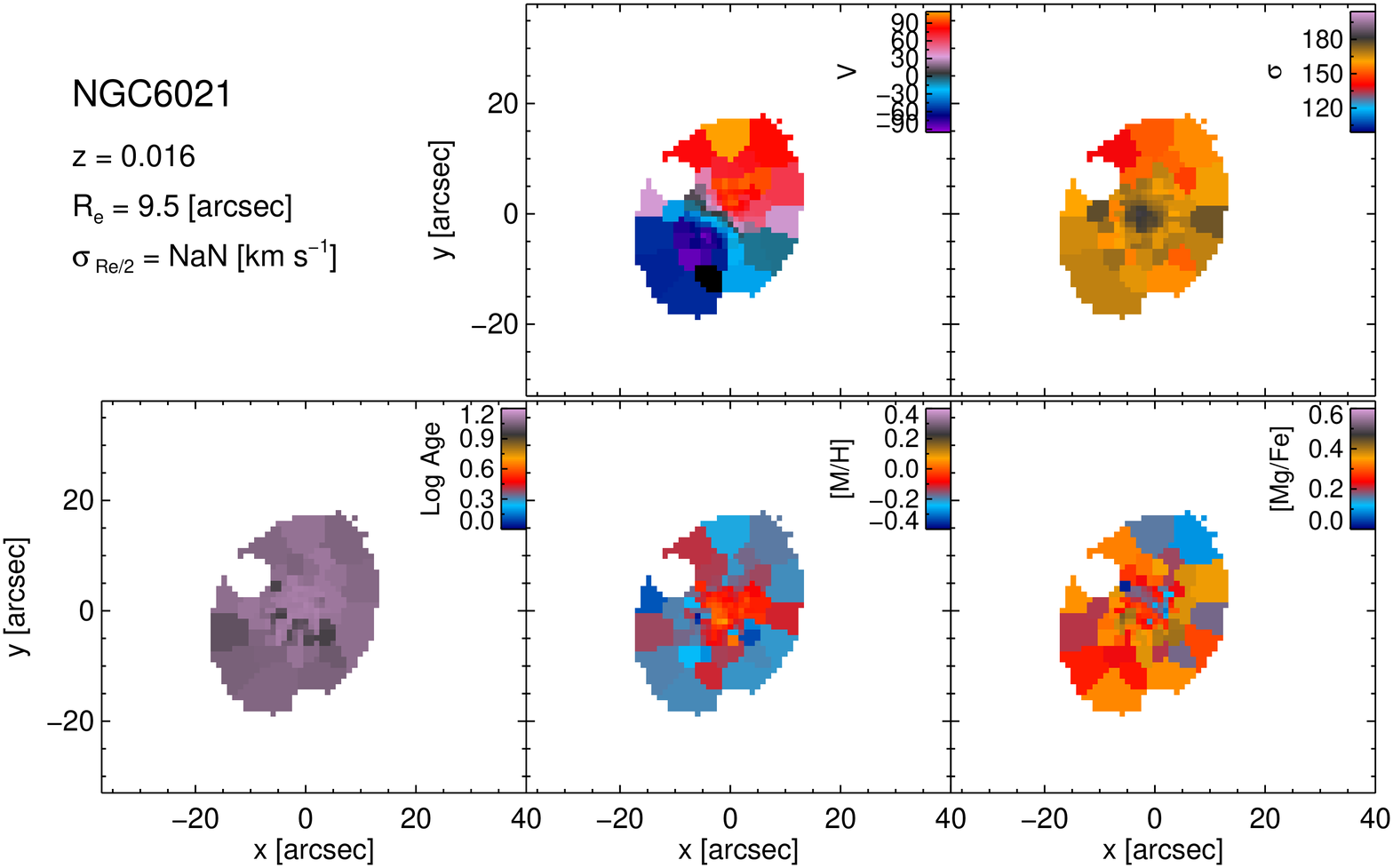}
\includegraphics[width=12cm]{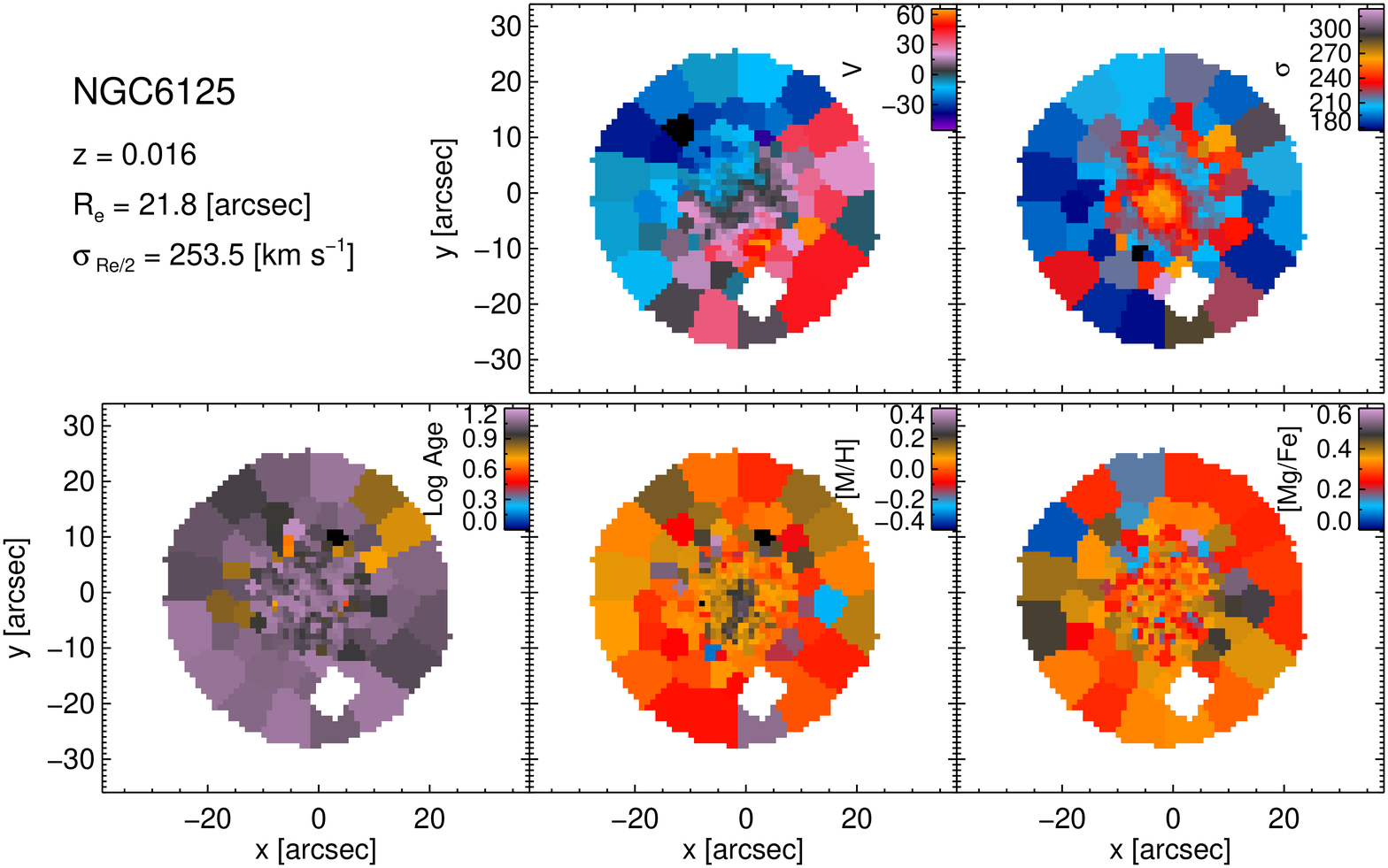}
\includegraphics[width=12cm]{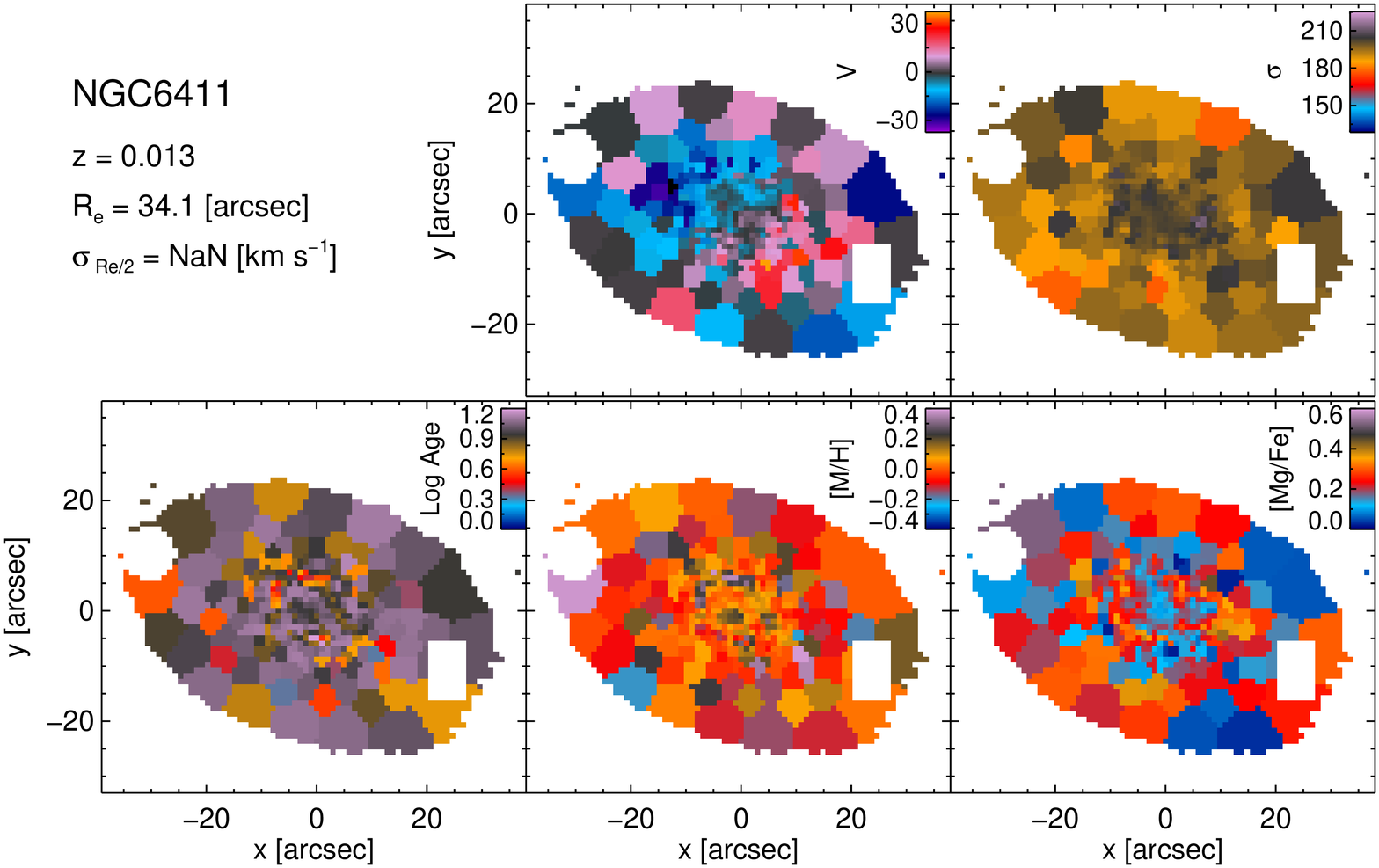}
\end{center}
\end{figure*}

\begin{figure*}
\begin{center}
\includegraphics[width=12cm]{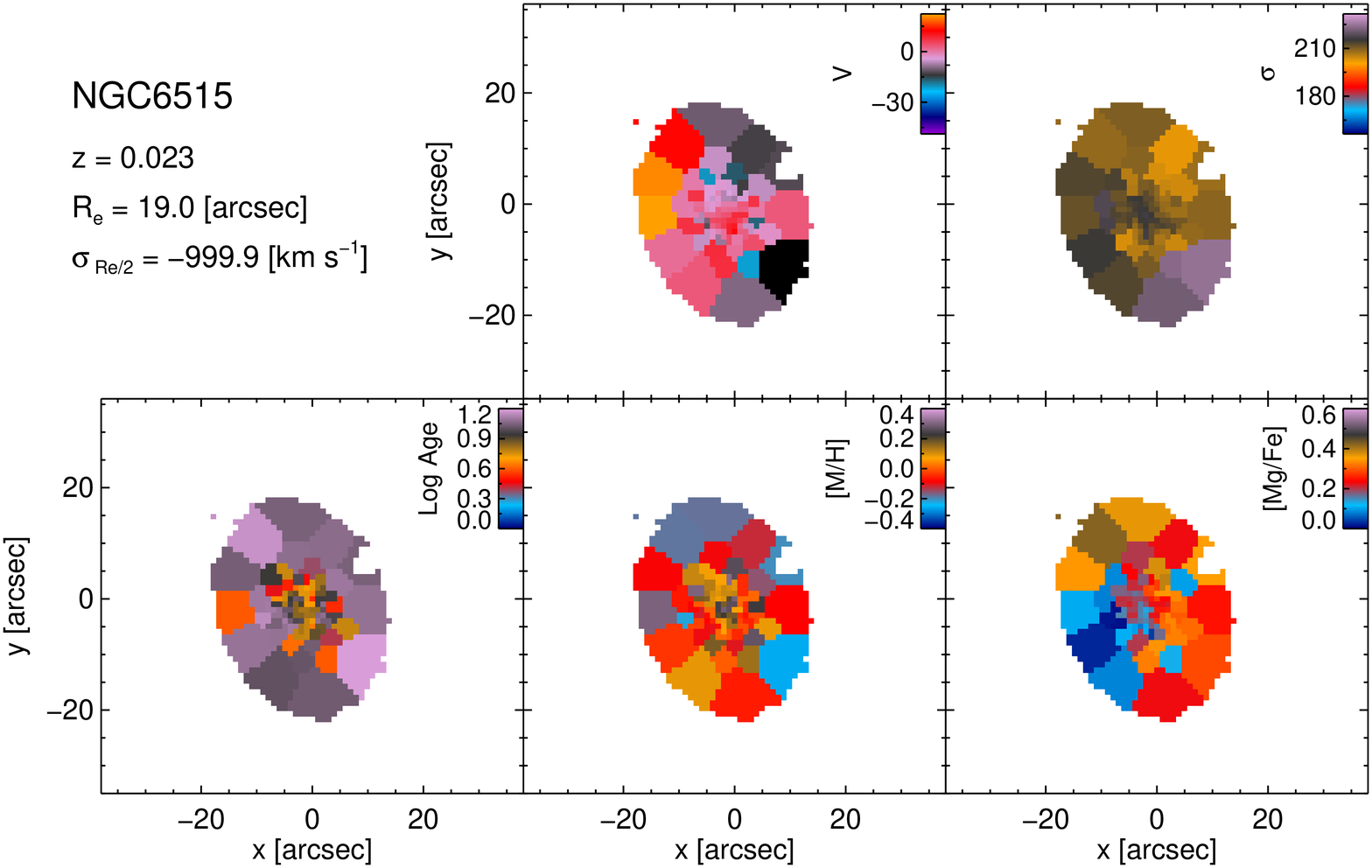}
\includegraphics[width=12cm]{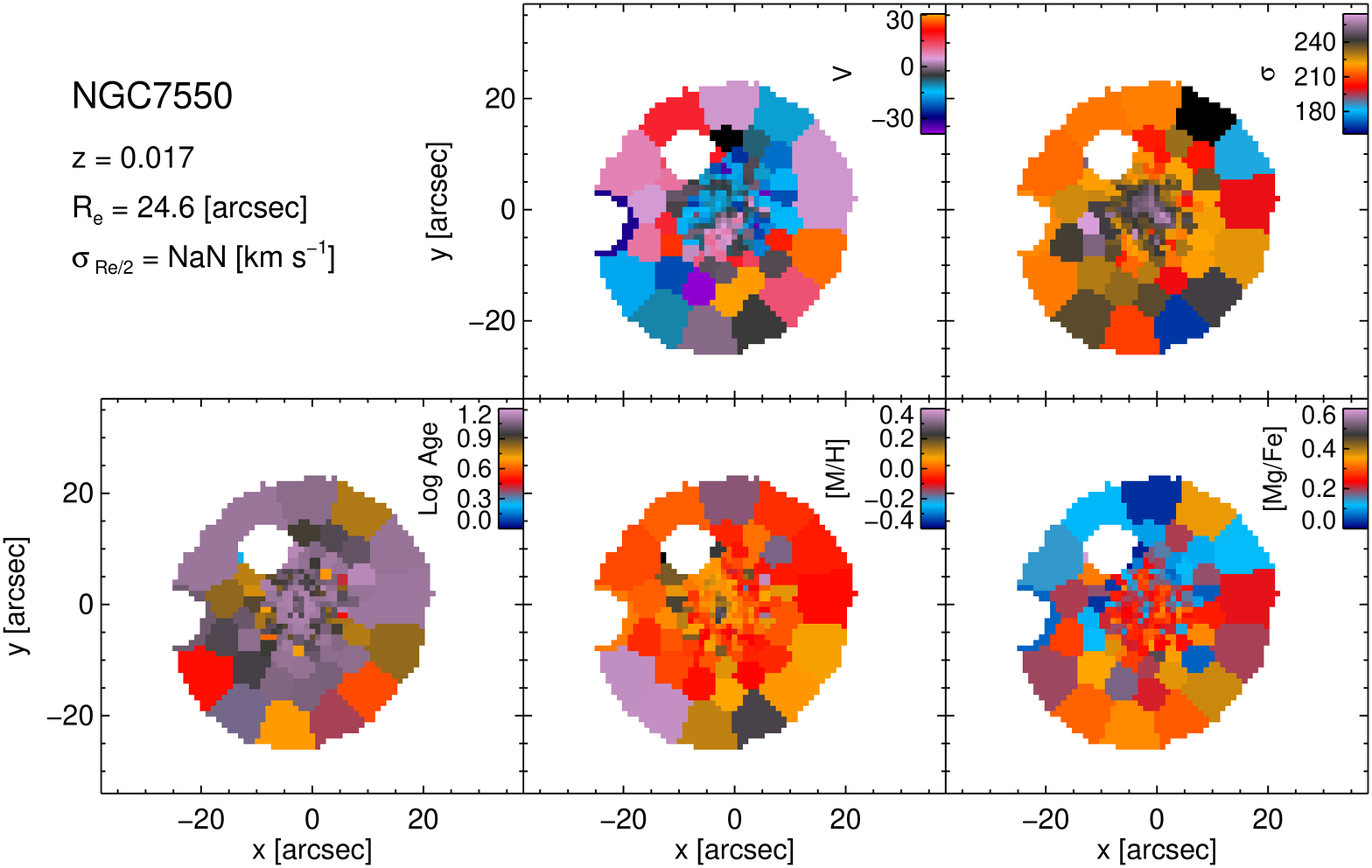}
\includegraphics[width=12cm]{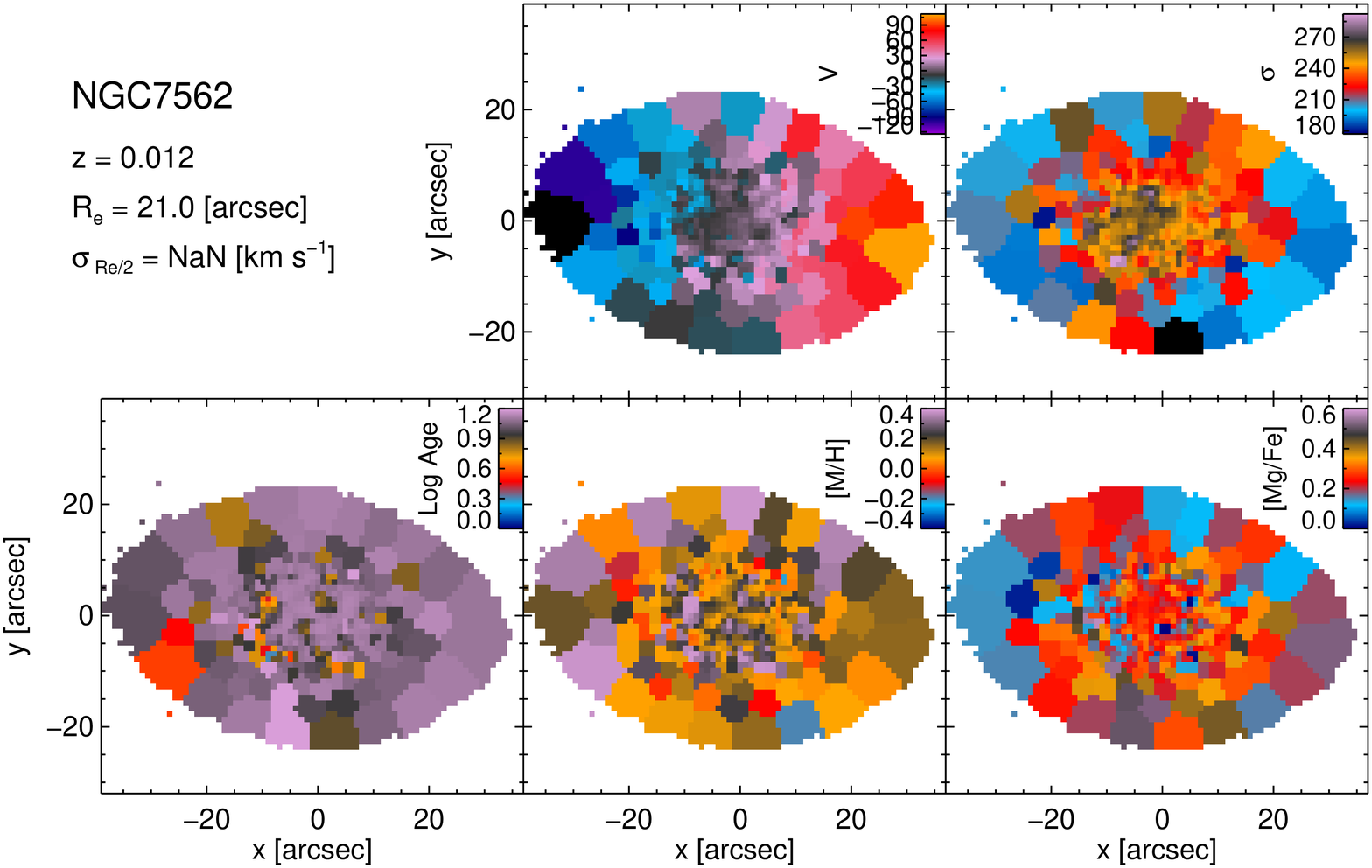}
\end{center}
\end{figure*}

\begin{figure*}
\begin{center}
\includegraphics[width=11.6cm]{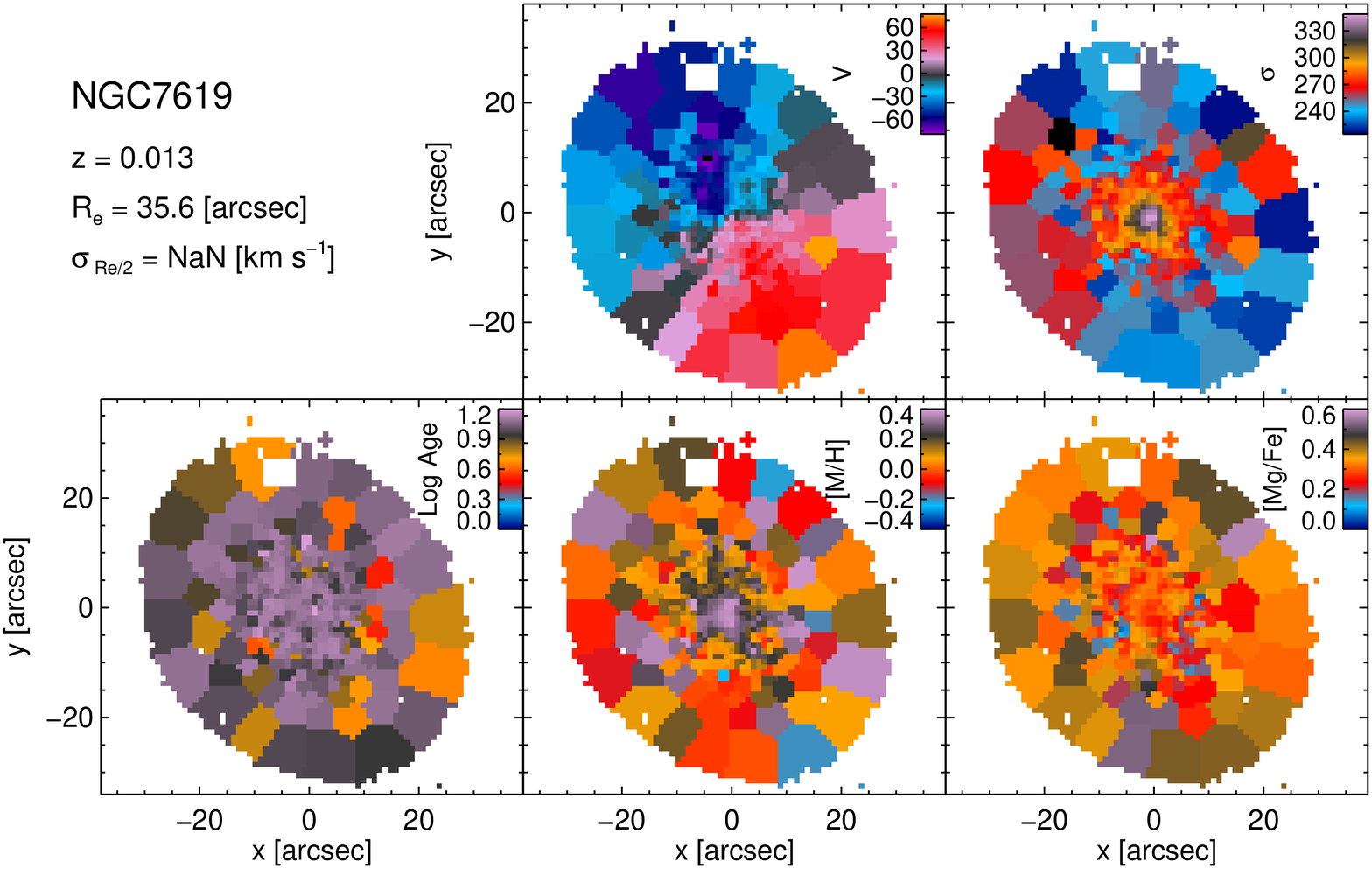}
\includegraphics[width=11.6cm]{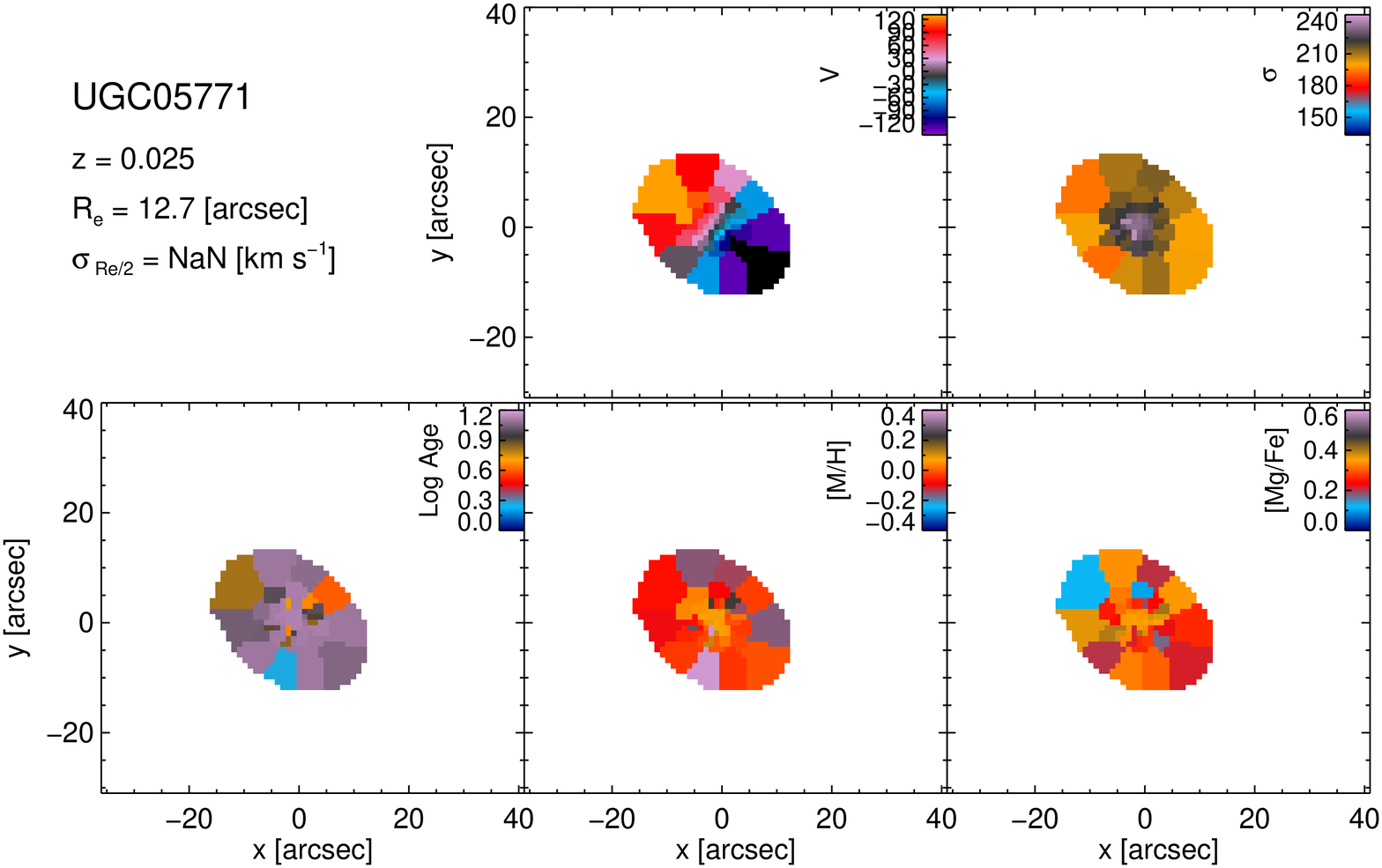}
\includegraphics[width=11.6cm]{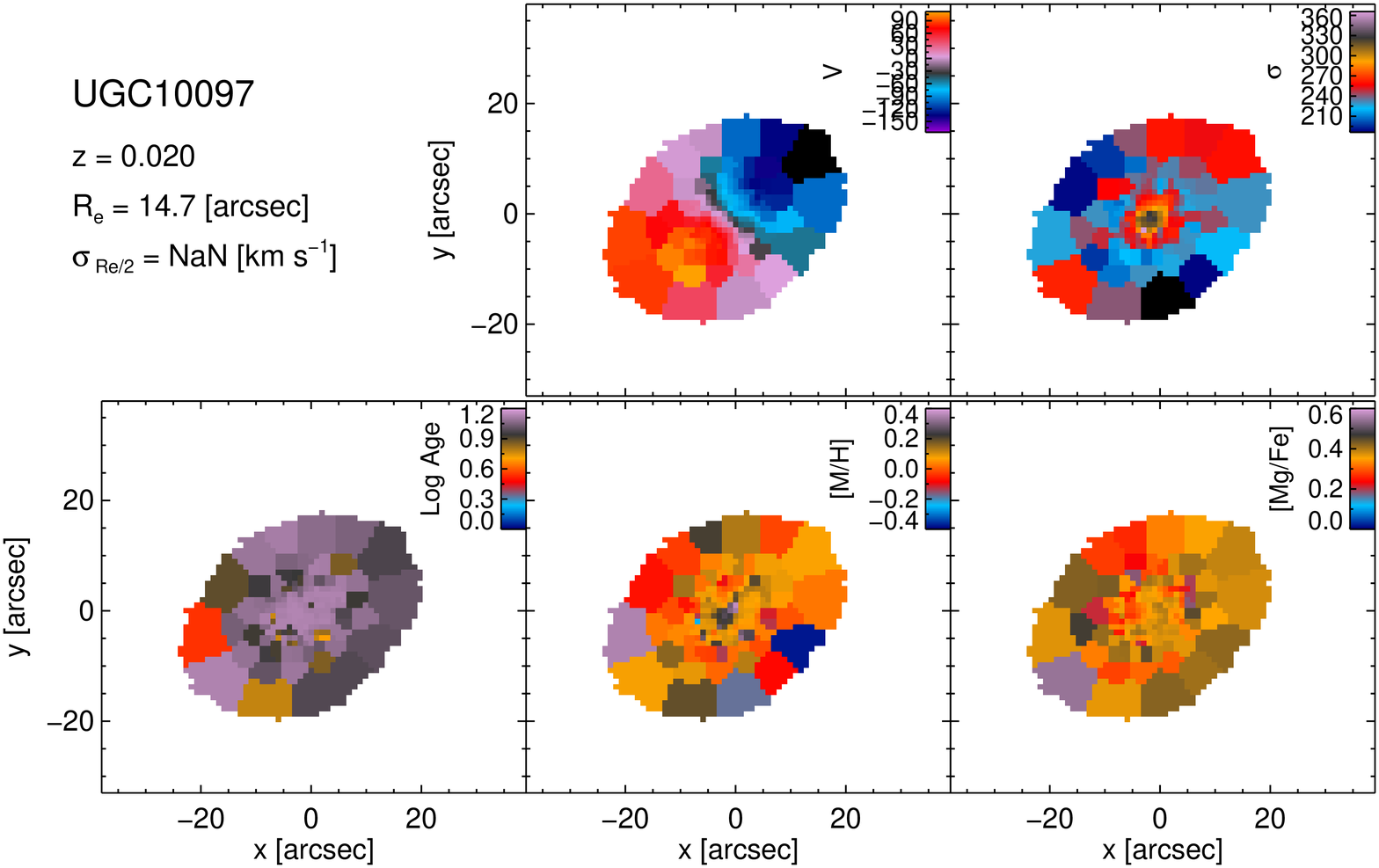}
\end{center}
\caption{Kinematical and stellar population parameters maps for the CALIFA sample}
\end{figure*}


\bsp	
\label{lastpage}
\end{document}